\pdfoutput=1  % serve per imporgli di andare in pdflatex, utile quando ci sono figure grosse in .eps, e arXiv chiede di convertirle in .pdf per farlo %compilare in pdflatex
\def\kl{\ton{\bds{\hat{S}}\bds\cdot\bds{\hat{l}}}}
\def\km{\ton{\bds{\hat{S}}\bds\cdot\bds{\hat{m}}}}
\def\kk{\ton{\bds{\hat{S}}\bds\cdot\bds{\hat{k}}}}
\def\kP{\ton{\bds{\hat{S}}\bds\cdot\bds{\hat{P}}}}
\def\kQ{\ton{\bds{\hat{S}}\bds\cdot\bds{\hat{Q}}}}
\def\gm{\textrm{gm}}

\def\masy{~\textrm{mas~yr}^{-1}}

\def\deg{~\textrm{deg}}

\def\kx{{\hat{S}}_x}
\def\ky{{\hat{S}}_y}
\def\kz{{\hat{S}}_z}

\def\nk{n_{\rm b}}

\def\Pb{P_{\rm b}}

\def\rfr#1{Equation~(\ref{#1})}
\def\rfrs#1#2{Equations~(\ref{#1})~to~(\ref{#2})}

\def\virg#1{``#1"}

\def\eqi{\begin{equation}}
\def\eqf{\end{equation}}
\def\eqia{\begin{eqnarray}}
\def\eqfa{\end{eqnarray}}

\def\rp#1#2{{#1\over#2}}
\def\lb#1{\label{#1}}

\def\bds#1{\boldsymbol{#1}}
\def\cO{\cos\Omega}
\def\sO{\sin\Omega}

\def\cI{\cos I}
\def\sI{\sin I}

\def\ton#1{\left(#1\right)}
\def\qua#1{\left[#1\right]}
\def\grf#1{\left\{#1\right\}}
\def\ang#1{\left\langle #1\right\rangle}
\documentclass[onecolumn]{aastex}
\usepackage{morefloats}
\usepackage[title]{appendix}
\usepackage{hyperref}
\usepackage{booktabs}
\usepackage[table,xcdraw]{xcolor}
\usepackage{multirow}
\usepackage{rotating,tabularx}
\usepackage{float}
\usepackage{enumerate}
\usepackage[polutonikogreek,english]{babel}
\usepackage{amsmath,starfont,textgreek,w-greek,wasysym}
\usepackage{amsthm}
\usepackage{amscd,lineno}
\usepackage{amssymb,dsfont}
\usepackage{graphicx,epsfig}
\usepackage{txfonts}
\usepackage{xr-hyper}
\RequirePackage{color}

\newcommand{\emaila}{lorenzo.iorio@libero.it}

\allowdisplaybreaks[1]

\begin{document}

\title{The post-Newtonian gravitomagnetic spin-octupole moment of an oblate rotating body and its effects on an orbiting test particle; are they measurable in the Solar system?}

\shortauthors{L. Iorio}

\author{Lorenzo Iorio\altaffilmark{1} }
\affil{Ministero dell'Istruzione, dell'Universit\`{a} e della Ricerca
(M.I.U.R.)-Istruzione
\\ Permanent address for correspondence: Viale Unit\`{a} di Italia 68, 70125, Bari (BA),
Italy}

\email{\emaila}

\begin{abstract}
We analytically work out the orbital effects induced by the \textcolor{black}{post-Newtonian} gravitomagnetic spin-octupole moment of an extended spheroidal rotating body endowed with angular momentum $\bds S$ and quadrupole mass moment $J_2$. Our results, proportional to $\textcolor{black}{G}S J_2 c^{-2}$,  hold for an arbitrary orientation of the body's symmetry axis $\bds{\hat{S}}$ and a generic orbital configuration of the test particle. Such effects may be measurable, in principle, with a dedicated spacecraft-based mission to Jupiter. For a moderately eccentric and fast path, the gravitomagnetic precessions of the node and the pericentre of a dedicated orbiter could be as large as $400~\textrm{milliarcseconds~per~year}$ or even $1,600-4,000~\textrm{milliarcseconds~per~year}$ depending on the orientation of its orbital plane in space. Numerical simulations of the Earth-probe range-rate signal confirm such expectations since its magnitude reaches the $\simeq 0.03-0.3~\textrm{millimetre~per~second}$ level after just 1 day. The precision of the current two-way Ka-band Doppler measurements of the spacecraft Juno, presently orbiting Jupiter, amounts to $\simeq 0.003~\textrm{millimetre~per~second}$ after $1,000$ seconds. \textcolor{black}{O}ther general relativistic effects \textcolor{black}{might} be measurable, including also those proportional to $GMJ_2c^{-2}$, never put to the test so far. Most of the competing Newtonian signals due to the classical multipoles of the planet's gravity field have quite different temporal signatures with respect to the post-Newtonian ones, making, thus, potentially easier disentangling them.
%Instead, the aforementioned effects for Juno, currently orbiting Jupiter, are too small amounting to $\simeq 0.2~\textrm{milliarcseconds~per~year}$.
\end{abstract}

keywords{
gravitation $-$ celestial mechanics $-$ space vehicles $-$ planets and satellites: individual: Jupiter
}

%keywords{Astrophysical studies of gravity; General relativity; Cosmological constant; Neutron stars \& pulsars; Classical black holes}
%

\section{Introduction}\lb{intro}
The Einstein's General Theory of Relativity (see, e.g., \citet{2015Univ....1...38I} and references therein) is currently the best description of the gravitational interaction at our disposal. It has successfully passed all the experimental and observational checks with which it has been put to the test so far \citep{2014LRR....17....4W} at different scales ranging from the Earth's surrounding \citep{2011PhRvL.106v1101E} and our Solar System \citep{Nordvedt2001} to extragalactic realms \citep{2018Sci...360.1342C}, including also compact stellar corpses \citep{2018IAUS..337..128K,2018Natur.559...73A} and the main sequence stars orbiting the supermassive black hole in our Galaxy \citep{2018A&A...615L..15G}, not to mention the recent direct discovery of the gravitational waves with Earth-based laser interferometers \citep{2016Univ....2...22C}. None the less, the still unexplained issues of the dark matter in galaxies and clusters of galaxies along with the observed accelerated expansion of the Universe may pose challenges to it \citep{2016Univ....2...23D,2016Univ....2...11V}.

Given its nature of fundamental pillar of our knowledge of the natural world, it is of the utmost importance to always submit under empirical scrutiny new parts of the theoretical structure of general relativity even where violations are, perhaps, least expected, as in the weak-field and slow-motion regime. \citet{Ginz59} wrote: \virg{[\ldots] the history of physics has seen no end of cases in which the certain has turned out to be false. A theory so fundamental to modern science must be rigorously verified if it is to be applied with complete confidence to the further development of cosmology and other areas of physics.}. To this aim, in this paper we will show that it may be possible, at least in principle, to bring  an aspect of the post-Newtonian approximation \citep{2014grav.book.....P} which has never been tested so far into the detectability domain within the Solar System.
%According to \citet{Ginz59}, \virg{The minuteness of all of the effects involved in the experimental verification of general relativity may perhaps cast doubt %upon [\ldots] the value of the effort required to verify it. [\ldots] the value of the theory can in no way be \virg{measured} by the magnitude of any %effects, especially those that can be observed within the confines of our solar system. [\ldots] The minuteness of the general relativity effects within our %tiny solar system should never lead us to regard the theory as trivial in the study of the universe}.

In the first post-Newtonian approximation, the metric tensor $g_{\sigma\nu},~\sigma,\nu = 0,1,2,3$ describing the gravitational field generated by a given mass-energy distribution made of a system of $N$ gravitationally interacting rotating bodies of arbitrary shape and composition is parameterized in terms of the so-called gravitoelectric potential $\phi$, which is a generalization of the Newtonian potential $U$, denoted sometimes by $w$, and the gravitomagnetic vector potential $\mathbf{w}$ \citep{1989NCimB.103...63B,1991PhRvD..43.3273D,2003AJ....126.2687S}. The latter one, on which we will focus, is  generated by matter current densities proportional to the off-diagonal components $T^{0j},~j=1,2,3$ of the energy-momentum tensor $T^{\sigma\nu},~\sigma,\nu = 0,1,2,3$ of the source.
Outside any body of the system, the gravitoelectric potentials admit multipole expansions in terms of certain  mass and spin multipole moments \citep{1989AIHPA..50..377B}; see, e.g., \citet{1998CQGra..15.1971B} and references therein for their role and importance in several branches of general relativity like gravitational waves.

Let us consider a single isolated extended rotating body at rest as source of the gravitational field.
The gravitomagnetic acceleration experienced by a test particle orbiting it is \citep{1994PhRvD..49..618D,2015CeMDA.123....1M}
\eqi
{\bds A}_\gm = \rp{\bds v}{c^2}\bds\times{\bds B}_\gm.\lb{accel}
\eqf
In the empty space outside the spinning body, its gravitomagnetic field ${\bds B}_\gm$ can be conveniently expressed in terms of a gravitomagnetic potential function $\phi_\gm$ as
\citep[Eq.~(30)]{2014CQGra..31x5012P}
\eqi
{\bds B}_\gm = -4\bds\nabla\bds\times\mathbf{w}= -{\bds\nabla}\phi_\gm.
\eqf
By assuming a uniformly rotating homogeneous oblate spheroid at rest, $\phi_\gm$ can be expanded in terms of its spin-multipole moments as
\citep[Eqs.~(31)-(32)]{2014CQGra..31x5012P}
\eqi
\phi_\gm = -\rp{30 G S}{r^2}\sum_{i=0}^{\infty}\rp{\ton{-1}^i}{\ton{2i + 3}\ton{2i + 5}}\ton{\rp{R_\textrm{e}\varepsilon}{r}}^{2i}P_{2i + 1}\ton{\xi}=-\rp{GS}{r^2}\qua{2\xi -\rp{6}{7}\ton{\rp{R_\textrm{e}\varepsilon}{r}}^2 P_3\ton{\xi} + \ldots }.\lb{phigm}
\eqf
According to \citet[Eq.~(27)]{2014CQGra..31x5012P}, the relation connecting the body's ellipticity $\varepsilon$ with the Newtonian even zonal harmonics
is
\eqi
J_{2i} = \rp{3\ton{-1}^i}{\ton{2i + 1}\ton{2i + 3}}\varepsilon^{2i},
\eqf
so that, for $i=1$, one has
\eqi
J_2 = -\rp{1}{5}\varepsilon^2.\lb{stronza}
\eqf
The term with $i = 0$ in \rfr{phigm} corresponds to the spin-dipole moment, and yields the usual Lense-Thirring orbital precessions \citep{1918PhyZ...19..156L,2012GReGr..44..719I} which have been the subject of intense experimental scrutiny so far; see, e.g, \citet{2013CEJPh..11..531R} and references therein.

Here, we will explicitly calculate the direct orbital effects arising from \rfr{accel} evaluated for the spin-octupole moment in \rfr{phigm}, corresponding to $i=1$. Then, we will show that, in principle, they could be detectable with a dedicated spacecraft-based mission to Jupiter.
To the present author's best knowledge, should his proposal be eventually successful, it would be the first time that a general relativistic higher spin multipole moment would be measured.

The paper is organized as follows. In Section~\ref{precess}, we analytically work out the rates of change, averaged over one orbital period, of the Keplerian orbital elements of a test particle affected by the post-Newtonian acceleration imparted by the gravitomagnetic spin-octupole moment of its primary, assumed uniformly rotating, homogeneous and spheroidal in shape. We treat it perturbatively by using the standard decomposition in radial, transverse and normal components and the Gauss equations for the variation of the osculating Keplerian orbital elements. We restrict a priori neither to any peculiar spatial orientation of $\bds{\hat{S}}$ nor to particular orbital configurations of the orbiter. We subsequently confirm the resulting analytical results by numerically integrating the equations of motion. The perspectives for measuring such effects around Jupiter, which is the fastest spinning and most oblate major body of the Solar System, are treated in Section~\ref{expe}. While Juno (Section~\ref{Juno}), currently orbiting the gaseous giant along a 53-day, highly eccentric orbit, is unsuitable because its expected post-Newtonian effects are too small, a hypothetical new Jovian probe, provisionally dubbed, with a touch of irony, IORIO (In-Orbit Relativity Iuppiter\footnote{\textit{Iupp\u{\i}t\u{e}r} is one of the forms of the Latin noun of the god Jupiter.} Observatory, or IOvis\footnote{\textit{I\u{o}vis} means \virg{of Jupiter} in Latin. } Relativity In-orbit Observatory), moving along a much faster, moderately eccentric orbit could, in principle, be successfully used (Section~\ref{IORIO}). Numerical simulations of the Earth-spacecraft range-rate measurements, which are the actual observable quantities in such an astronomical scenario, preliminarily confirm such expectations. We investigate also other general relativistic features of motion impacting the probe's range-rate. In Section~\ref{multipoles}, we assess the consequences of the mismodeling in the Newtonian potential coefficients of Jupiter, while Section~\ref{poleposition} is devoted to the impact of the uncertainty in the Jovian spin axis orientation.
For each of such sources of systematic errors, we quantitatively evaluate the level of improvement with respect to their current accuracies required to bring their range-rate signatures at least to the same level of the various post-Newtonian signatures of interest. Furthermore, we look also at the temporal patterns of the competing Newtonian signals with respect to the relativistic ones.
We summarize our findings and offer our conclusions in Section~\ref{conclu}.
For the benefit of the reader, Appendix~\ref{appena} contains a list of symbols and definitions of the quantities used throughout the text, while tables and figures are grouped in Appendix~\ref{appenb}. Finally, it is appropriate to note that the present work should be regarded just as a concept study preliminarily investigating to a certain level of detail a scenario which may be potentially able to measure the investigated effect, not as a formal mission proposal.
\section{The long-term orbital precessions}\lb{precess}
For the sake of simplicity, let us, first, assume a coordinate system whose fundamental $\grf{x,~y}$ plane coincides with the body's equator, so that its symmetry axis $\bds{\hat{S}}$ is aligned with the reference $z$ axis.
The long-term rates of change of the osculating Keplerian orbital elements of the test particle, obtained by averaging   over one orbital revolution the right-hand-sides of the standard Gauss equations \citep{2011rcms.book.....K,2014grav.book.....P} evaluated onto the Keplerian ellipse as reference trajectory, turn out to be
\begin{align}
\dot a_\gm & = 0\lb{adot}, \\ \nonumber \\
\dot e_\gm & = \rp{225 e G S R^2 J_2\cos I\sin^2 I\sin 2\omega}{28 c^2 a^5\ton{1-e^2}^{5/2}}\lb{edot}, \\ \nonumber \\
\dot I_\gm & = -\rp{225 e^2 G S R^2 J_2\cos^2 I\sin I\sin 2\omega}{28 c^2 a^5\ton{1-e^2}^{7/2}}\lb{Idot}, \\ \nonumber \\
\dot\Omega_\gm & = \rp{45  G S R^2 J_2}{112 c^2 a^5\ton{1-e^2}^{7/2}}\qua{-2 \ton{2 + 3 e^2}\ton{3 + 5 \cos 2I} +
 5 e^2 \ton{1 + 3 \cos 2I} \cos 2\omega}\lb{Odot}, \\ \nonumber \\
\dot\omega_\gm & = -\rp{225  G S R^2 J_2\cos I}{112 c^2 a^5\ton{1-e^2}^{7/2}}\grf{ -2 e^2 - 2 \ton{8 + 7 e^2} \cos 2 I
+ \qua{-2 - 3 e^2 + \ton{2 + 7 e^2} \cos 2I} \cos 2\omega }.\lb{odot}
\end{align}
For the sake of simplicity, here and in the following we omit the brackets $\ang{\ldots}$ denoting the average over one orbital period.

Let us, now, remove the limitation on the orientation of the primary's spin axis allowing it to be arbitrarily directed in space. The  resulting long-term rates of change of the Keplerian orbital elements are
\begin{align}
\dot a_\gm &=0,\lb{dota} \\ \nonumber \\
\dot e_\gm &= \rp{225 e G J_2 R^2 S}{14 a^5 c^2 \ton{1 - e^2}^{5/2}} \mathcal{E}\ton{I,~\Omega,~\omega;~\bds{\hat{S}}},\lb{dote} \\ \nonumber \\
\dot I_\gm  &= \rp{45 G J_2 R^2 S}{224 a^5 c^2 \ton{1 - e^2}^{7/2}}\mathcal{I}\ton{I,~\Omega,~\omega;~\bds{\hat{S}}},\lb{dotI} \\ \nonumber \\
\dot \Omega_\gm &= \rp{45 G J_2 R^2 S}{56 a^5 c^2 \ton{1 - e^2}^{7/2}} \mathcal{N}\ton{I,~\Omega,~\omega;~\bds{\hat{S}}},\lb{dotO} \\ \nonumber \\
\dot\omega_\gm &= \rp{45 G J_2 R^2 S}{56 c^2 a^5\ton{1 - e^2}^{7/2}}\mathcal{P}\ton{I,~\Omega,~\omega;~\bds{\hat{S}}}.\lb{doto}
\end{align}
with
\begin{align}
\mathcal{E} & = \kk\kP\kQ,\lb{Ecoef}\\ \nonumber\\
\mathcal{I} \nonumber &= 5 e^2 \cos 3I \sin 2 \omega \ton{\ky \cos\Omega -\kx \sin\Omega} \qua{1 -7 \kz^2 + \ton{-\kx^2 + \ky^2} \cos 2 \Omega - 2 \kx \ky \sin 2 \Omega} -\\ \nonumber \\
\nonumber & -5 \cos 2I \ton{4 + 6 e^2 +e^2 \cos 2 \omega}\kl\qua{-1 + 3 \kz^2 + \ton{\kx^2 - \ky^2} \cos 2 \Omega + 2 \kx \ky \sin 2 \Omega} - \\ \nonumber \\
\nonumber &- 5 e^2 \kz \sin 3I \sin 2 \omega \qua{-3 + 5 \kz^2 + 3 \ton{\kx^2 - \ky^2} \cos 2 \Omega + 6 \kx \ky \sin 2 \Omega} + \\ \nonumber \\
\nonumber & + 5 e^2 \kz \sin I \sin 2 \omega \qua{3 -5 \kz^2 + 5 \ton{\kx^2 - \ky^2}\cos 2 \Omega + 10 \kx \ky \sin 2 \Omega} - \\ \nonumber \\
\nonumber &- 5 e^2 \cos I \sin 2 \omega \ton{-\ky \cos\Omega + \kx \sin\Omega}\qua{3 - 5 \kz^2 + 5 \ton{\kx^2 - \ky^2} \cos 2 \Omega + 10 \kx \ky \sin 2 \Omega} - \\ \nonumber \\
\nonumber &- 10 \kz \ton{4 + 6 e^2 + e^2 \cos 2 \omega} \sin 2I \qua{-2 \kx \ky \cos 2 \Omega + \ton{\kx^2 - \ky^2} \sin 2 \Omega} + \\ \nonumber \\
\nonumber & + \kl \grf{5 e^2 \cos 2 \omega \qua{-1 - 5 \kz^2 + 5 \ton{\kx^2 - \ky^2} \cos 2 \Omega + 10 \kx \ky \sin 2 \Omega} + \right.\\ \nonumber \\
&\left. + 2 \ton{2 + 3 e^2} \qua{-1 -5 \kz^2 + 5 \ton{\kx^2 - \ky^2} \cos 2 \Omega + 10 \kx \ky \sin 2 \Omega}},\lb{Icoef} \\ \nonumber \\
\mathcal{N} \nonumber & =  -5 \cos^2 I \cot I\ton{-4 - 6 e^2 + 3 e^2 \cos 2 \omega} \ton{\ky \cos\Omega - \kx \sin\Omega}^3 + \\ \nonumber \\
\nonumber & + 5 \cos I \cot I \ton{\ky \cos\Omega - \kx \sin\Omega}^2\qua{3 \kz \ton{4 + 6 e^2 - 3 e^2 \cos 2 \omega} \sin I + 4 e^2 \kl\sin 2 \omega } + \\ \nonumber \\
\nonumber & + 10 e^2 \csc I \sin 2\omega \kl \qua{2 \kz^2 \sin^2 I + \kl^2} + \\ \nonumber \\
\nonumber & + 5 \kz \cos^2\omega \qua{\ton{4 + 5 e^2} \kx^2 \cos^2\Omega + \ton{4 + 3 e^2} \kz^2 \sin^2 I + \ton{4 + 5 e^2} \ky \ton{\kx \sin 2\Omega + \ky \sin^2\Omega}} + \\ \nonumber \\
\nonumber & + 5 \kz \sin^2\omega \qua{\ton{4 + 7 e^2} \kx^2 \cos^2\Omega + \ton{4 + 9 e^2} \kz^2 \sin^2 I  + \ton{4 + 7 e^2} \ky \ton{\kx \sin 2\Omega + \ky \sin^2\Omega}} - \\ \nonumber \\
\nonumber & - 2 \csc I \grf{8 \kz \sin I - \cos I \ton{-8 - 12 e^2 + 5 e^2 \cos 2 \omega} \ton{\ky \cos\Omega - \kx \sin\Omega} + \right.\\ \nonumber \\
\nonumber & \left. + e^2 \qua{\kz \sin I\ton{12 - 5 \cos 2 \omega}  + 5\kl \sin 2 \omega }} + \\ \nonumber \\
\nonumber & + 5 \cot I \ton{\ky \cos\Omega - \kx \sin\Omega} \grf{8 e^2 \kz \kl \sin I \sin 2 \omega  + \right.\\ \nonumber \\
\nonumber &\left. + \cos^2\omega \qua{\ton{4 + 5 e^2} \kx^2 \cos^2\Omega + 3 \ton{4 + 3 e^2} \kz^2 \sin^2 I + \ton{4 + 5 e^2} \ky \ton{\kx \sin 2\Omega + \ky \sin^2\Omega}} + \right. \\ \nonumber \\
&\left. + \sin^2\omega\qua{\ton{4 + 7 e^2} \kx^2 \cos^2\Omega + 3 \ton{4 + 9 e^2} \kz^2 \sin^2 I + \ton{4 + 7 e^2} \ky \ton{\kx \sin 2\Omega + \ky \sin^2\Omega}}},\lb{Ocoef}\\ \nonumber \\
\mathcal{P} \nonumber & =
-40 e^2 \kz \kl\cos^2 I \sin 2\omega \ton{\ky \cos\Omega -\kx \sin\Omega}  - 20 \ton{1 + 2 e^2} \kl \km \kk\sin 2\omega  -\\ \nonumber \\
\nonumber &- 5 \cos^2 I \cot I \ton{\ky \cos\Omega - \kx \sin\Omega}^2 \qua{3 \kz \ton{4 + 6 e^2 - 3 e^2 \cos 2\omega} \sin I + 4 e^2 \kl\sin 2\omega } - \\ \nonumber \\
\nonumber &- 10 e^2 \cot I \sin 2\omega \kl \ton{2 \kz^2 \sin^2 I + \kl^2} -5 \cos I \cot I\cos^2 \omega  \ton{\ky \cos\Omega -\kx \sin\Omega}\times\\ \nonumber \\
\nonumber &\times \qua{\ton{4 +5 e^2} \kx^2 \cos^2 \Omega +3 \ton{4 + 3 e^2} \kz^2 \sin^2 I + \ton{4 +5 e^2} \ky \sin\Omega \ton{2 \kx \cos\Omega +\ky \sin\Omega}} -\\ \nonumber \\
\nonumber &- 5 \kz \cos I \sin^2 \omega \qua{\ton{4 +7 e^2} \kx^2 \cos^2 \Omega + \right.\\ \nonumber\\
\nonumber &\left. + \ton{4 + 9 e^2} \kz^2 \sin^2 I + \ton{4 +7 e^2} \ky\ton{\kx \sin 2\Omega +\ky \sin^2\Omega}} -\\ \nonumber \\
\nonumber &- 5 \cos I \cot I \sin^2 \omega \ton{\ky \cos\Omega -\kx \sin\Omega} \qua{\ton{4 +7 e^2} \kx^2 \cos^2 \Omega +3 \ton{4 + 9 e^2} \kz^2 \sin^2 I +\right.\\ \nonumber\\
\nonumber &\left. + \ton{4 +7 e^2} \ky\ton{\kx \sin 2\Omega +\ky \sin^2\Omega}} -\\ \nonumber\\
\nonumber &- 10 \kk\cos^2 \omega  \grf{\qua{\ton{7 + 6 e^2} \kx^2 + \ton{5 + 2 e^2} \ky^2 \cos^2 I} \cos^2 \Omega + \ton{5 + 2 e^2} \kz^2 \sin^2 I +\right.\\ \nonumber \\
\nonumber &\left. + 2 \ky \cos\Omega \grf{\ton{5 + 2 e^2} \kz \cos I \sin I +\kx \qua{7 + 6 e^2 - \ton{5 + 2 e^2} \cos^2 I} \sin\Omega} +\right.\\ \nonumber \\
\nonumber &+\left. \sin\Omega \qua{-\ton{5 + 2 e^2} \kx \kz \sin 2I + \ton{\ton{7 + 6 e^2} \ky^2 + \ton{5 + 2 e^2} \kx^2 \cos^2 I} \sin\Omega}} -\\ \nonumber \\
\nonumber &- 10\kk \sin^2 \omega  \grf{\qua{\ton{5 + 2 e^2} \kx^2 + \ton{7 + 6 e^2} \ky^2 \cos^2 I} \cos^2 \Omega + \ton{7 + 6 e^2} \kz^2 \sin^2 I +\right.\\ \nonumber \\
\nonumber &+\left. 2 \ky \cos\Omega \grf{\ton{7 + 6 e^2} \kz \cos I \sin I +\kx \qua{5 + 2 e^2 - \ton{7 + 6 e^2} \cos^2 I} \sin\Omega} +\right.\\ \nonumber \\
\nonumber &+\left. \sin\Omega \qua{-\ton{7 + 6 e^2} \kx \kz \sin 2I + \ton{\ton{5 + 2 e^2} \ky^2 + \ton{7 + 6 e^2} \kx^2 \cos^2 I} \sin\Omega}} + \\ \nonumber \\
\nonumber & + 2 \csc I \grf{-\cos^2 I \ton{-8 - 12 e^2 +5 e^2 \cos 2\omega} \ton{\ky \cos\Omega -\kx \sin\Omega} + \right.\\ \nonumber \\
\nonumber & +\left. 4 \ton{3 + 2 e^2} \sin^2 I \ton{-\ky \cos\Omega + \kx \sin\Omega} + \right.\\ \nonumber \\
\nonumber & +\left. 5 \cos I \qua{\kz \ton{4 + 4 e^2 - e^2 \cos 2\omega} \sin I +e^2\kl \sin 2\omega }} -\\ \nonumber \\
\nonumber &- 5 \cot I \grf{-\cos^3 I \ton{-4 - 6 e^2 +3 e^2 \cos 2\omega} \ton{\ky \cos\Omega -\kx \sin\Omega}^3 + \right.\\ \nonumber \\
\nonumber &\left. + \kz \sin I \cos^2 \omega  \qua{\ton{4 + 5 e^2} \kx^2 \cos^2 \Omega + \ton{4 +3 e^2} \kz^2 \sin^2 I + \right.\right.\\ \nonumber \\
&\left.\left. + \ton{4 +5 e^2} \ky \sin\Omega \ton{2 \kx \cos\Omega +\ky \sin\Omega}}}.\lb{ocoef}
\end{align}
It can be noted that \rfrs{dota}{doto}, along with \rfrs{Ecoef}{ocoef}, reduce to \rfrs{adot}{odot} for $\kx=\ky=0,~\kz=1$.

Our analytical results are fully confirmed by a numerical integration of the equations of motions, as shown by Figure~\ref{fig0}. Indeed, the analytically computed annual shifts, calculated with \rfrs{dota}{ocoef} for an arbitrary orbital configuration referred to the Earth's mean equator at the epoch J2000.0 of a fictitious test particle orbiting a hypothetical primary with the same physical characteristic of Jupiter, agree with the numerically produced time series of the Keplerian orbital elements obtained by  integrating the equations of motion including the acceleration of \rfr{accel} evaluated with \rfr{phigm} for $i=1$.

Finally, we recall that orbital effects proportional to $GSJ_2c^{-2}$ \citep{2015IJMPD..2450067I} arise also from the interplay between the well known Newtonian quadrupolar acceleration due to $J_2$ and the post-Newtonian Lense-Thirring acceleration proportional to $GS c^{-2}$. Their order of magnitude is the same of the direct rates of change treated in the present Section. None the less, such indirect, mixed effects are likely unmeasurable in actual data reductions since they cannot be expressed in terms of a dedicated, solve-for scaling parameter which could be explicitly estimated. It is so because, contrary to the direct effects derived from \rfr{accel}, they do not come from a distinct acceleration which can be suitably parameterized.
\section{Perspectives of measuring the post-Newtonian gravitomagnetic orbital precessions due to the spin-octupole moment of Jupiter}\lb{expe}
\subsection{Juno}\lb{Juno}
The spacecraft Juno is currently orbiting Jupiter, whose relevant physical parameters are reported in Table~\ref{tavolaJup}, along a highly elliptical trajectory characterized by the orbital parameters listed in Table~\ref{tavolaJuno}.
The huge oblateness of the gaseous giant and the large eccentricity of the probe may suggest, at first sight, to look at such a system as a unique opportunity, in principle, to put to the test for the first time the gravitomagnetic effects due to the spin-octupole moment of an extended body. Unfortunately, the resulting orbital precessions of Juno turn out to be too small, as shown by Table~\ref{tavolaJuno} and Figure~\ref{fig1} displaying the simulated Earth-spacecraft range-rate signatures at the perijove passages PJ03, PJ06. In fact, the directly observable quantity of Juno used to mapping the Jovian gravity field is the two-way Ka-band Doppler shift. The frequent maneuvers required to keep the alignment of the transmitting antenna with the Earth tend to destroy the dynamical coherence of the orbit, not allowing to obtain steady time series of the spacecraft's orbital elements. Thus, the analytical calculation based on them should be regarded just as useful and easily understandable tools to perform a-priori sensitivity analyses. The same considerations hold, in principle, also for any other spacecraft orbiting Jupiter and communicating with the Earth. The signatures in Figure~\ref{fig1} were obtained as follows. For each perijove passes, we numerically integrated the equations of motion of the Earth, Jupiter and Juno in Cartesian rectangular coordinates referred to the \textcolor{black}{International Celestial Reference Frame (ICRF)} with and without the disturbing post-Newtonian acceleration  under investigation.
More specifically, in our simplified model the Earth is subjected to the Newtonian acceleration due to the Sun, while Jupiter feels only the Newtonian acceleration of the Sun; the equations of motions of both the planets were integrated in a Solar System barycentric coordinate system. The equations of motion of Juno were integrated in a Jovicentric coordinate system; they include the Newtonian accelerations of Jupiter and the Sun and the post-Newtonian acceleration of \rfr{accel}.
For each perijove passes, both the runs shared the same set of initial conditions which were retrieved from the WEB interface HORIZONS maintained by JPL, NASA, for given initial epochs which, in the present case, are December 11, 2016, h: 13:00 (PJ03) and May 19, 2017, h: 02:00 (PJ06), respectively. After each run, a numerical time series of the Earth-probe range-rate $\dot\rho(t)$ was produced by projecting the Juno's velocity vector onto the Earth-Jupiter unit vector; $\dot\rho_\textrm{pert}(t)$ includes also the effect of the perturbing gravitomagnetic acceleration, while $\dot\rho_\textrm{N}(t)$ is the purely classical one due to only the Newtonian monopoles of the Sun and Jupiter. In order to single out the effect of the post-Newtonian acceleration of interest, the differences of both the time series were computed obtaining the curves for $\Delta\dot\rho(t)=\dot\rho_\textrm{pert}(t) - \dot\rho_\textrm{N}(t)$ displayed in Figure~\ref{fig3}. Our method\footnote{In actual data reductions, the appropriate time and spatial coordinates transformations between the Solar System Barycentric coordinate system and the suitably constructed planetocentric coordinate systems for Jupiter and the Earth \citep{1989NCimB.103...63B} are fully modeled and implemented, among other things,  according to the most recent IAU resolutions \citep{2011rcms.book.....K}.}, which will be used also in Section~\ref{IORIO} for other Newtonian and post-Newtonian accelerations, was successfully tested by reproducing the Newtonian range-rate signatures due to the odd zonals $J_3,~J_5,~J_7,~J_9$ at PJ03, PJ06 displayed in  \citet[Extended Data Fig.~3]{2018Natur.555..220I}.
%It may be interesting to recall that the Period Reduction Maneuver, originally scheduled for October 19, 2016 in order to reduce the orbital period of the %spacecraft from\footnote{Juno entered orbit around Jupiter on July 5, 2016 along a 53-d trajectory (see the Object Data Page of Juno on the WEB interface %HORIZONS maintained by the JPL).} 52 d to 14 d, could not be implemented because of a technical problem with the helium valves which are important during main %engine burns \citep{2017juno}. By the way, also in that case the  precessions would have been likely too small, amounting to $\simeq 0.8\masy$ at most.
\subsection{A dedicated, new spacecraft}\lb{IORIO}
However, Jupiter can still be considered as a viable scenario to try to measure its \textcolor{black}{post-Newtonian} gravitomagnetic spin-octupole  effects. Indeed, by keeping a hypothetical new spacecraft at about the same distance from it along a much faster orbit, it is possible to select suitable values for $I,~\Omega,~\omega$ allowing for quite large precessions. Tables~\ref{tavola1}~to~\ref{tavola2}, which refer to a Jovian equatorial coordinate system, deal with two different orbital configurations yielding nominal precessions for the node and the pericentre as large as $\simeq 10^2-10^3\masy$, which are remarkably large values. More specifically, for a mildly eccentric orbit with $r\simeq 1.015~\textcolor{black}{R}$ with $I=\omega=90\deg$, the gravitomagnetic node precession would be as large as $\dot\Omega_\gm=400\masy$, while for $I=360\deg,~\omega=270\deg$ and the same orbit radius as before one has even $\dot\Omega_\gm=-1,600\masy,~\dot\omega_\gm = 4,000\masy$. Such an insight is confirmed by some numerical simulations of the Earth-probe range-rate signature. Indeed, by adopting the \textcolor{black}{ICRF} and a Juno-like spatial orientation for the previously considered almost circular, fast jovicentric orbit of the proposed spacecraft, Figure~\ref{fig2} shows that the size of its relativistic signature would reach the $\simeq 0.03~\textrm{mm~s}^{-1}$ level after just 1 d. It should be recalled that the Doppler measurement accuracy of Juno is $\simeq 0.003~\textrm{mm~s}^{-1}$ after $1,000$ s. Figure~\ref{fig3} preliminarily investigates the sensitivity to the individual orbital elements. It turns out that, while the gravitomagnetic range-rate signature is rather insensitive to the eccentricity, at least for small values of it, the pericentre and the true anomaly, the size of the orbit and the orientation of its orbital plane in space have a major impact. Indeed, if, on the one hand, a \textcolor{black}{sufficiently} low orbit is mandatory to increase the signal of interest, on the other hand, certain values of the inclination and the node may push it up to the $\simeq 0.3~\textrm{mm~s}^{-1}$ level for $a = 1.015~R,~e = 0.0049$.

However, caution is in order since dedicated studies will be required to further investigate our idea in terms of its actual feasibility from a practical and engineering point of view. We mention the threat posed, in principle, to the electronics of Jovian probes by the Io plasma torus\footnote{The intense volcanic activity of Io, which is the dominant source of plasma at Jupiter, pours material into Io's atmosphere which is lost to the Jovian magnetosphere near Io. Such a material is then ionized and trapped by the magnetic field forming a torus of plasma around Jupiter. The torus consists of different regions extending from $\simeq 4~R$ to $\simeq 10~R$ \citep{2017AGUFMSM33C2669H}.} and the potentially quite large $\Delta v$ required to implement a successful  orbit insertion. Another crucial issue is represented by the impact of other competing dynamical effects, which would act as source of systematic errors potentially biasing the recovery of the relativistic effect of interest. In this regard, we remark that the proposed scenario would benefit of the notable improvement of our knowledge about both the Jupiter's spin pole position and the Newtonian part of its gravity field arising from the analysis of the full data record of Juno, which is scheduled to deorbit into the planet on\footnote{See https://www.jpl.nasa.gov/missions/juno/ on the Internet.} July 2021. Suffice it to say that, until now, just 2 (PJ03 and PJ06) out of a total of expected 25 perijove passages dedicated to gravity field determination have been fully analyzed \citep{2018Natur.555..220I}, while the results from PJ08,~PJ10,~PJ11 should be publicly released soon \citep{2018EGUGA..20.9150D}.  Table~\ref{tavolaJup} displays, among other things, the best estimates and the associated realistic uncertainties for the even and odd zonal coefficients $J_{\ell},~\ell=2,3,4,\ldots,12$, and the tesseral and sectorial multipoles $C_{2,1},~S_{2,1},~C_{2,2},~S_{2,2}$. The RA and Dec. of $\bds{\hat{S}}$ are currently known to an accuracy of $\simeq 100~\textrm{mas}$, as shown in Table~\ref{tavolaJup}, while their rates of change are accurate to $\simeq 50\masy$ \citep{2018EGUGA..20.9150D}. As far as the first even zonal harmonic of the Jovian gravity field, from Fig.~2 of the poster presented by \citet{2018EGUGA..20.9150D} it seems that its most recent accuracy is $\simeq 4\times 10^{-9}$, corresponding to a relative accuracy of $\simeq 3\times 10^{-7}$. Moreover, it is not unrealistic to assume that the measurement accuracy $\upsigma_{\dot\rho}$ may be better than that of Juno, whose measurements are mostly taken only at its perijove passages, because of the comparatively much larger number $\mathrm{N}$ of data points due to the higher orbital frequency and lower eccentricity. Indeed,  $\upsigma_{\dot\rho}$ scales as $1/\sqrt{\mathrm{N}}$.
In the following, we want to quantitatively assess such issues in connection with the full potential of the proposed mission concept as a tool to measure even more general relativistic features of motion ranging from the standard Schwarzschild-like one proportional to $GM c^{-2}$, to the so far never tested gravitoelectric effect proportional to $GMJ_2 c^{-2}$ \citep{1988CeMec..42...81S,Sof89,1991ercm.book.....B}, including also the gravitomagnetic Lense-Thirring frame-dragging \citep{1918PhyZ...19..156L} proportional to $GS c^{-2}$.
\subsubsection{The impact of the mismodeling in the Jovian gravity field's multipoles}\lb{multipoles}
Figures~\ref{figJ2}~to~\ref{figS22}, obtained with the same computational method previously outlined in Section~\ref{Juno}, depict the numerically simulated Newtonian \textcolor{black}{(blue dashed curves)} and post-Newtonian \textcolor{black}{(red continuous curves)} range-rate time series for a given orbital configuration of the probe which, as it will be shown below, should make the detection of the relativistic signals more favorable.
In order to better visualize the temporal patterns of the various effects, the classical signatures were produced by using fictitious values $\mathcal{C}^\ast$ of the Newtonian gravity field coefficients able to make their magnitudes roughly equal to those of the post-Newtonian time series of interest. If such figures $\mathcal{C}^\ast$  for the Jovian multipoles are smaller than their present-day uncertainties listed in Table~\ref{tavolaJup}, they can be interpreted as a measure of how much they should still be improved with respect to their current levels of accuracy in order to make the size of the Newtonian signatures at least equal to the relativistic ones. If, instead, $\mathcal{C}^\ast$ are larger than their present mismodeling, they can be viewed as a measure of the relative accuracy with which a given relativistic signal would be impacted right now. See Table~\ref{tabres} for a complete list of such improvement factors for all the Newtonian multipoles considered here in connection with the various relativistic effects. It turns out that the largest improvements-of the order of $\simeq \textcolor{black}{10}-500$, with a peak of $1,000$ for $J_{10}$-would be required to bring the Newtonian signals to the level of the post-Newtonian gravitomagnetic effect proportional to $GSJ_2 c^{-2}$. \textcolor{black}{A} much smaller improvement would be required to make the size of the classical multipole signatures comparable with the post-Newtonian gravitoelectric and gravitomagnetic effects proportional to $GMJ_2c^{-2},~GSc^{-2}$. As far as the Schwarzschild-type signature is concerned, the current level of accuracy in almost all the Jovian multipoles, with the exception of \textcolor{black}{$J_{10},J_{11},~J_{12},~S_{2,1},~S_{2,2}$},  would yield a bias at the $\simeq 1-10$ per cent level. A very important feature of all the curves displayed in Figures~\ref{figJ2}~to~\ref{figS22} is that the relativistic ones exhibit neatly different temporal patterns with respect to the Newtonian ones, making, thus, easier to detect them. It would not be so for different orbital geometries of the probe.
\subsubsection{The impact of the uncertainty in the Jupiter's pole position}\lb{poleposition}
The position of the Jovian spin axis, determined by its right ascension $\alpha$ and declination $\delta$ with respect to the ICRF \citep{2018EGUGA..20.9150D}, enters the Newtonian accelerations induced by the gravity field multipoles in a nonlinear way. It can be easily realized, e.g.,  by inspecting the analytical expressions of the long-term precessions of the Keplerian orbital elements due to some even and odd zonal harmonics calculated by \citet{2011PhRvD..84l4001I,2013JApA...34..341R,2014Ap&SS.352..493R} for an arbitrary orientation of $\bds{\hat{S}}$. Thus, the uncertainties $\upsigma_\alpha,~\upsigma_\delta$ have an impact on the general relativistic effects of interest through the Newtonian multipolar signatures. The latest determinations of $\alpha,~\delta$ along with the associated realistic uncertainties, of the order of $\upsigma_\alpha,~\upsigma_\delta\simeq 0.1~\textrm{arcsec}$ \citep{2018EGUGA..20.9150D}, are listed in Table~\ref{tavolaJup}.

Figure~\ref{figRADEC} depicts the numerically simulated mismodeled range-rate signals due to the first four even zonals of Jupiter induced by the present-day errors $\upsigma_\alpha,~\upsigma_\delta$.
They were obtained as described in the previous Section by using the nominal values of the even zonals and taking the differences between the time series computed with $\delta_\textrm{max} = \delta + \upsigma_\delta,~\delta_\textrm{min} = \delta - \upsigma_\delta$ (\textcolor{black}{green dashed} curves) and $\alpha_\textrm{max} = \alpha + \upsigma_\alpha,~\alpha_\textrm{min} = \alpha - \upsigma_\alpha$ (\textcolor{black}{orange continuous} curves), respectively.
It turns out that the largest residual signals are due to the uncertainty in the declination. The largest one occurs for $J_2$, with an amplitude which can reach $\Delta\dot\rho_{\upsigma_{\delta}}^{J_2}\lesssim 60~\textrm{mm~s}^{-1}$. The signatures of the odd zonals are completely negligible. It can be shown that an improvement of $\upsigma_\delta$ by a factor of 100 with respect to the current value of Table~\ref{tavolaJup} would bring the size of the Newtonian $J_2$-induced range-rate time series to the same level of the post-Newtonian one proportional to $GSJ_2c^{-2}$. Such an improvement seems to be quite feasible in view of the fact that it already occurred from the analysis of PJ03, PJ06 \citep[Tab.~1]{2018Natur.555..220I} to that of  PJ08, PJ10, PJ11  \citep{2018EGUGA..20.9150D}. In any case, as already noticed in the previous Section, the temporal pattern of the classical $J_2$ signal is different from the relativistic ones.
\section{Summary and overview}\lb{conclu}
We analytically worked out the long-term rates of change of the Keplerian orbital elements of a test particle orbiting an extended spheroidal rotating body induced by its general relativistic gravitomagnetic spin-octupole moment to the first post-Newtonian order. We neither assumed a preferred orientation for the body's symmetry axis nor adopted a particular orbital configuration for the test particle. Thus, our results have a general validity, being applicable, in principle, to whatsoever astronomical and astrophysical scenario of interest. We successfully checked them numerically by integrating the equations of motion.

We applied them to Jupiter, which is the fastest spinning and most oblate major body of the Solar System, and some existing or hypothetical spacecraft orbiting it. While for Juno the gravitomagnetic precessions, of the order of $\simeq 0.2\masy$, are too small to be detectable, for a putative new probe orbiting the gaseous giant in, say, $0.12~\textrm{d}$ along a moderately eccentric orbit with $r\simeq 1.015~\textcolor{black}{R}$, the spin-quadrupole effects may be as large as $400-4,000\masy$ depending on the orbital geometry, within the measurability threshold with the current tracking technologies. We confirmed such expectations by numerically calculating in the ICRF the signature induced by the general relativistic spin-octupole moment of Jupiter on the Earth-satellite range-rate measurements which, in a real data analysis, would represent the actual observable quantity. Indeed, by conservatively assuming a range-rate experimental precision of $\simeq 0.003~\textrm{mm~s}^{-1}$ over 1,000 s, as for Juno, it turns out  that the post-Newtonian effect of interest could overcome such a level after just 1 full orbital revolution reaching, say, $0.03-0.3~\textrm{mm~s}^{-1}$ after 1 d depending mainly on the orientation of the orbital plane in space. Furthermore, in order to explore the full potential of the proposed mission concept, we looked also at the post-Newtonian gravitoelectric effects proportional to $GMJ_2c^{-2}$, which have never been put to the test so far, and at the standard Lense-Thirring and Schwarzschild signatures, proportional to $GSc^{-2},~GMc^{-2}$, respectively.

The experimental uncertainties in the values of both the Newtonian coefficients of the multipolar expansion of the Jovian gravity field and in the orientation of the spin axis of Jupiter would induce mismodeled range-rate signatures in the Doppler measurements of the spacecraft acting as sources of competing systematic biases for the post-Newtonian signals of interest. At present, just 5 of the planned 25 perijove passes dedicated to mapping the planet's gravity field of the ongoing Juno mission, scheduled to end in July 2021, have been analyzed so far. Thus, if and when the proposed mission will be finally implemented, it will benefit of the analysis of the entire Juno data record yielding a much more accurate determination of the Jovian gravity field coefficients and pole position than now.

For a given orbital configuration of the spacecraft, we numerically simulated its mismodeled Newtonian range-rate signatures due to the gravity field coefficients and the spin axis position of Jupiter currently determined by Juno, and the predicted post-Newtonian signals. We determined the level of improvement of the Jovian multipoles and pole position with respect to their present-day accuracies still required to bring the competing classical effects to the level of the various relativistic ones. It turned out that the most demanding requirements pertain the measurability  of the $GSJ_2c^{-2}$ signature, implying improvements by a factor of $\simeq \textcolor{black}{10}-500$ for most of the Jovian gravity coefficients considered, with a peak of $1,000$ for $J_{10}$.
The other relatively small post-Newtonian effects, proportional to $GMJ_2c^{-2},~GSc^{-2}$, require less demanding improvements by a factor of just $\simeq 5-50$ or less. The Schwarzschild signature would be measurable right now at a $\simeq 1-10\%$ level, apart from the impact of
\textcolor{black}{$J_{10},~ J_{11},~ J_{12},~S_{2,1},~S_{2,2}$}.
As far as the Jupiter's spin axis is concerned, an improvement by a factor of 100 would be required for its declination $\delta$ to make the size of the $J_2$-induced signature to the same level of the post-Newtonian $GSJ_2c^{-2}$ one. The uncertainty in the declination $\alpha$ is less important. The range-rate signals due to the odd zonals are affected by the errors in the pole position at a negligible level. A fundamental outcome of our analysis consists of the fact that the temporal patterns of the relativistic signatures turned out to be quite different from the classical ones, making, thus, easier, in principle, to separate the post-Newtonian from the Newtonian effects.

Finally, we remark once more that the present work is not a formal mission proposal; instead, it should be regarded just as a sort of expanded mission concept which need further, dedicated studies concerning, e.g., the practical feasibility of the suggested scenario taking into account several important technological and engineering issues.
%Further, dedicated studies should explore in detail the actual feasibility of such a scenario by investigating the impact of other competing dynamical effects %which may bias the recovery of the relativistic signal we are interested in. However, if and when our proposal will be finally implemented, it will benefit of %the analysis of the full data record of Juno, expected to end its mission in 2021 by plunging into Jupiter, which, among other things, will greatly increase %our knowledge of the planet's spin axis orientation and gravity field. Moreover, the measurement accuracy should be better than for Juno because of the %expected larger number of measurements due to the shorter orbital period and lower eccentricity.

%Finally, it should be remarked that the proposed spacecraft would allow to accurately measure also other general relativistic effects, which will be the %subject of forthcoming researches.
\section*{Acknowledgements}
I am grateful to D. Durante for useful information.
%-----------------------------------------
\begin{appendices}
\section{Notations and definitions}\lb{appena}
Here, some basic notations and definitions used throughout the text are presented \citep{1991ercm.book.....B,2003ASSL..293.....B,2011rcms.book.....K,2014grav.book.....P}.
\begin{description}
\item[] $G:$ Newtonian constant of gravitation
\item[] $c:$ speed of light in vacuum
\item[] $g_{\sigma\nu}:$ spacetime metric tensor
\item[] $\phi,~w:$ gravitoelectric potential
\item[] $U:$ Newtonian gravitational potential
\item[] $\mathbf{w}:$ gravitomagnetic potential
\item[] $T^{\sigma\nu}:$ energy-momentum tensor of the source
%\item[] $\epsilon:$ mean obliquity
%\item[] $\upmu_0:$ magnetic permeability of vacuum
\item[] $M:$ mass of the primary
\item[] $\mu\doteq GM:$ gravitational parameter of the primary
\item[] $S:$ magnitude of the angular momentum of the primary
\item[] ${\bds{\hat{S}}} = \grf{\kx,~\ky,~\kz}:$ spin axis of the primary in some coordinate system
\item[] $\alpha:$ right ascension (RA) of the primary's spin axis with respect to the Earth's mean equator at  epoch J2000.0
\item[] $\delta:$ declination (DEC) of the primary's spin axis with respect to the Earth's mean equator at epoch J2000.0
\item[] $\kx = \cos\delta\cos\alpha:$ $x$ component of the primary's spin axis with respect to the Earth's mean equator at epoch J2000.0
\item[] $\ky = \cos\delta\sin\alpha:$ $y$ component of the primary's spin axis with respect to the Earth's mean equator at epoch J2000.0
\item[] $\kz = \sin\delta:$ $z$ component of the primary's spin axis with respect to the Earth's mean equator at epoch J2000.0
\item[] $R_\textrm{e}:$ equatorial radius of the primary
\item[] $R_\textrm{p}:$ polar radius radius of the primary
\item[] $\varepsilon\doteq\sqrt{1 - \ton{\rp{R_\textrm{p}}{R_\textrm{e}}}^2}:$ ellipticity of the oblate primary
\item[] $J_\ell,~\ell=2,~3,~4,\ldots:$ Newtonian zonal multipole mass moments of the primary's gravity field
\item[] $C_{2,1},~S_{2,1},~C_{2,2},~S_{2,2}:$ tesseral $\textcolor{black}{(m=1)}$ and sectorial $\textcolor{black}{(m=2)}$ multipole mass moments of degree $\ell = 2$ of the primary's gravity field
\item[] ${\bds B}_\gm:$ post-Newtonian gravitomagnetic field in the empty space surrounding the rotating primary
\item[] $\phi_\gm:$ gravitomagnetic potential function in the empty space surrounding the rotating primary
\item[] ${\bds A}_\gm:$ post-Newtonian gravitomagnetic acceleration experienced by the test particle
%\item[] $\alpha_\textrm{X}:$ right ascension (RA) of the 3rd body's spin axis
%\item[] $\delta_\textrm{X}:$ declination (DEC) of the 3rd body's spin axis
%\item[] $\kx^\textrm{eq}=\cos\delta_\textrm{X}\cos\alpha_\textrm{X}:$ component of the 3rd body's spin axis w.r.t. the reference $x$ axis of an equatorial %coordinate system
%\item[] $\ky^\textrm{eq}=\cos\delta_\textrm{X}\sin\alpha_\textrm{X}:$ component of the 3rd body's spin axis w.r.t. the reference $y$ axis of an equatorial %coordinate system
%\item[] $\kz^\textrm{eq}=\sin\delta_\textrm{X}:$ component of the 3rd body's spin axis w.r.t. the reference $z$ axis of an equatorial coordinate system
\item[] ${\bds r}:$ instantaneous position vector of the test particle with respect to the primary
\item[] $r_\textrm{min}:$ pericentre distance of the test particle with respect to the primary
\item[] $r_\textrm{min}:$ apocentre distance of the test particle with respect to the primary
\item[] $r:$ instantaneous distance of the test particle from the primary
\item[] ${\bds{\hat{r}}}\doteq {\bds r}/r:$ versor of the position vector of the test particle
\item[] $\xi\doteq \bds{\hat{S}}\bds\cdot\bds{\hat{r}}:$ cosine of the angle between the primary's spin axis and the position vector of the test particle
\item[] $P_{2i + 1}\ton{\xi}:$ Legendre polynomial of degree $2i + 1$
\item[] $\bds v:$ velocity vector of the test particle
\item[] $f:$ true anomaly of the test particle's orbit
\item[] $a:$  semimajor axis of the test particle's orbit
\item[] $\nk \doteq \sqrt{\mu/a^3}:$  Keplerian mean motion of the test particle's orbit
\item[] $\Pb\doteq 2\uppi/\nk:$ orbital period of the test particle's orbit
\item[] $e:$  eccentricity of the test particle's orbit
\item[] $I:$  inclination of the orbital plane of the test particle's orbit to the reference $\grf{x,~y}$ plane of some coordinate system
\item[] $\Omega:$  longitude of the ascending node  of the test particle's orbit referred to the reference $\grf{x,~y}$ plane of some coordinate system
\item[] $\omega:$  argument of pericentre  of the test particle's orbit referred to the reference $\grf{x,~y}$ plane of some coordinate system
\item[] $\bds{\hat{l}}\doteq\grf{\cO,~\sO,~0}:$ unit vector directed along the line of the nodes toward the ascending node
\item[] $\bds{\hat{m}}\doteq\grf{-\cI\sO,~\cI\cO,~\sI}:$ unit vector directed transversely to the line of the nodes in the orbital plane
\item[] $\bds{\hat{k}}\doteq\grf{\sI\sO,~-\sI\cO,~ \cI}:$ unit vector perpendicular to the orbital plane directed along the orbital angular momentum
\item[] $\bds{\hat{P}}\doteq \bds{\hat{l}}\cos\omega + \bds{\hat{m}}\sin\omega:$ unit vector in the orbital plane directed along the line of apsides towards the pericentre
\item[] $\bds{\hat{Q}}\doteq -\bds{\hat{l}}\sin\omega + \bds{\hat{m}}\cos\omega:$ unit vector in the orbital plane directed transversely to the line of apsides
\end{description}
\section{Tables and figures}\lb{appenb}
\begin{table*}
\caption{Relevant physical parameters of Jupiter. Most of the reported values come from \citet{2003AJ....126.2687S,2010ITN....36....1P,2018Natur.555..220I,2018EGUGA..20.9150D} and references therein. In particular, the values and the uncertainties of $\alpha,~\delta$ determining the Jovian pole position at the epoch J2017.0 come from \citet{2018EGUGA..20.9150D}, while the multipoles of the gravity potential are retrieved from \citet[Tab.~1]{2018Natur.555..220I}.
}\lb{tavolaJup}
\begin{center}
\begin{tabular}{|l|l|l|}
  \hline
Parameter  & Units & Numerical value \\
%\hline
%$G$ & $\textrm{kg}^{-1}~\textrm{m}^3~\textrm{s}^{-2}$ & $6.67259\times 10^{-11} $\\
%
%$c$ & $\textrm{m~s}^{-1}$ & $2.99792458\times 10^8$\\
\hline
$\mu$ & $\textrm{m}^3~\textrm{s}^{-2}$ & $1.26713\times 10^{17}$\\
$S$ & $\textrm{kg~m}^2$~$\textrm{s}^{-1}$ & $6.9\times 10^{38}$\\
$\alpha$ & $\textrm{deg}$ & $268.057132\pm 0.000036$\\
$\delta$ & $\textrm{deg}$ & $64.497159 \pm 0.000045$\\
$R$ & $\textrm{km}$ & $71,492$\\
$J_2$ & $\ton{\times 10^{-6}}$ & $14,696.572\pm 0.014$\\
$J_3$ & $\ton{\times 10^{-6}}$ & $-0.042 \pm 0.010$\\
$J_4$ & $\ton{\times 10^{-6}}$ & $-586.609\pm 0.004$\\
$J_5$ & $\ton{\times 10^{-6}}$ & $-0.069 \pm 0.008$\\
$J_6$ & $\ton{\times 10^{-6}}$ & $34.198 \pm 0.009$\\
$J_7$ & $\ton{\times 10^{-6}}$ & $0.124 \pm 0.017$\\
$J_8$ & $\ton{\times 10^{-6}}$ & $-2.426 \pm 0.025$\\
$J_9$ & $\ton{\times 10^{-6}}$ & $-0.106 \pm 0.044$\\
$J_{10}$ & $\ton{\times 10^{-6}}$ & $0.172 \pm 0.069$\\
$J_{11}$ & $\ton{\times 10^{-6}}$ & $0.033 \pm 0.112$\\
$J_{12}$ & $\ton{\times 10^{-6}}$ & $0.047 \pm 0.178$\\
$C_{2,1}$ & $\ton{\times 10^{-6}}$ & $-0.013\pm 0.015$ \\
$S_{2,1}$ & $\ton{\times 10^{-6}}$ & $-0.003 \pm 0.026$\\
$C_{2,2}$ & $\ton{\times 10^{-6}}$ & $0.000\pm 0.008$ \\
$S_{2,2}$ & $\ton{\times 10^{-6}}$ & $0.000\pm 0.011$ \\
\hline
\end{tabular}
\end{center}
\end{table*}
\begin{table*}
\caption{Relevant orbital parameters of the spacecraft Juno currently orbiting Jupiter.
%Most of the reported values come from \citet{2003AJ....126.2687S,2007CeMDA..98..155S,2010ITN....36....1P} and references therein.
Here, $R$ is meant as the equatorial radius $R_\textrm{e}$ of Jupiter. The source for the orbital elements of Juno, referred to the Jovian equator, is the freely consultable database JPL HORIZONS on the Internet at https://ssd.jpl.nasa.gov/?horizons from which they were retrieved by choosing the time of writing this paper as input epoch. The values of the post-Newtonian gravitomagnetic precessions of Juno due to the spin-octupole moment of Jupiter, calculated by means of \rfrs{edot}{odot}, are listed as well.
}\lb{tavolaJuno}
\begin{center}
\begin{tabular}{|l|l|l|}
  \hline
Parameter  & Units & Numerical value \\
%\hline
%$G$ & $\textrm{kg}^{-1}~\textrm{m}^3~\textrm{s}^{-2}$ & $6.67259\times 10^{-11} $\\
%
%$c$ & $\textrm{m~s}^{-1}$ & $2.99792458\times 10^8$\\
\hline
%
%$\mu_{\jupiter}$ & $\textrm{m}^3~\textrm{s}^{-2}$ & $1.26713\times 10^{17}$\\
%
%$S_{\jupiter}$ & $\textrm{kg~m}^2$~$\textrm{s}^{-1}$ & $6.9\times 10^{38}$\\
%
%$\alpha_{\jupiter}$ & $\textrm{deg}$ & $268.056595$\\
%
%$\delta_{\jupiter}$ & $\textrm{deg}$ & $64.495303$\\
%
%$J_2^{\jupiter}$ & $-$ & $1.4696572\times 10^{-2}$\\
%
%$R_{\jupiter}$ & $\textrm{km}$ & $71,492$\\
%
$a$ & $R$ & $56.7633$\\
$e$ & $-$ & $0.9818125961521484$\\
$r_\textrm{min}$ & $R$ & $1.03238$\\
$r_\textrm{max}$ & $R$ & $112.494$\\
$I$ & $\textrm{deg}$ & $98.98696267273439$ \\
$\Omega$ & $\textrm{deg}$ & $270.7926907554042$ \\
$\omega$ & $\textrm{deg}$ & $163.1267988695804$ \\
$\Pb$ & $\textrm{d}$ & $52.8133$\\
\hline
$\dot e_\gm$ & $\textrm{yr}^{-1}$ & $5\times 10^{-12}$ \\
$\dot I_\gm$ & $\masy$ & $0.004$\\
$\dot \Omega_\gm$ & $\masy$ & $0.172$\\
$\dot \omega_\gm$ & $\masy$ & $0.190$\\
\hline
\end{tabular}
\end{center}
\end{table*}
\begin{table*}
\caption{Relevant orbital parameters for a hypothetical spacecraft, referred to the Jovian equator, and its post-Newtonian gravitomagnetic precessions  due to the spin-octupole moment of Jupiter, calculated by means of \rfrs{edot}{odot}. Cfr. with the other orbital configuration proposed in Table~\ref{tavola2}.
}\lb{tavola1}
\begin{center}
\begin{tabular}{|l|l|l|}
  \hline
Parameter  & Units & Numerical value \\
\hline
$a$ & $R$ & $1.015$\\
$e$ & $-$ & $0.0049$\\
$r_\textrm{min}$ & $R$ & $1.01$\\
$r_\textrm{max}$ & $R$ & $1.02$\\
$\Pb$ & $\textrm{d}$ & $0.12$\\
$I$ & $\textrm{deg}$ & $90$ \\
%
%$\Omega$ & $\textrm{deg}$ & $270.7926907554042$ \\
%
$\omega$ & $\textrm{deg}$ & $90$ \\
\hline
$\dot e_\gm$ & $\textrm{yr}^{-1}$ & $0.0$ \\
$\dot I_\gm$ & $\masy$ & $0.0$\\
$\dot \Omega_\gm$ & $\masy$ & $400$\\
$\dot \omega_\gm$ & $\masy$ & $0.0$\\
\hline
\end{tabular}
\end{center}
\end{table*}
\begin{table*}
\caption{Relevant orbital parameters for a hypothetical spacecraft, referred to the Jovian equator, and its post-Newtonian gravitomagnetic precessions  due to the spin-octupole moment of Jupiter, calculated by means of \rfrs{edot}{odot}. Cfr. with the other orbital configuration proposed in Table~\ref{tavola1}.
}\lb{tavola2}
\begin{center}
\begin{tabular}{|l|l|l|}
  \hline
Parameter  & Units & Numerical value \\
\hline
$a$ & $R$ & $1.015$\\
$e$ & $-$ & $0.0049$\\
$r_\textrm{min}$ & $R$ & $1.01$\\
$r_\textrm{max}$ & $R$ & $1.02$\\
$\Pb$ & $\textrm{d}$ & $0.12$\\
$I$ & $\textrm{deg}$ & $360$ \\
%
%$\Omega$ & $\textrm{deg}$ & $270.7926907554042$ \\
%
$\omega$ & $\textrm{deg}$ & $270$ \\
\hline
$\dot e_\gm$ & $\textrm{yr}^{-1}$ & $0.0$ \\
$\dot I_\gm$ & $\masy$ & $0.0$\\
$\dot \Omega_\gm$ & $\masy$ & $-1,600$\\
$\dot \omega_\gm$ & $\masy$ & $4,000.2$\\
\hline
\end{tabular}
\end{center}
\end{table*}
\begin{table*}
\caption{IORIO scenario: improvement factors $\textcolor{black}{\kappa}$ (if $>1$) required to each of the Jovian multipole coefficients with respect to their current accuracy levels (see \citet[Tab.~1]{2018Natur.555..220I} and Table~\ref{tavolaJup}) to make the size of the corresponding Newtonian range-rate signatures equal to the magnitude of the general relativistic ones; see Figures~\ref{figJ2}~to~\ref{figS22}. If, in a given row, $\textcolor{black}{\kappa}<1$, the current level of accuracy in the multipole of that row would allow right now to measure the corresponding relativistic effects with the relative accuracies as good as $\textcolor{black}{\kappa}$ themselves. For example, in the second row corresponding to $J_3$, there are two figures smaller than 1; it means that the present-day accuracy in $J_3$ would yield a mismodeled Newtonian signal impacting, say, the Schwarzschild-like one at $1.67\%$. Instead, the accuracy of $J_3$ should be improved by a factor of $12.5$ with respect to its current level in order to induce a mismodeled Newtonian signature having, at least, the same magnitude of the relativistic effect proportional to $GSJ_2c^{-2}$. From Table~\ref{tavolaJup}, it should be noted that the values of $J_{11},~J_{12},~C_{2,1},~S_{2,1},~C_{2,2},~S_{2,2}$ are statistically compatible with zero.
}\lb{tabres}
\begin{center}
\begin{tabular}{|l|l|l|l|l|}
  \hline
Multipole  & $GSJ_2c^{-2}$ & $GMJ_2 c^{-2}$ & $GSc^{-2}$ & $GMc^{-2}$ \\
\hline
$J_2$ & $70$ & $5$ & $2.5$ & $0.11$ \\
$J_3$ & $12.5$ & $1.1$ & $0.5$ & $0.0167$ \\
$J_4$ & $33$ & $3.3$ & $1.4$ & $0.033$ \\
$J_5$ & $10$ & $0.8$ & $0.58$ & $0.01$ \\
$J_6$ & $100$ & $5$ & $2.5$ & $0.12$ \\
$J_7$ & $50$ & $4.5$ & $3.3$ & $0.067$ \\
$J_8$ & $33$ & $4$ & $2.2$ & $0.05$ \\
$J_9$ & $100$ & $10$ & $3.3$ & $0.15$ \\
$J_{10}$ & $1,000$ & $33$ & $33$ & $0.98$ \\
$J_{11}$ & $500$ & $28.6$ & $20$ & $0.7$ \\
$J_{12}$ & $500$ & $50$ & $28.6$ & $1.1$ \\
$C_{2,1}$ & $50$ & $4$ & $2.8$ & $0.1$ \\
$S_{2,1}$ & $500$ & $20$ & $12.5$ & $\textcolor{black}{1}$ \\
$C_{2,2}$ & $333$ & $28.6$ & $18.2$ & $0.4$ \\
$S_{2,2}$ & $500$ & $33$ & $20$ & $1$ \\
\hline
\end{tabular}
\end{center}
\end{table*}
\clearpage
\begin{figure*}
\begin{center}
\centerline{
\vbox{
\begin{tabular}{cc}
\epsfysize= 5.2 cm\epsfbox{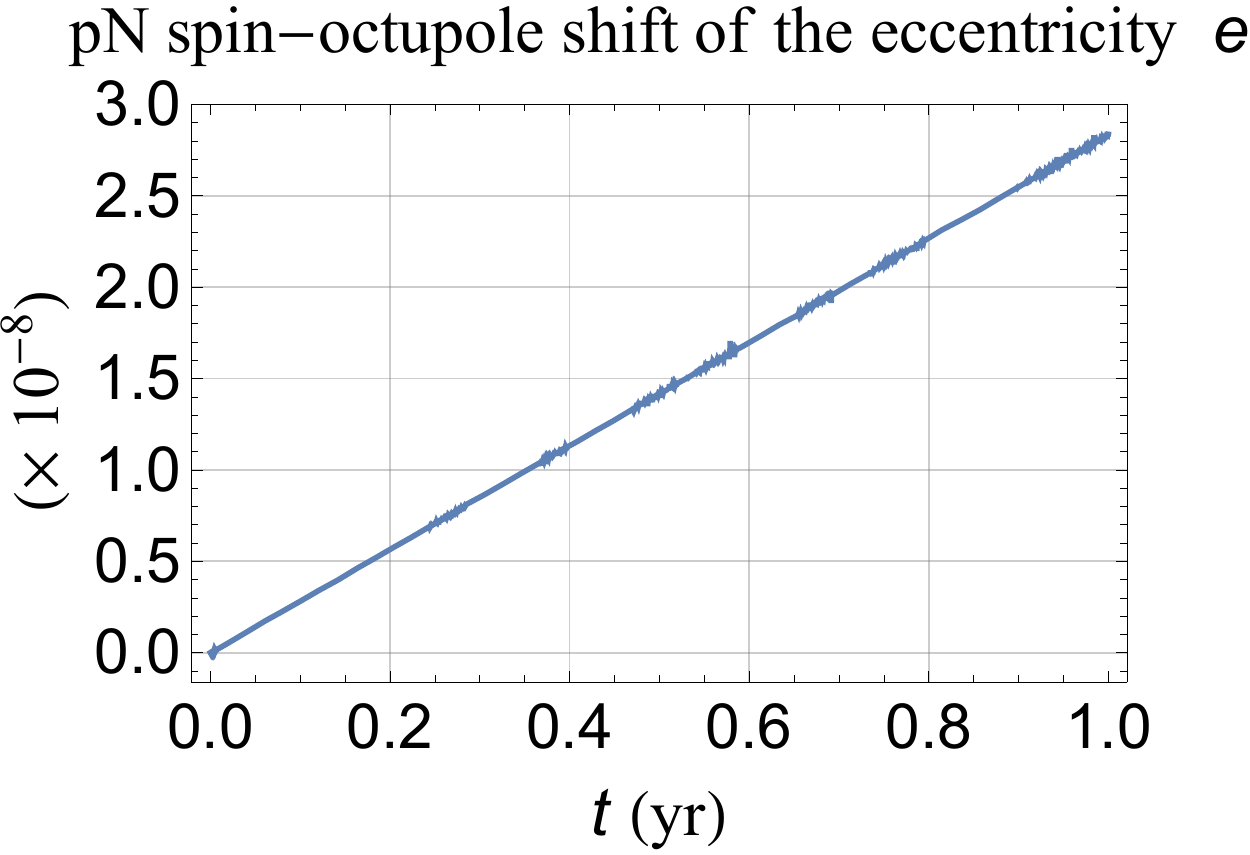}&\epsfysize= 5.2 cm\epsfbox{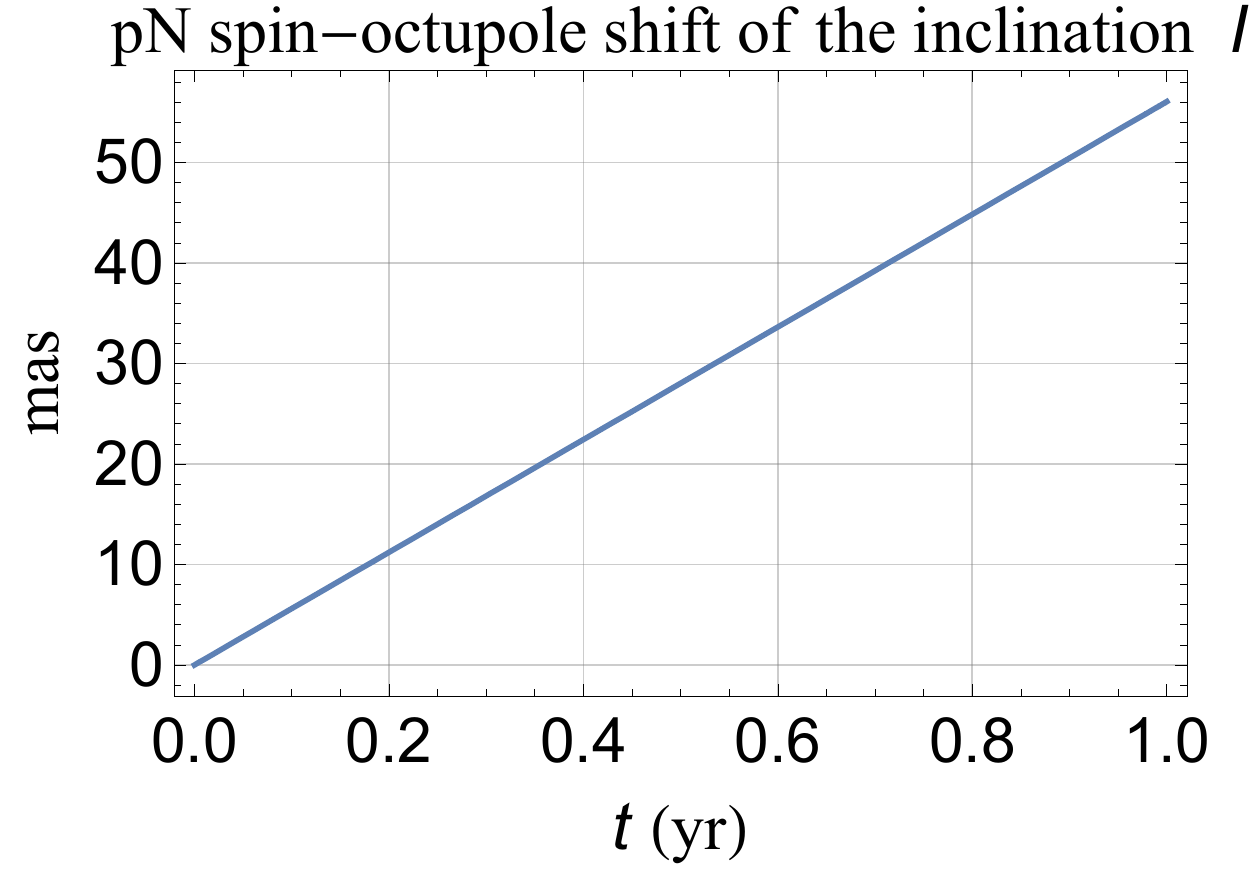}\\
\epsfysize= 5.2 cm\epsfbox{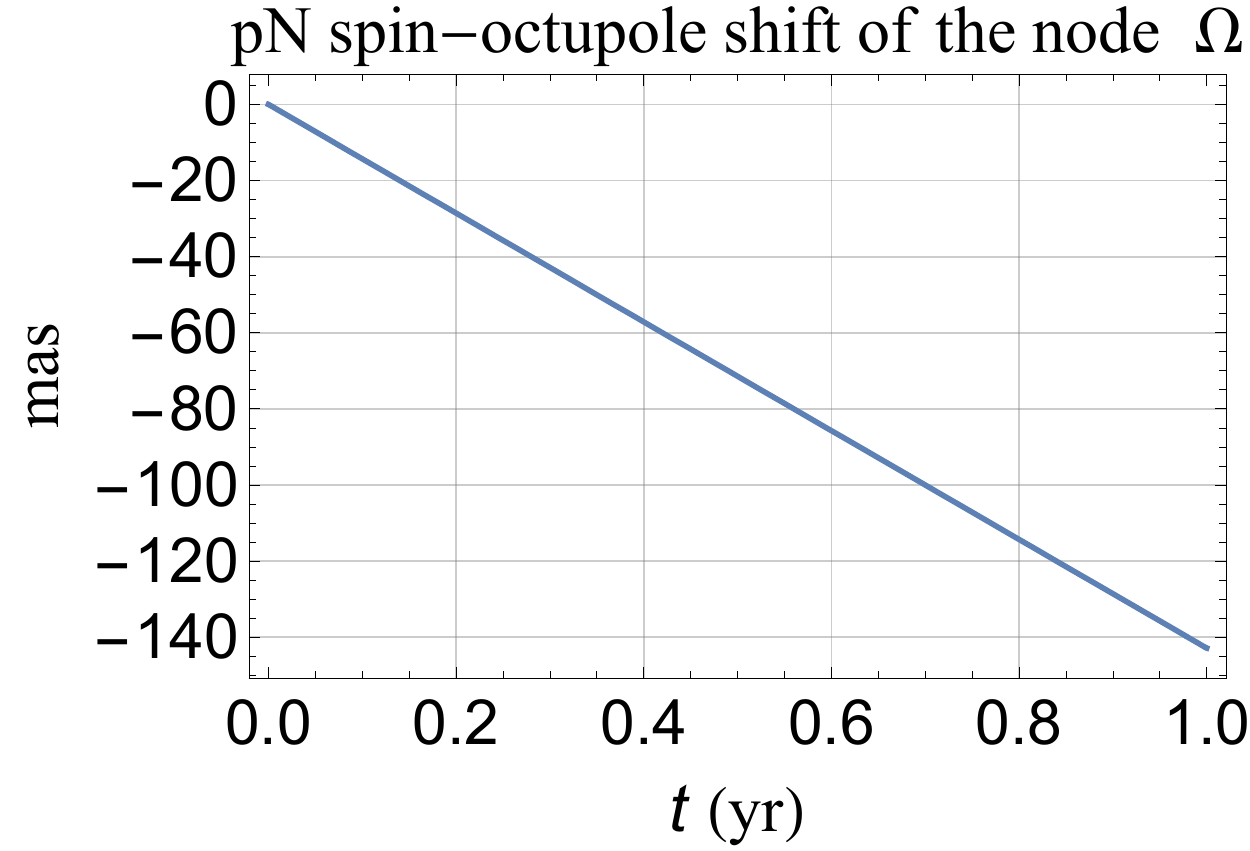}&\epsfysize= 5.2 cm\epsfbox{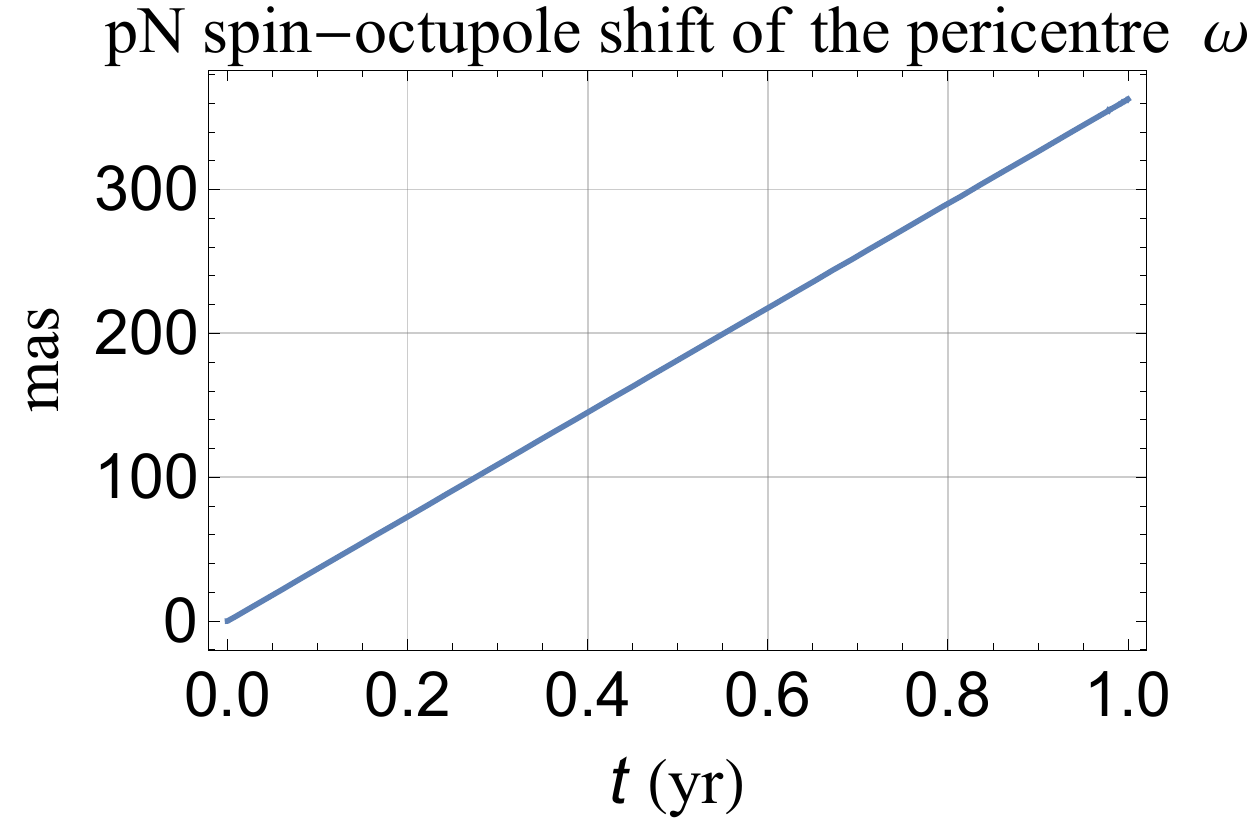}\\
\end{tabular}
}
}
\caption{Numerically computed time series of the post-Newtonian shifts experienced by the eccentricity $e$, inclination $I$, node $\Omega$ and pericentre $\omega$ of a fictitious test particle induced by the gravitomagnetic spin-octupole moment of a putative central body characterized by the same physical properties of Jupiter (see Table~\ref{tavolaJup}). They were obtained by numerically integrating the equations of motion of the orbiter in Cartesian rectangular coordinates referred to the Earth's mean equator at the epoch J2000.0 with and without the acceleration of \rfr{accel} calculated with \rfr{phigm} for $i=1$. Both runs shared the same set of arbitrary initial conditions $a_0 = 1.5~R,~e_0 = 0.3,~I_0 = 45\deg,~\Omega_0 = 30\deg,~\omega_0 = 50\deg,~f_0 = 45\deg$. For each Keplerian orbital element, its time series calculated  from the purely Newtonian run was subtracted from that obtained from the post-Newtonian integration in order to obtain the signatures displayed here. The resulting rates of change, in \textcolor{black}{yr$^{-1}$ and} $\masy$, turn out to agree with the analytically computed ones in \rfrs{dote}{doto} with \rfrs{Ecoef}{ocoef} which are $\dot e_\gm = 2.835\times 10^{-8}~\textrm{yr}^{-1},~\dot I_\gm = 56.05\masy,~\dot\Omega_\gm = -142.89\masy,~\dot\omega_\gm = 362.74 \masy$.}\label{fig0}
\end{center}
\end{figure*}
\begin{figure*}
\begin{center}
\centerline{
\vbox{
\begin{tabular}{c}
\epsfysize= 7.5 cm\epsfbox{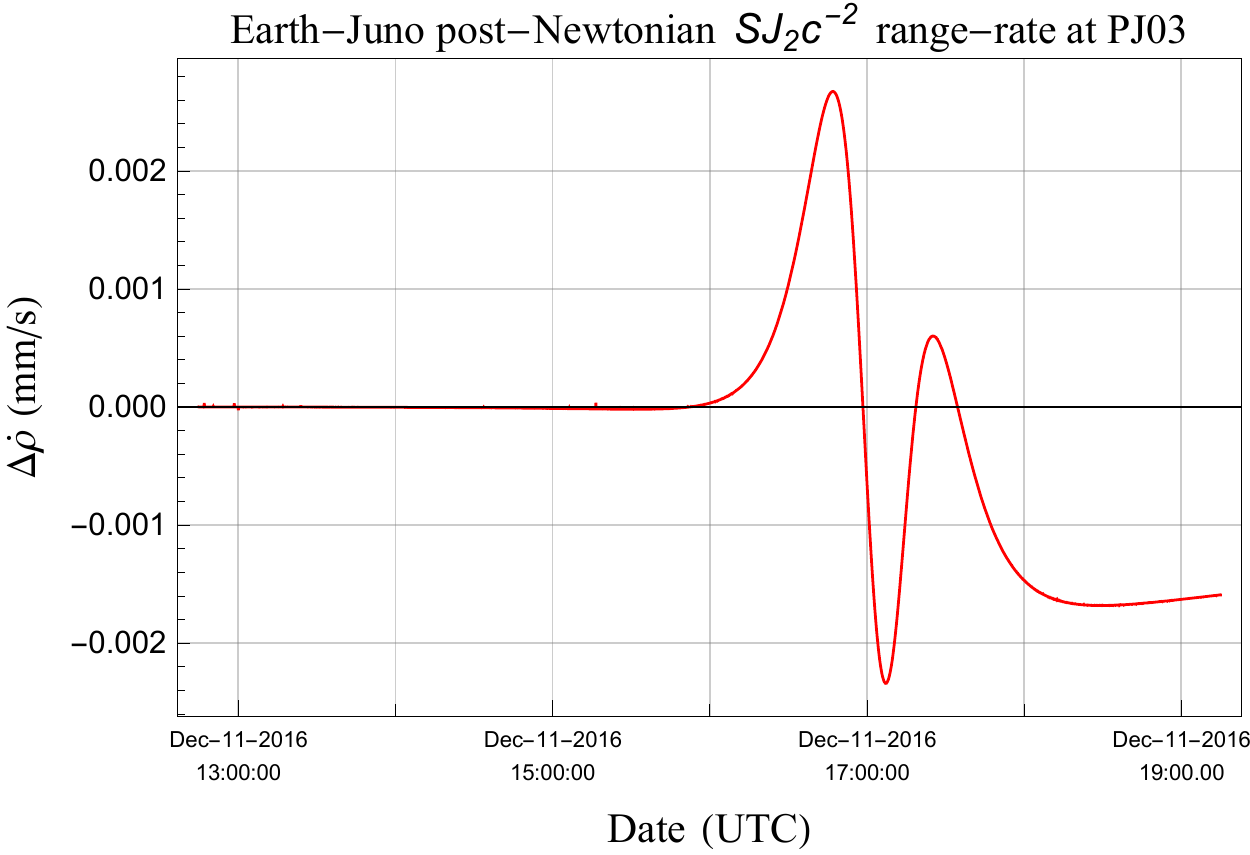}\\
\epsfysize= 7.5 cm\epsfbox{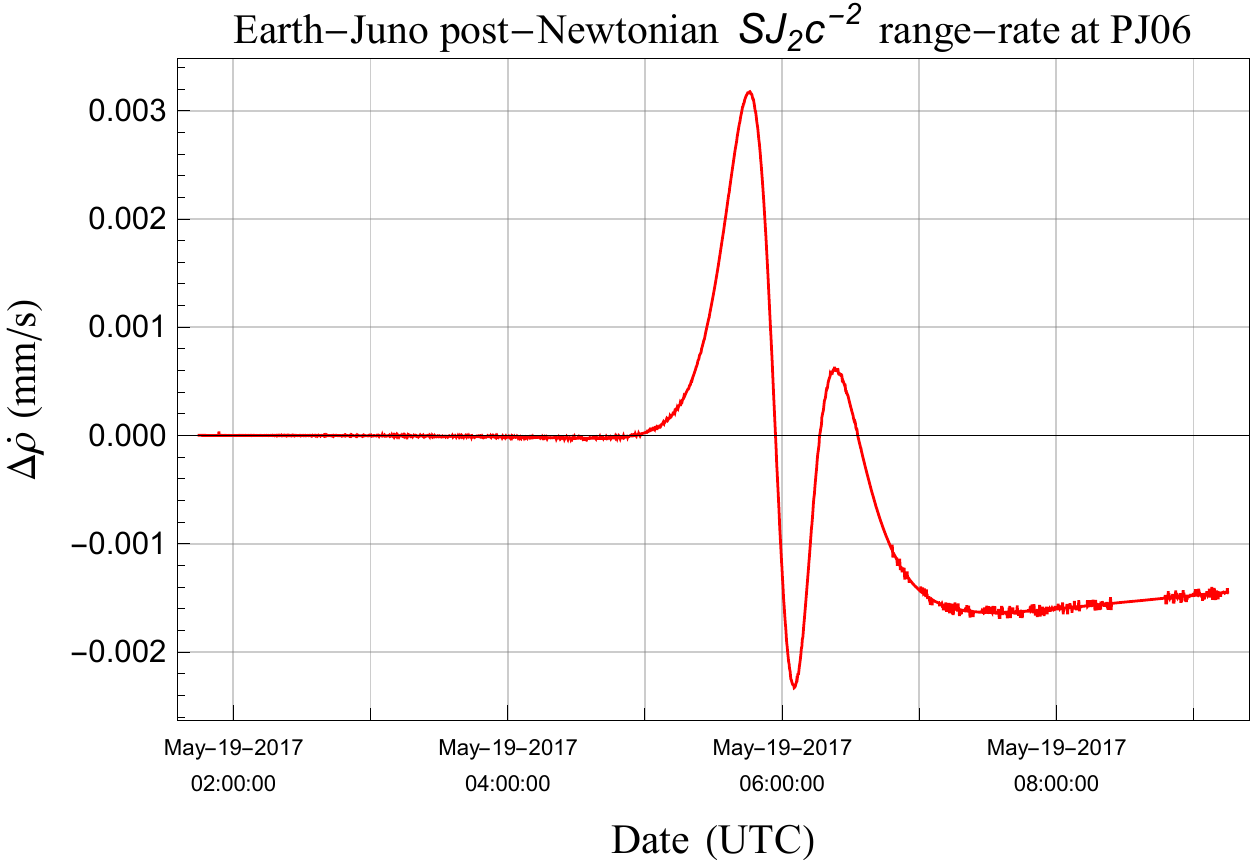}\\
\end{tabular}
}
}
\caption{Simulated range-rate signatures $\Delta\dot\rho$, in $\textrm{mm~s}^{-1}$, of Juno due to the post-Newtonian gravitomagnetic spin-octupole moment of Jupiter at the first two perijove passages PJ03 (December 11, 2016) and PJ06 (May 19, 2017) dedicated to gravity science. They were obtained by numerically integrating the equations of motion of the Earth, Jupiter and Juno in Cartesian rectangular coordinates referred to the ICRF with and without the general relativistic acceleration of \rfr{accel}, calculated for \rfr{phigm} with $i=1$, starting from the same set of initial conditions retrieved from the Web interface HORIZONS maintained by JPL. Then, for each perijove passage, the range-rate time series computed from the purely Newtonian run was subtracted from that obtained from the post-Newtonian integration in order to yield the curves displayed here. Cfr. with the two-way Ka-band range-rate residuals of Juno for the same perijove passes displayed in the Extended Data Figure~1 of \citet{2018Natur.555..220I} whose ranges of variation amount to $\simeq 0.050~\textrm{mm~s}^{-1}$, with a root-mean-square value of $\simeq 0.015~\textrm{mm~s}^{-1}$.}\label{fig1}
\end{center}
\end{figure*}
\begin{figure*}
\begin{center}
\centerline{
\vbox{
\begin{tabular}{c}
\epsfysize= 10.0 cm\epsfbox{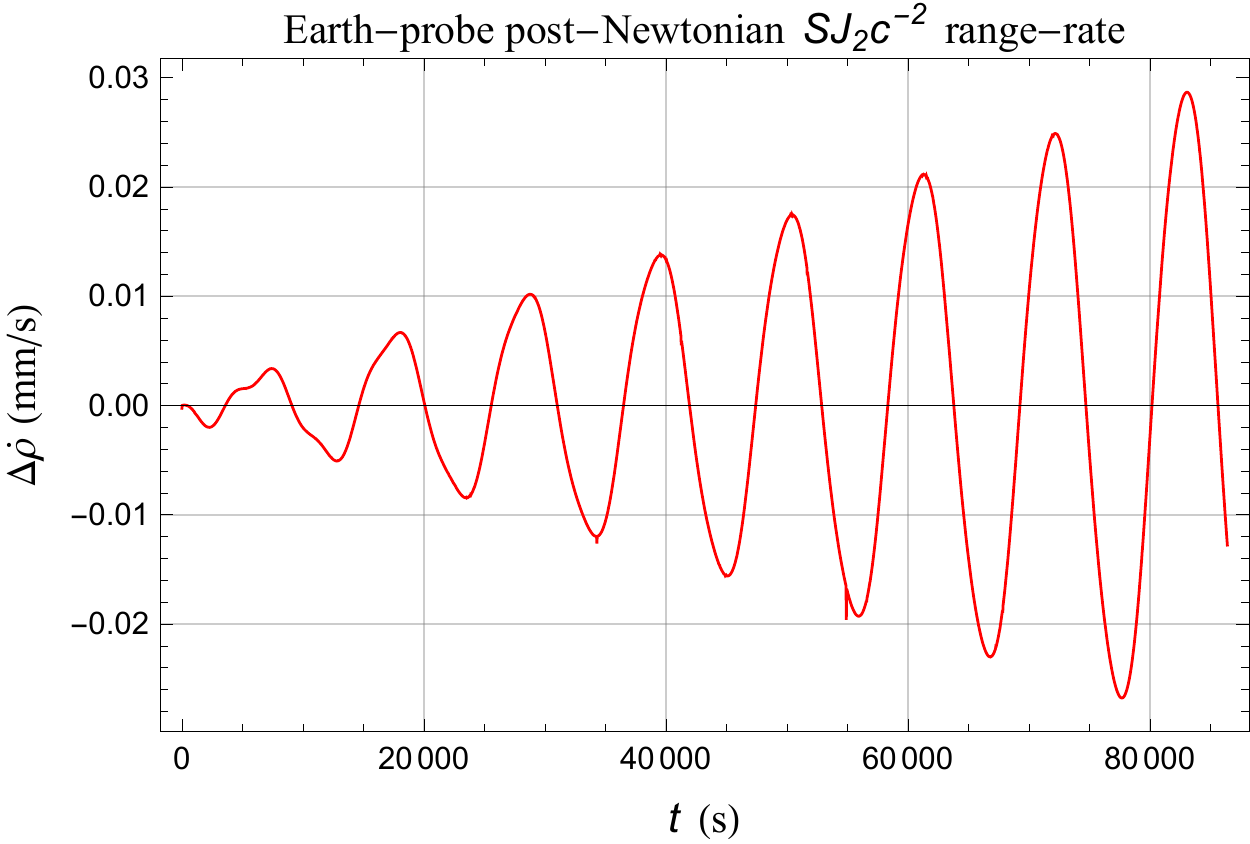}\\
\end{tabular}
}
}
\caption{Simulated range-rate signature $\Delta\dot\rho$, in $\textrm{mm~s}^{-1}$, of a hypothetical Jovian orbiter characterized by the ICRF-related orbital configuration $a = 1.015~\textcolor{black}{R},~e = 0.0049,~I = 90.63\deg,~\Omega = 268.85\deg,~\omega = 149.43\deg,~f_0 = 228.32\deg$ induced by the post-Newtonian gravitomagnetic spin-octupole moment of Jupiter after 1 d. It was obtained as described in the caption of Figure~\ref{fig1}.}\label{fig2}
\end{center}
\end{figure*}
\begin{figure*}
\begin{center}
\centerline{
\vbox{
\begin{tabular}{cc}
\epsfysize= 5.4 cm\epsfbox{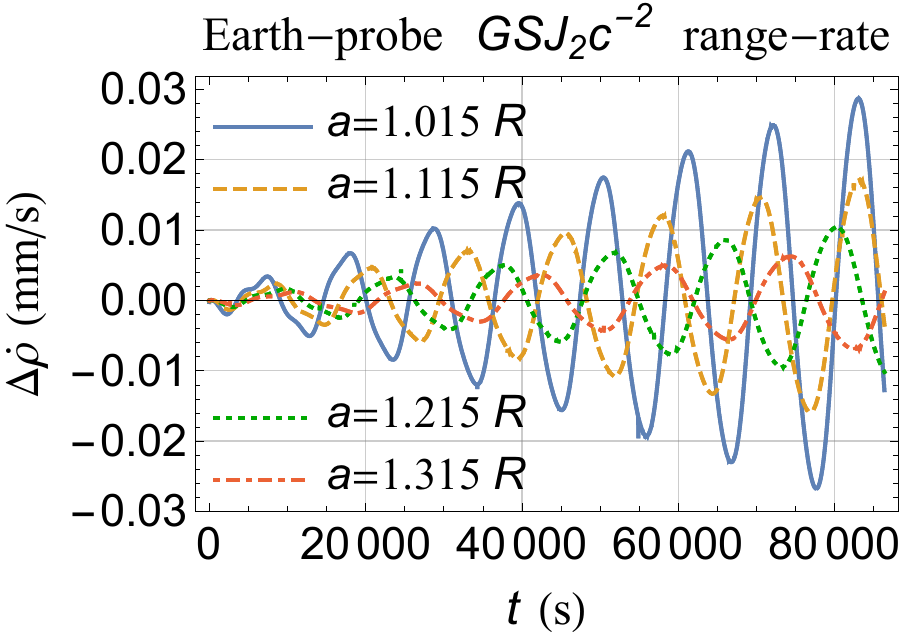}&\epsfysize= 5.4 cm\epsfbox{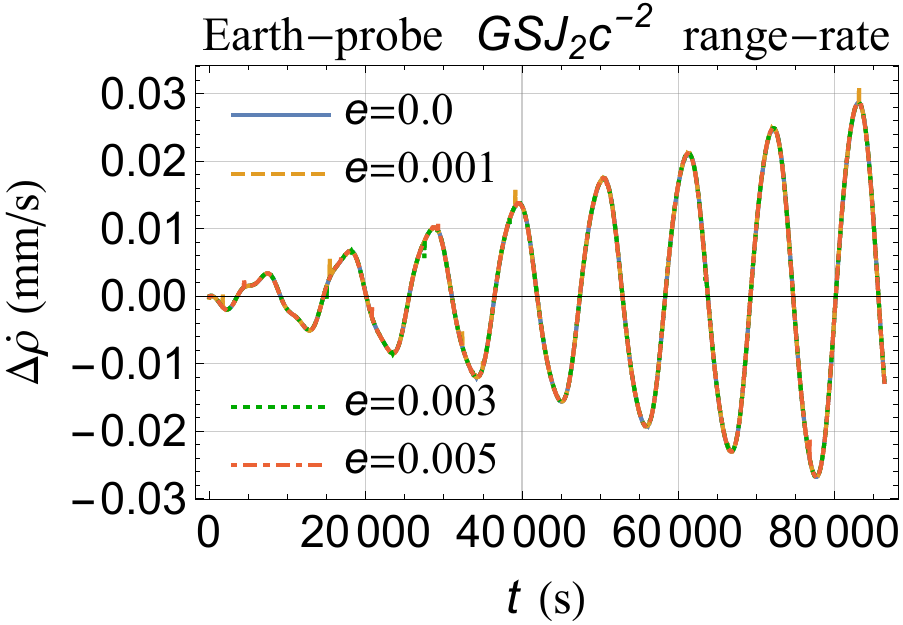}\\
\epsfysize= 5.4 cm\epsfbox{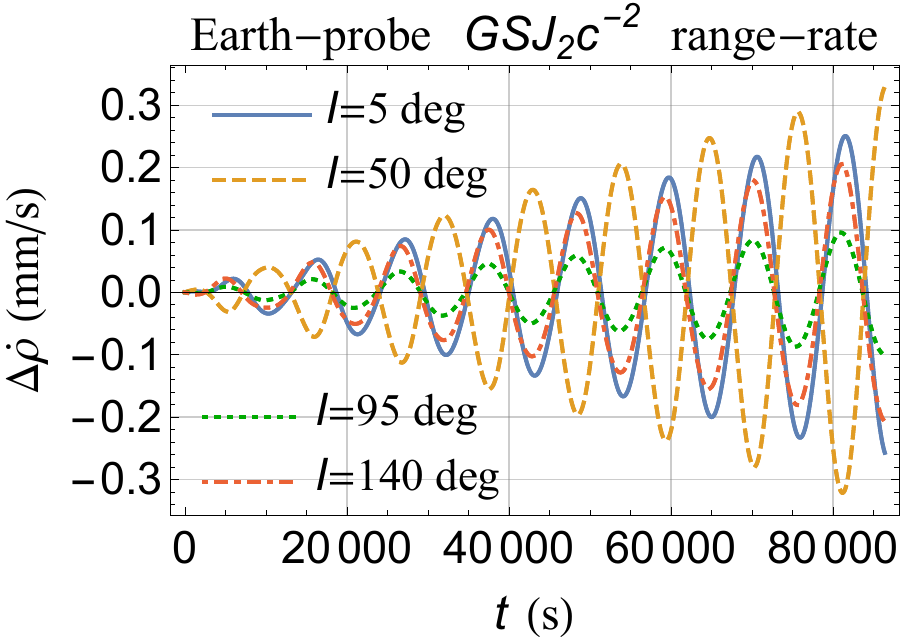}&\epsfysize= 5.4 cm\epsfbox{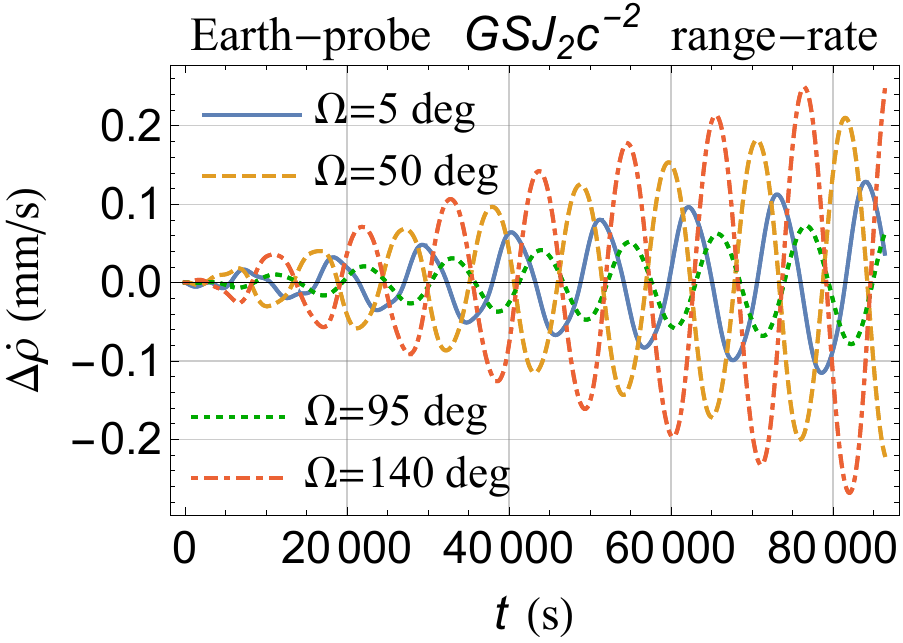}\\
\epsfysize= 5.4 cm\epsfbox{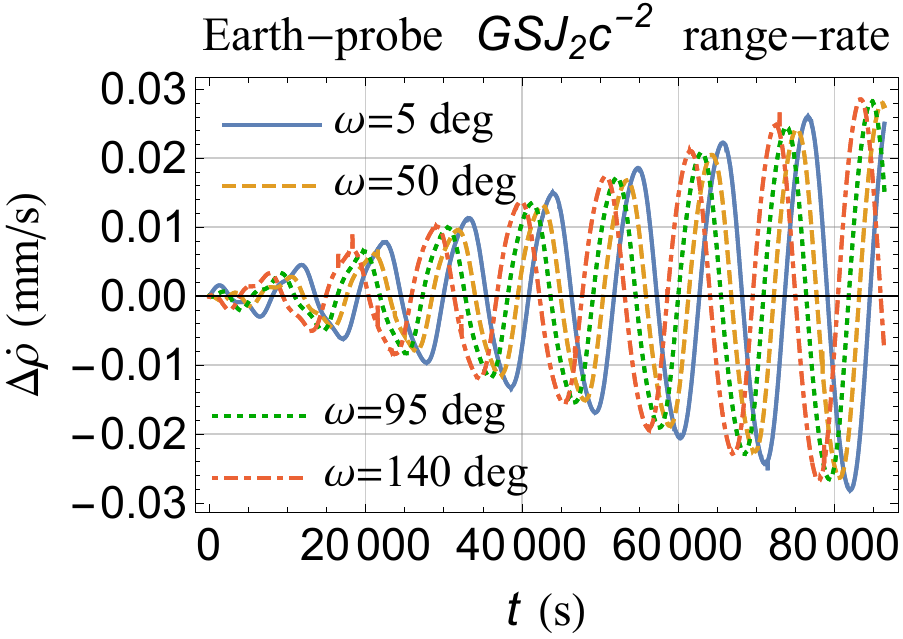}&\epsfysize= 5.4 cm\epsfbox{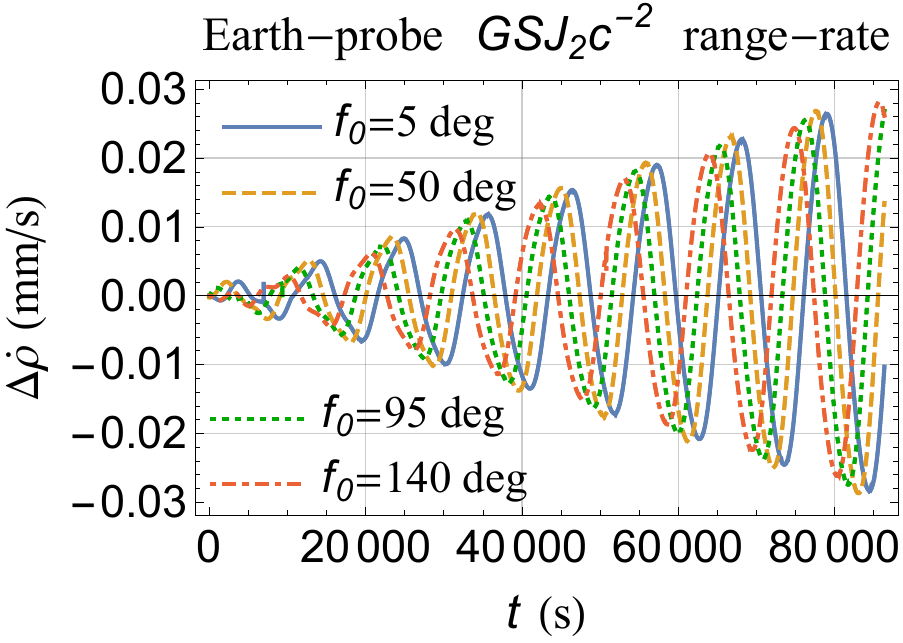}\\
\end{tabular}
}
}
\caption{Simulated range-rate signatures $\Delta\dot\rho$, in $\textrm{mm~s}^{-1}$, of a hypothetical Jovian orbiter  induced by the post-Newtonian gravitomagnetic spin-octupole moment of Jupiter after 1 d. They were obtained as described in the caption of Figure~\ref{fig1} by allowing the orbital elements of the spacecraft to vary within certain ranges of values with respect to the reference orbital configuration  used to produce Figure~\ref{fig2}.}\label{fig3}
\end{center}
\end{figure*}
\begin{figure*}
\begin{center}
\centerline{
\vbox{
\begin{tabular}{cc}
\epsfysize= 5.4 cm\epsfbox{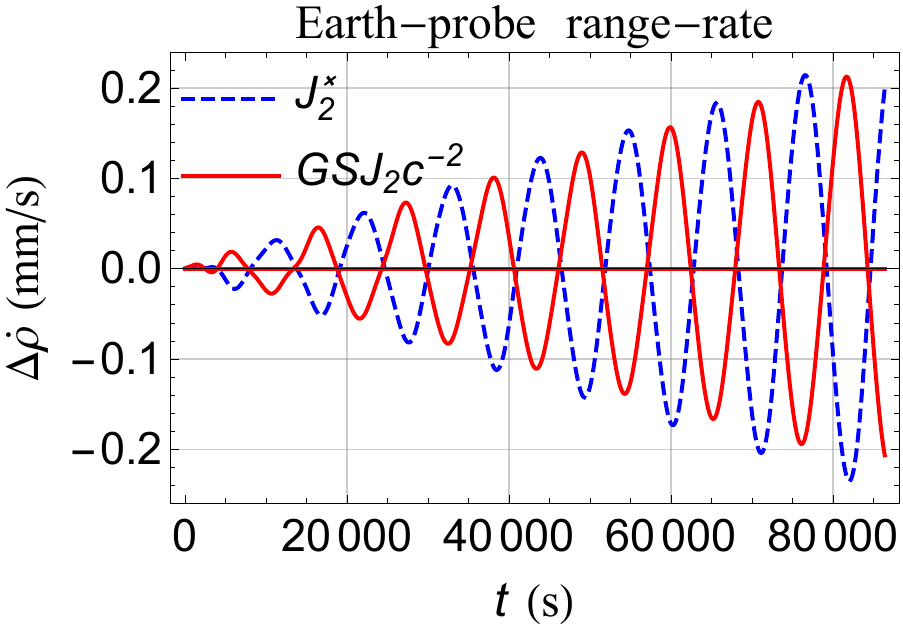}&\epsfysize= 5.4 cm\epsfbox{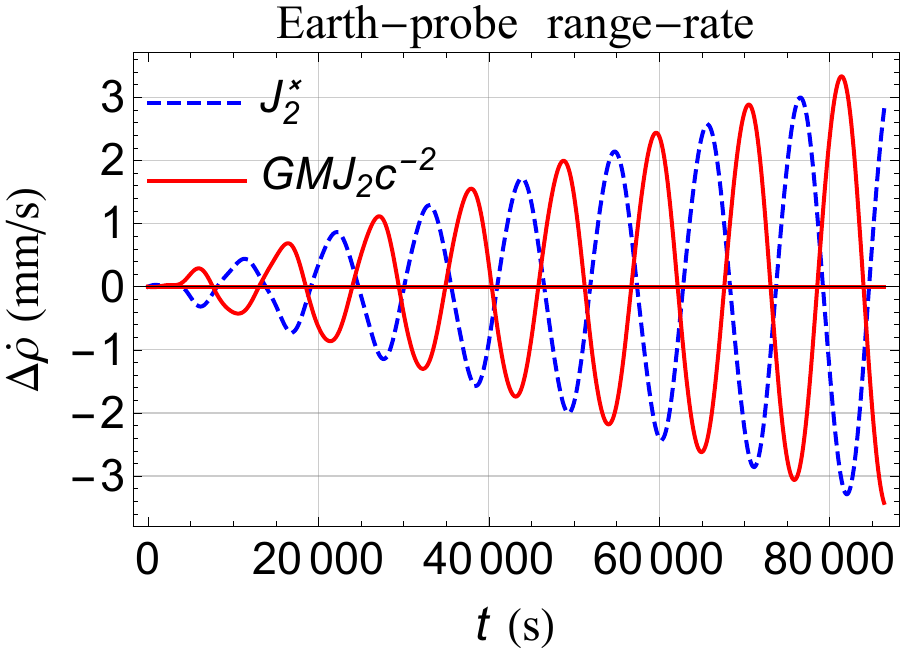}\\
\epsfysize= 5.4 cm\epsfbox{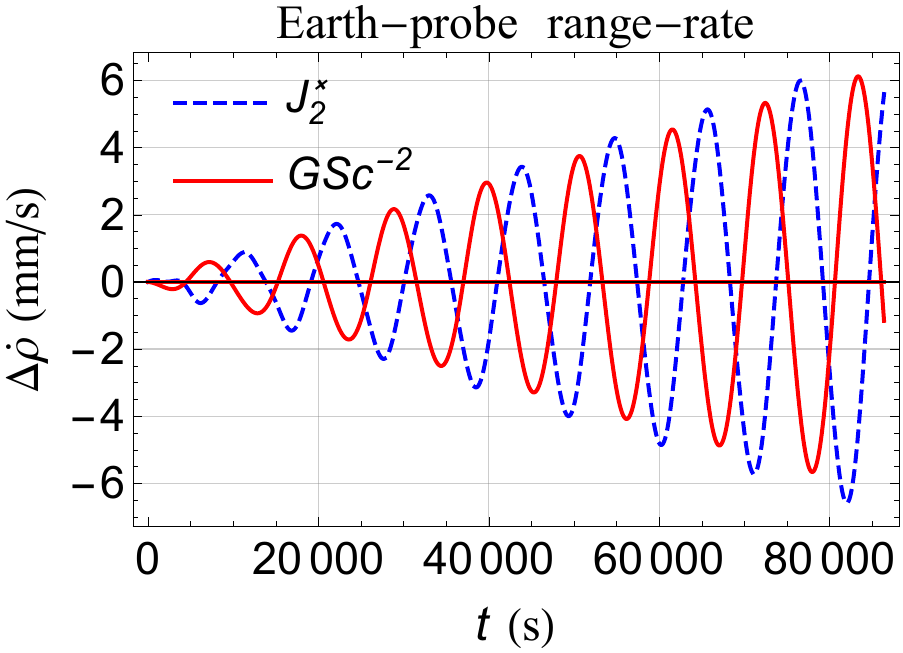}&\epsfysize= 5.4 cm\epsfbox{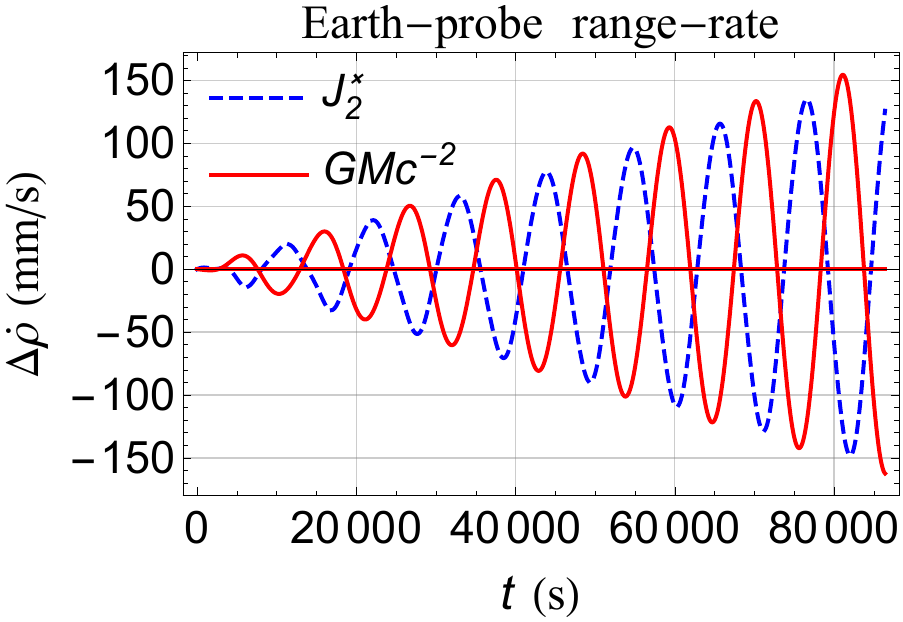}\\
\end{tabular}
}
}
\caption{Simulated range-rate signatures $\Delta\dot\rho$, in $\textrm{mm~s}^{-1}$, of a hypothetical Jovian orbiter  induced by the nominal post-Newtonian accelerations considered in the text and by the Newtonian first even zonal harmonic $J_2$ of Jupiter after 1 d. In each panel, a fictitious value $J_2^\ast$  is used in the Newtonian signature just for illustrative and comparative purposes. Indeed, it is suitably tuned from time to time in order to bring the associated classical signature to the level of the nominal post-Newtonian effect of interest, for which the actual value of $J_2$ is, instead, used, so to inspect the mutual (de)correlations of their temporal patterns more easily. Upper-left corner: post-Newtonian gravitomagnetic spin-octupole moment $\left(GSJ_2c^{-2};~J_2^\ast = 2.0\times 10^{-10}\right)$. Upper-right corner: post-Newtonian gravitoelectric moment $\left(GMJ_2 c^{-2};~J_2^\ast = 2.8\times 10^{-9}\right)$. Lower-left corner: Lense-Thirring effect $\left(GS c^{-2};~J_2^\ast = 5.6\times 10^{-9}\right)$. Lower-right corner: Schwarzschild $\left(GM c^{-2};~J_2^\ast = 1.26\times 10^{-7}\right)$. The present-day actual uncertainty in the Jovian first even zonal is $\upsigma_{J_2} = 1.4\times 10^{-8}$ \citep[Tab.~1]{2018Natur.555..220I}. The adopted orbital configuration for the probe is $a_0 = 1.015~R,~e_0 = 0.0049,~I_0 = 50\deg,~\Omega_0 = 140\deg,~\omega_0 = 149.43\deg,~f_0 = 228.32\deg$ }\label{figJ2}
\end{center}
\end{figure*}
\begin{figure*}
\begin{center}
\centerline{
\vbox{
\begin{tabular}{cc}
\epsfysize= 5.4 cm\epsfbox{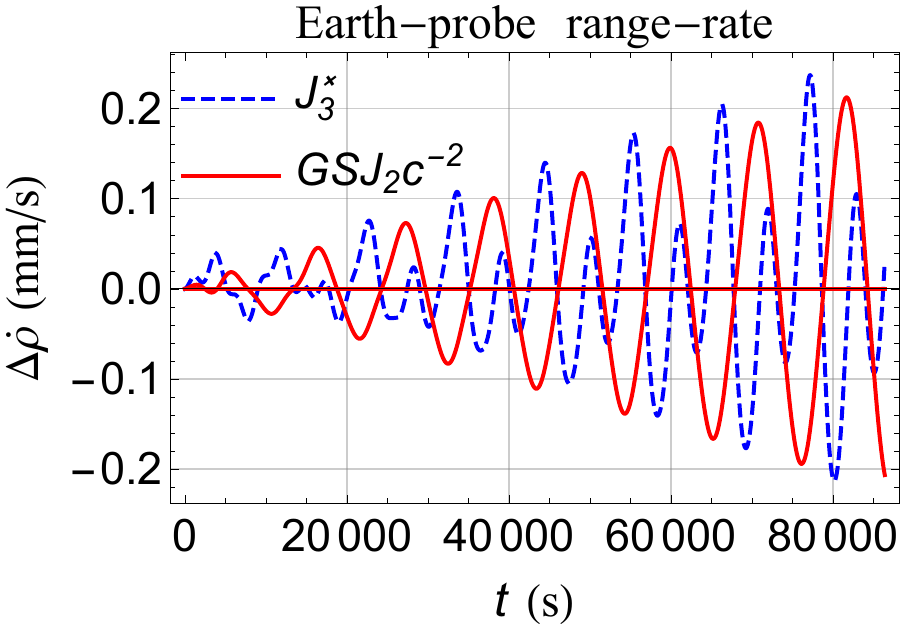}&\epsfysize= 5.4 cm\epsfbox{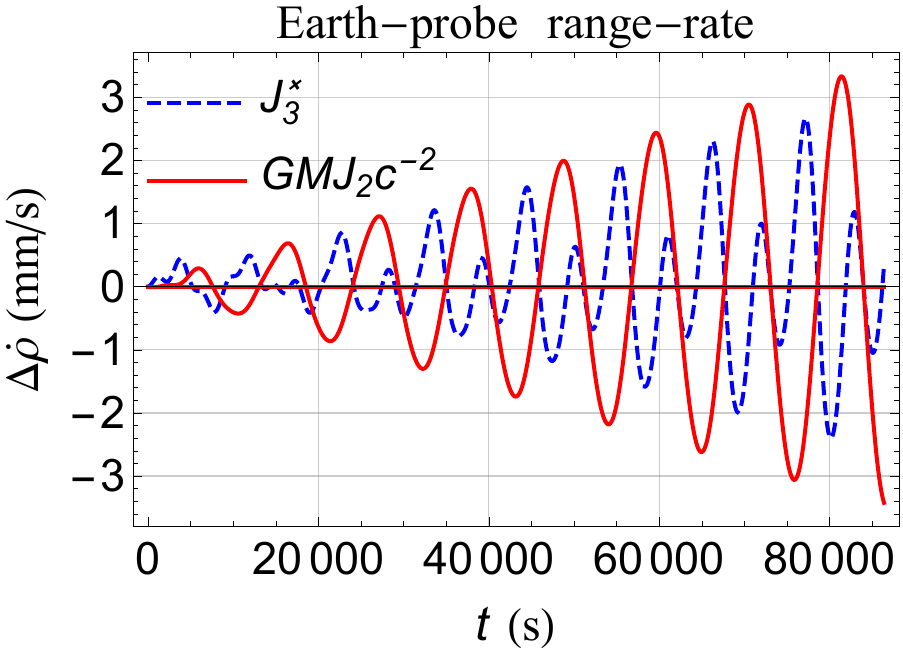}\\
\epsfysize= 5.4 cm\epsfbox{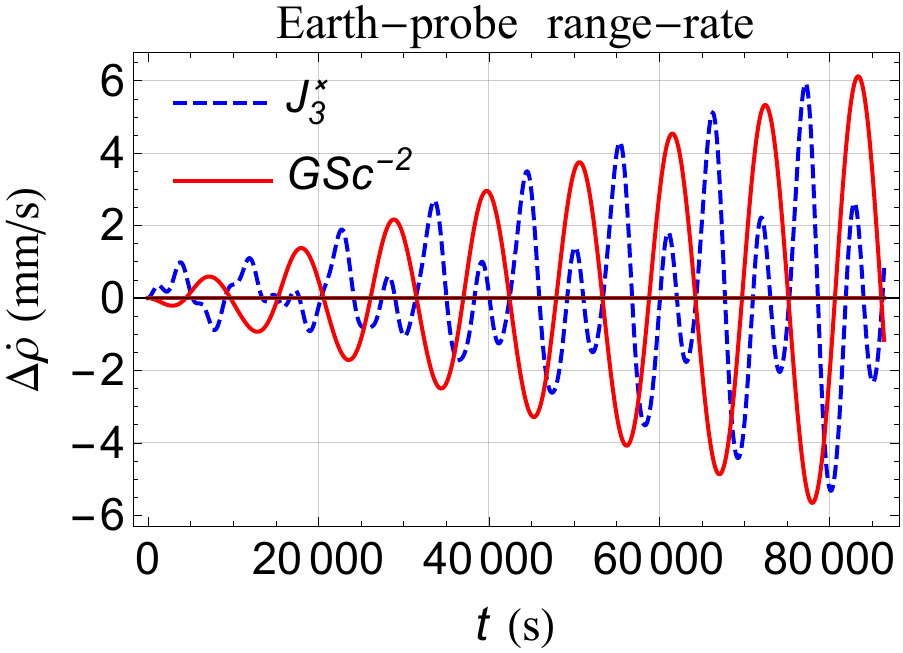}&\epsfysize= 5.4 cm\epsfbox{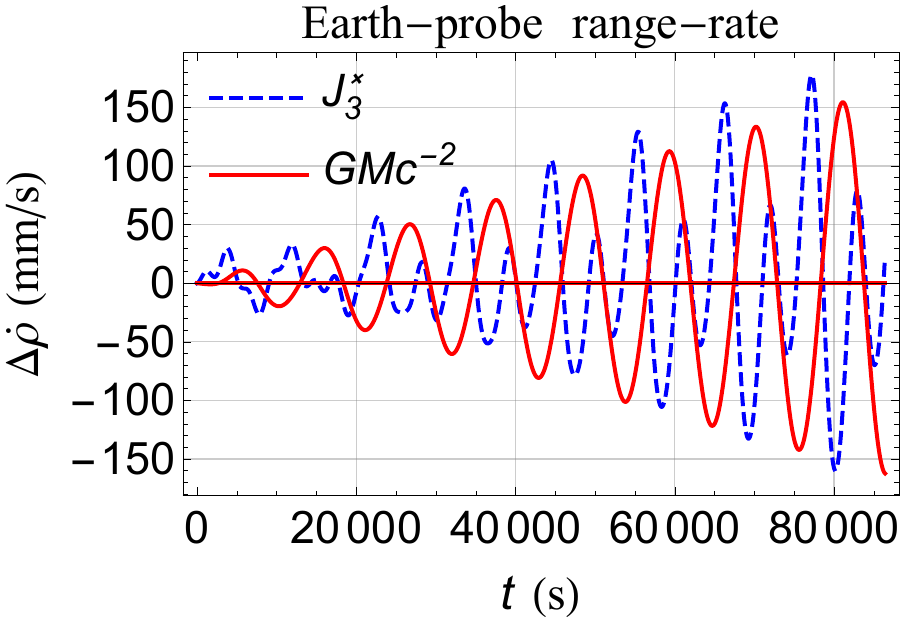}\\
\end{tabular}
}
}
\caption{Simulated range-rate signatures $\Delta\dot\rho$, in $\textrm{mm~s}^{-1}$, of a hypothetical Jovian orbiter  induced by the nominal post-Newtonian accelerations considered in the text and by the Newtonian first odd zonal harmonic $J_3$ of Jupiter after 1 d. In each panel, a fictitious value $J_3^\ast$  is used in the Newtonian signature just for illustrative and comparative purposes. Indeed, it is suitably tuned from time to time in order to bring the associated classical signature to the level of the nominal post-Newtonian effect of interest, so to inspect the mutual (de)correlations of their temporal patterns more easily. Upper-left corner: post-Newtonian gravitomagnetic spin-octupole moment $\left(GSJ_2c^{-2};~J_3^\ast = 8.0\times 10^{-10}\right)$. Upper-right corner: post-Newtonian gravitoelectric moment $\left(GMJ_2 c^{-2};~J_3^\ast = 9.0\times 10^{-9}\right)$. Lower-left corner: Lense-Thirring effect $\left(GS c^{-2};~J_3^\ast = 2.0\times 10^{-8}\right)$. Lower-right corner: Schwarzschild $\left(GM c^{-2};~J_3^\ast = 6.0\times 10^{-7}\right)$. The present-day actual uncertainty in the Jovian first odd zonal is $\upsigma_{J_3} = 1.0\times 10^{-8}$ \citep[Tab.~1]{2018Natur.555..220I}. The adopted orbital configuration for the probe is $a_0 = 1.015~R,~e_0 = 0.0049,~I_0 = 50\deg,~\Omega_0 = 140\deg,~\omega_0 = 149.43\deg,~f_0 = 228.32\deg$ }\label{figJ3}
\end{center}
\end{figure*}
\begin{figure*}
\begin{center}
\centerline{
\vbox{
\begin{tabular}{cc}
\epsfysize= 5.4 cm\epsfbox{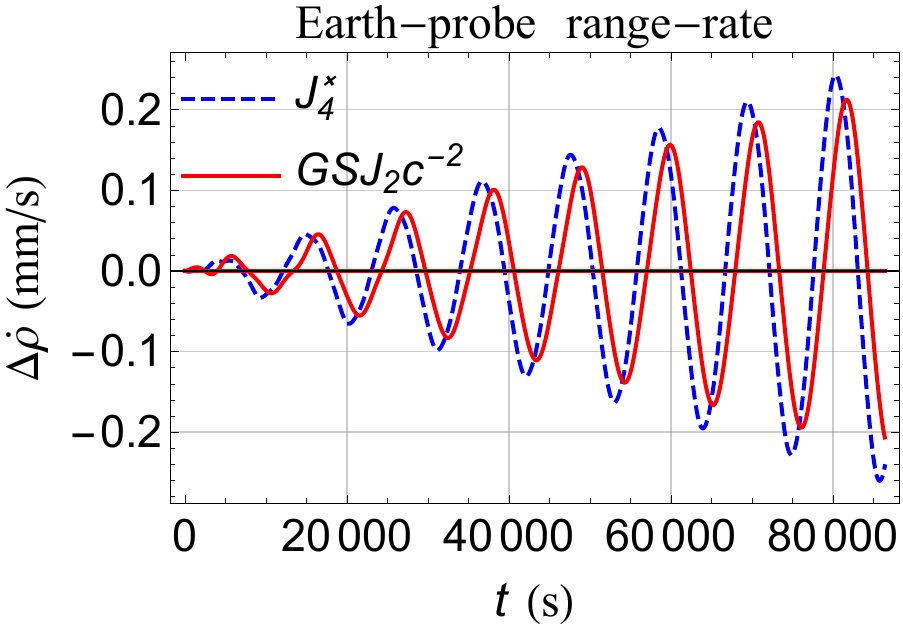}&\epsfysize= 5.4 cm\epsfbox{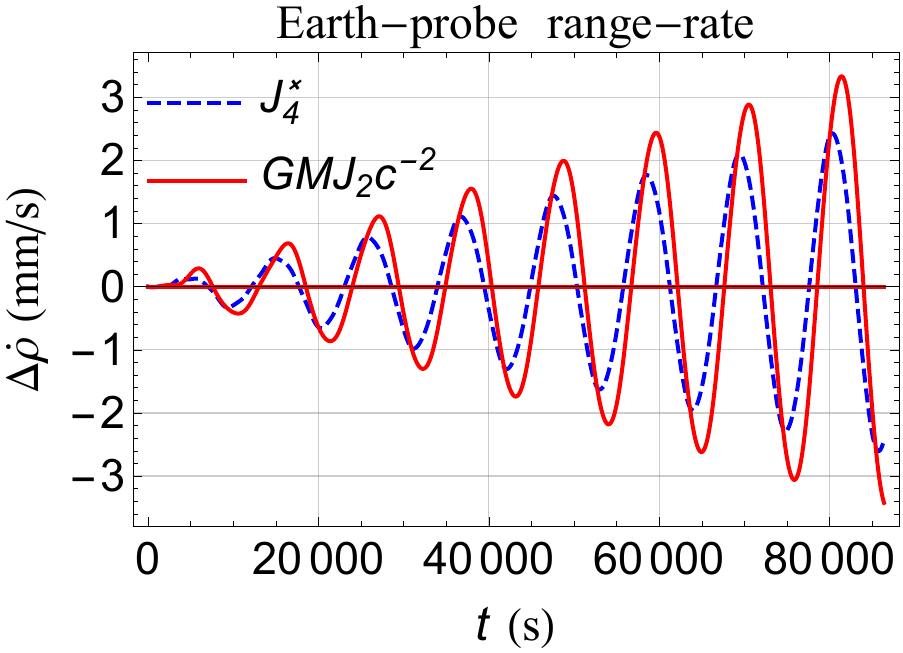}\\
\epsfysize= 5.4 cm\epsfbox{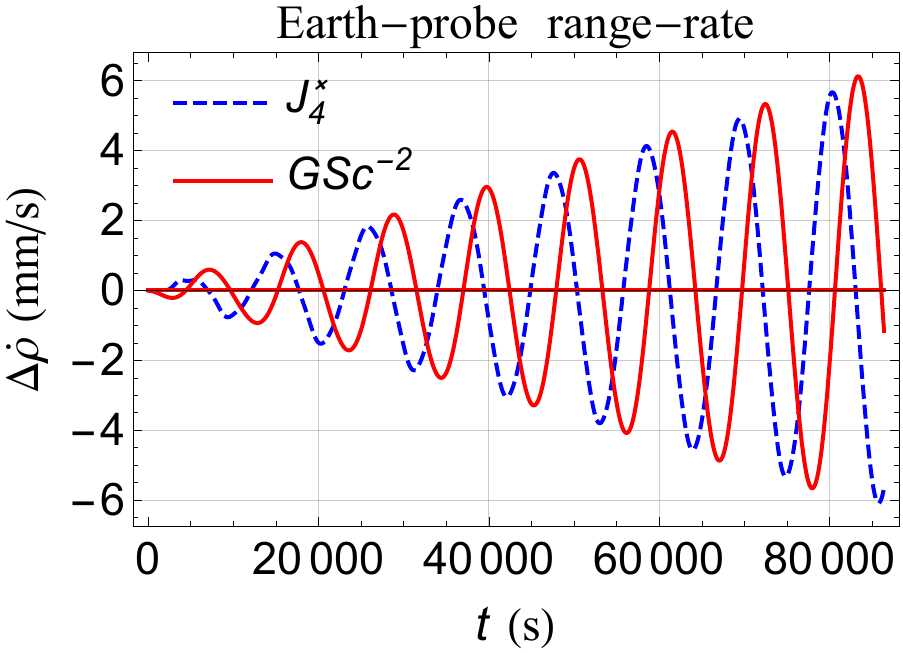}&\epsfysize= 5.4 cm\epsfbox{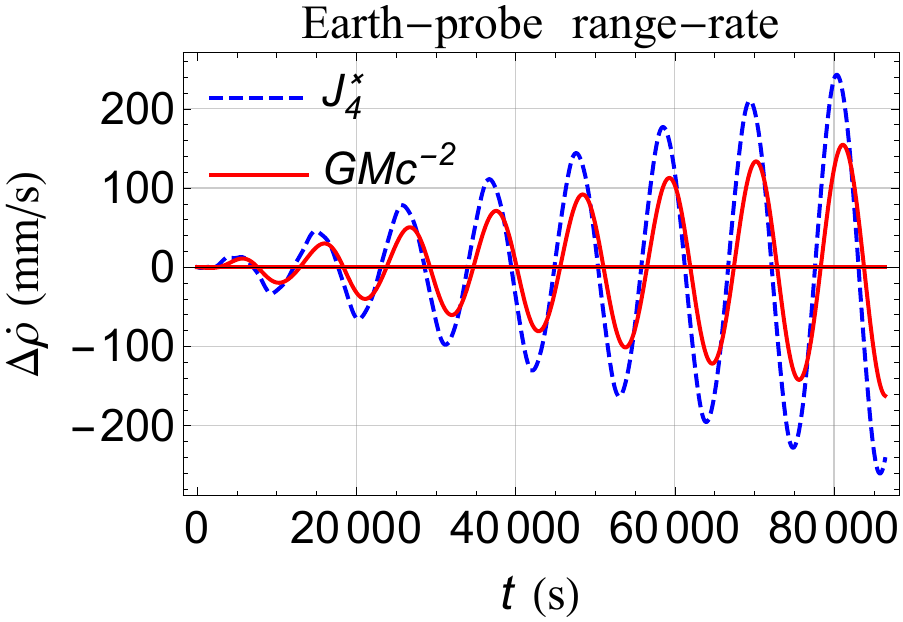}\\
\end{tabular}
}
}
\caption{Simulated range-rate signatures $\Delta\dot\rho$, in $\textrm{mm~s}^{-1}$, of a hypothetical Jovian orbiter  induced by the nominal post-Newtonian accelerations considered in the text and by the Newtonian second even zonal harmonic $J_4$ of Jupiter after 1 d. In each panel, a fictitious value $J_4^\ast$  is used in the Newtonian signature just for illustrative and comparative purposes. Indeed, it is suitably tuned from time to time in order to bring the associated classical signature to the level of the nominal post-Newtonian effect of interest,  so to inspect the mutual (de)correlations of their temporal patterns more easily. Upper-left corner: post-Newtonian gravitomagnetic spin-octupole moment $\left(GSJ_2c^{-2};~J_4^\ast = 1.2\times 10^{-10}\right)$. Upper-right corner: post-Newtonian gravitoelectric moment $\left(GMJ_2 c^{-2};~J_4^\ast = 1.2\times 10^{-9}\right)$. Lower-left corner: Lense-Thirring effect $\left(GS c^{-2};~J_4^\ast = 2.8\times 10^{-9}\right)$. Lower-right corner: Schwarzschild $\left(GM c^{-2};~J_4^\ast = 1.2\times 10^{-7}\right)$. The present-day actual uncertainty in the Jovian second even zonal is $\upsigma_{J_4} = 4\times 10^{-9}$ \citep[Tab.~1]{2018Natur.555..220I}. The adopted orbital configuration for the probe is $a_0 = 1.015~R,~e_0 = 0.0049,~I_0 = 50\deg,~\Omega_0 = 140\deg,~\omega_0 = 149.43\deg,~f_0 = 228.32\deg$ }\label{figJ4}
\end{center}
\end{figure*}
\begin{figure*}
\begin{center}
\centerline{
\vbox{
\begin{tabular}{cc}
\epsfysize= 5.4 cm\epsfbox{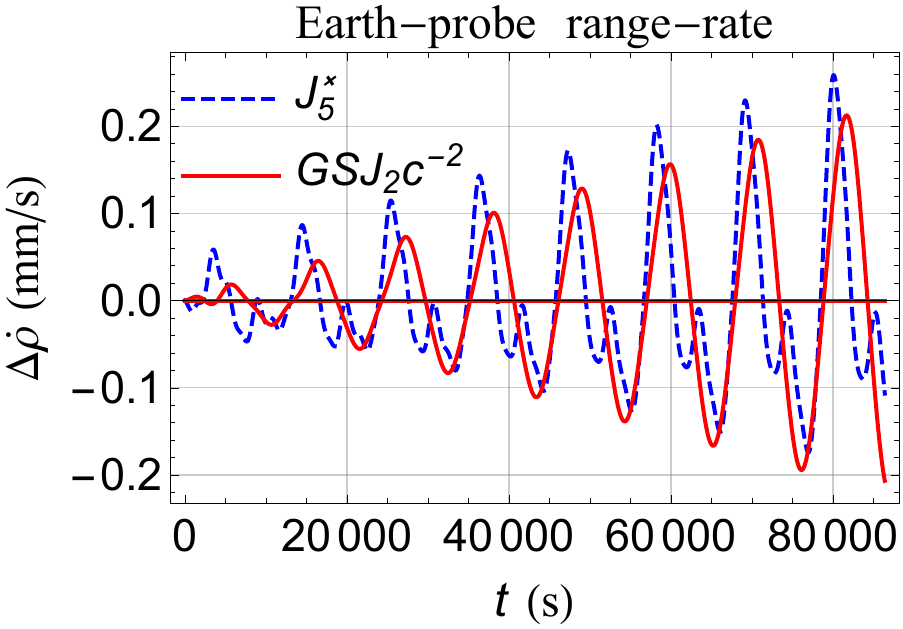}&\epsfysize= 5.4 cm\epsfbox{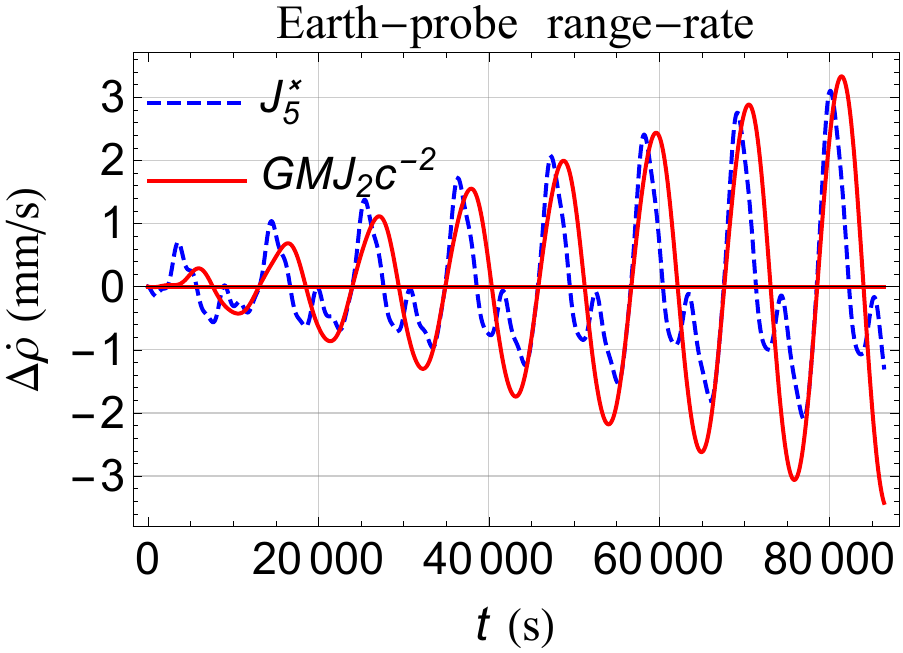}\\
\epsfysize= 5.4 cm\epsfbox{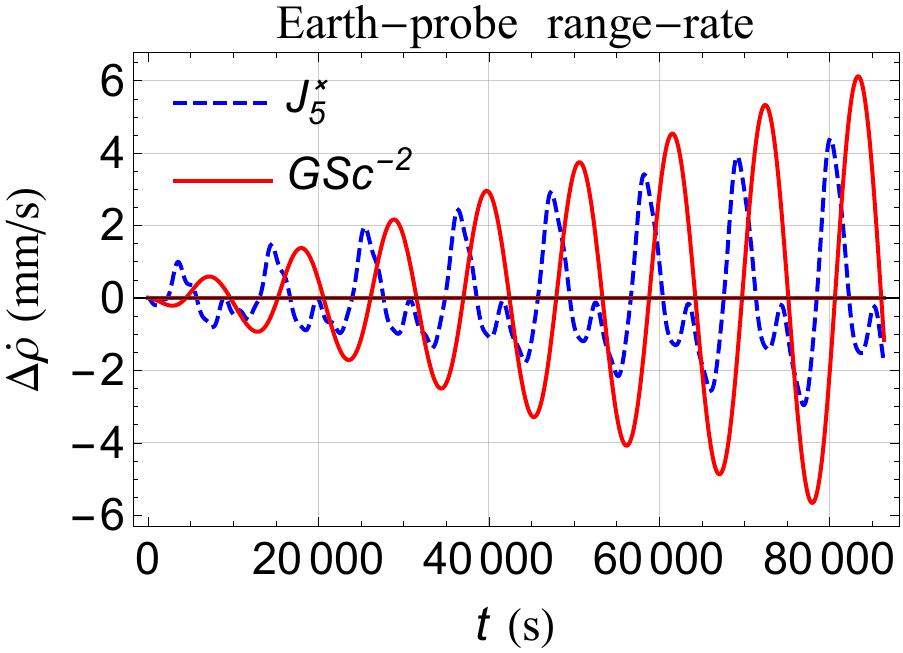}&\epsfysize= 5.4 cm\epsfbox{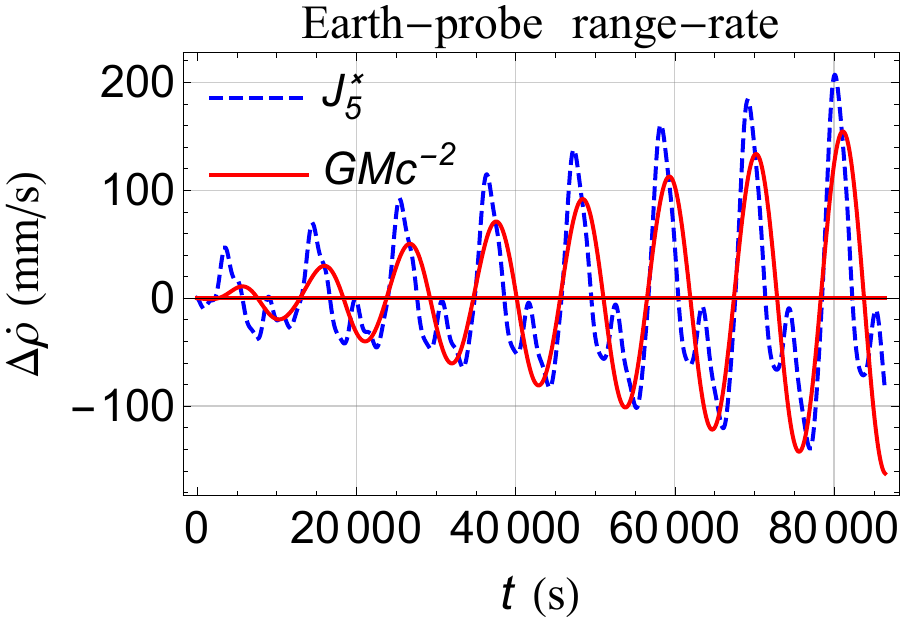}\\
\end{tabular}
}
}
\caption{Simulated range-rate signatures $\Delta\dot\rho$, in $\textrm{mm~s}^{-1}$, of a hypothetical Jovian orbiter  induced by the nominal post-Newtonian accelerations considered in the text and by the Newtonian second odd zonal harmonic $J_5$ of Jupiter after 1 d. In each panel, a fictitious value $J_5^\ast$  is used in the Newtonian signature just for illustrative and comparative purposes. Indeed, it is suitably tuned from time to time in order to bring the associated classical signature to the level of the nominal post-Newtonian effect of interest,  so to inspect the mutual (de)correlations of their temporal patterns more easily. Upper-left corner: post-Newtonian gravitomagnetic spin-octupole moment $\left(GSJ_2c^{-2};~J_5^\ast = 8.0\times 10^{-10}\right)$. Upper-right corner: post-Newtonian gravitoelectric moment $\left(GMJ_2 c^{-2};~J_5^\ast = 9.6\times 10^{-9}\right)$. Lower-left corner: Lense-Thirring effect $\left(GS c^{-2};~J_5^\ast = 1.36\times 10^{-8}\right)$. Lower-right corner: Schwarzschild $\left(GM c^{-2};~J_5^\ast = 6.4\times 10^{-7}\right)$. The present-day actual uncertainty in the Jovian second odd zonal is $\upsigma_{J_5} = 8\times 10^{-9}$ \citep[Tab.~1]{2018Natur.555..220I}. The adopted orbital configuration for the probe is $a_0 = 1.015~R,~e_0 = 0.0049,~I_0 = 50\deg,~\Omega_0 = 140\deg,~\omega_0 = 149.43\deg,~f_0 = 228.32\deg$ }\label{figJ5}
\end{center}
\end{figure*}
\begin{figure*}
\begin{center}
\centerline{
\vbox{
\begin{tabular}{cc}
\epsfysize= 5.4 cm\epsfbox{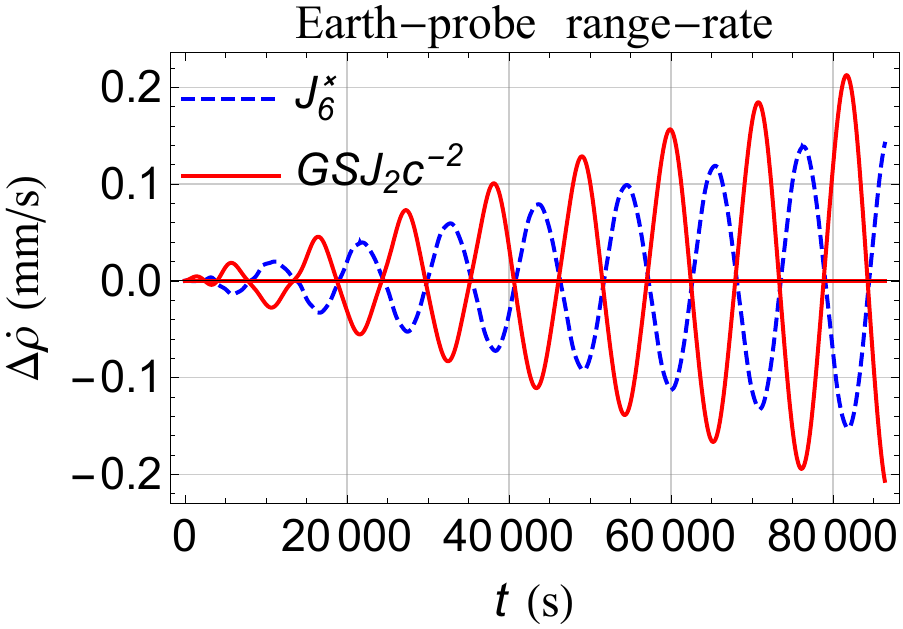}&\epsfysize= 5.4 cm\epsfbox{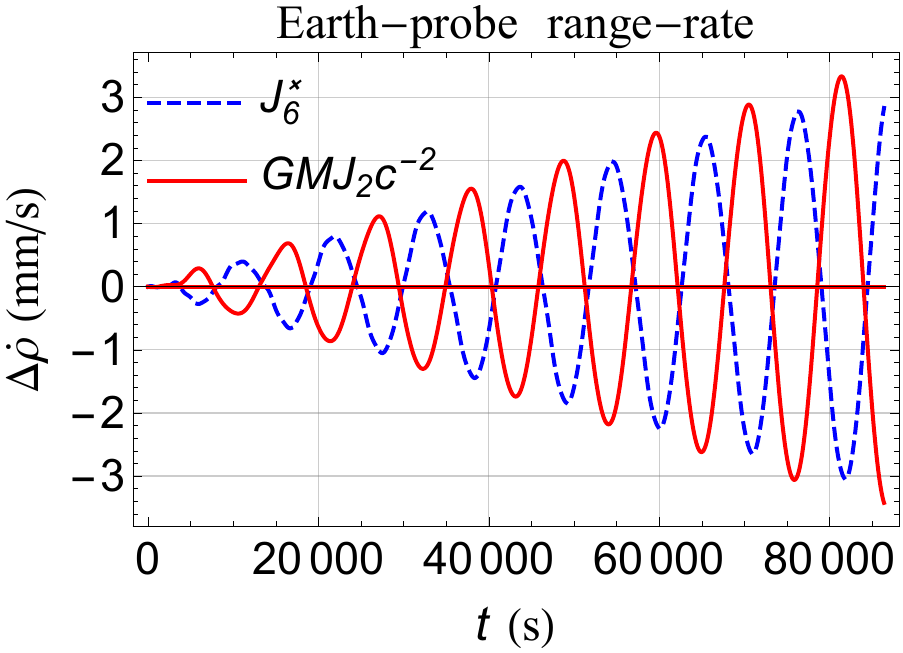}\\
\epsfysize= 5.4 cm\epsfbox{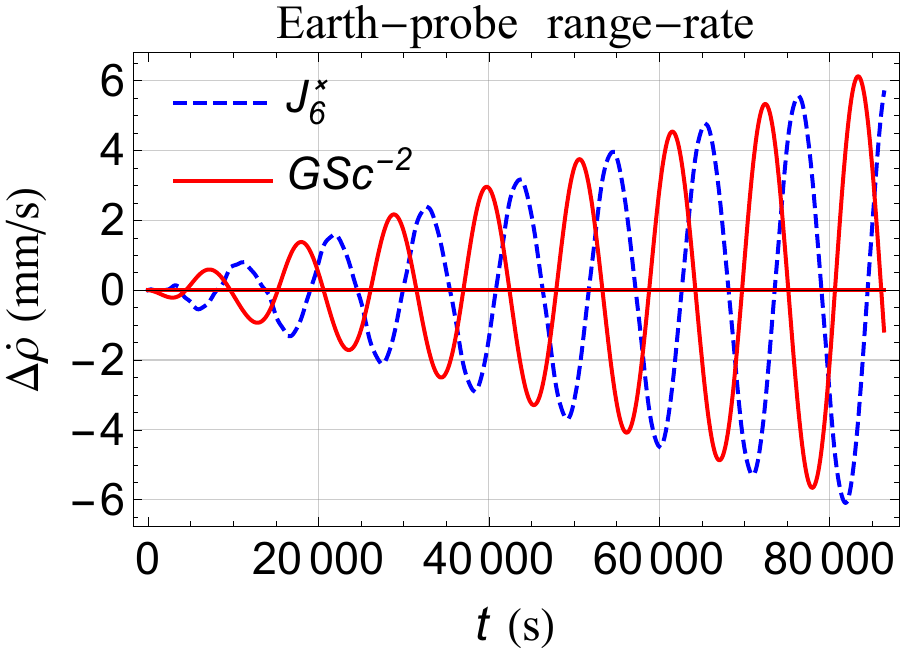}&\epsfysize= 5.4 cm\epsfbox{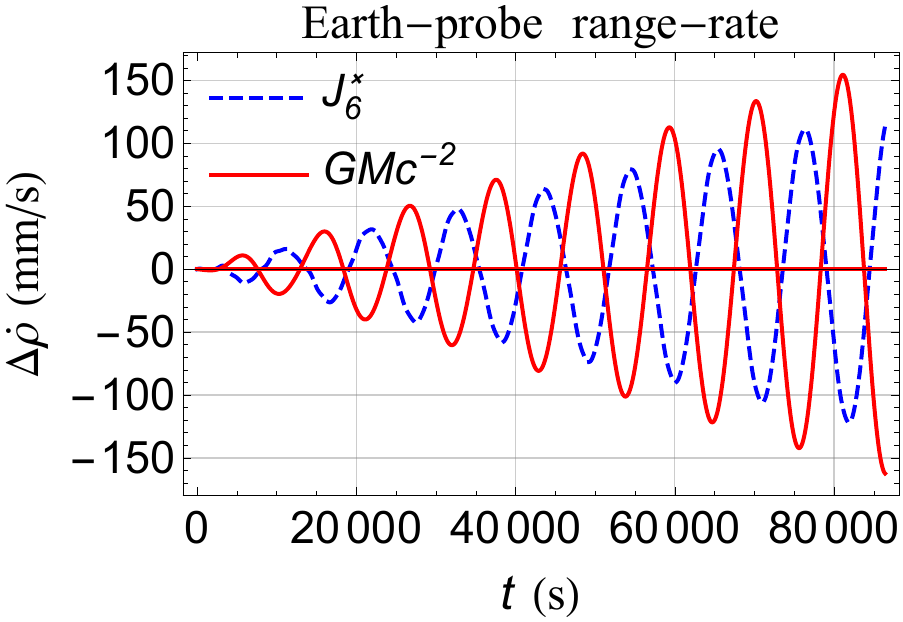}\\
\end{tabular}
}
}
\caption{Simulated range-rate signatures $\Delta\dot\rho$, in $\textrm{mm~s}^{-1}$, of a hypothetical Jovian orbiter  induced by the nominal post-Newtonian accelerations considered in the text and by the Newtonian third even zonal harmonic $J_6$ of Jupiter after 1 d. In each panel, a fictitious value $J_6^\ast$  is used in the Newtonian signature just for illustrative and comparative purposes. Indeed, it is suitably tuned from time to time in order to bring the associated classical signature to the level of the nominal post-Newtonian effect of interest,  so to inspect the mutual (de)correlations of their temporal patterns more easily. Upper-left corner: post-Newtonian gravitomagnetic spin-octupole moment $\left(GSJ_2c^{-2};~J_6^\ast = 9.0\times 10^{-11}\right)$. Upper-right corner: post-Newtonian gravitoelectric moment $\left(GMJ_2 c^{-2};~J_6^\ast = 1.8\times 10^{-9}\right)$. Lower-left corner: Lense-Thirring effect $\left(GS c^{-2};~J_6^\ast = 3.6\times 10^{-9}\right)$. Lower-right corner: Schwarzschild $\left(GM c^{-2};~J_6^\ast = 7.2\times 10^{-8}\right)$. The present-day actual uncertainty in the Jovian third even zonal is $\upsigma_{J_6} = 9\times 10^{-9}$ \citep[Tab.~1]{2018Natur.555..220I}. The adopted orbital configuration for the probe is $a_0 = 1.015~R,~e_0 = 0.0049,~I_0 = 50\deg,~\Omega_0 = 140\deg,~\omega_0 = 149.43\deg,~f_0 = 228.32\deg$ }\label{figJ6}
\end{center}
\end{figure*}
\begin{figure*}
\begin{center}
\centerline{
\vbox{
\begin{tabular}{cc}
\epsfysize= 5.4 cm\epsfbox{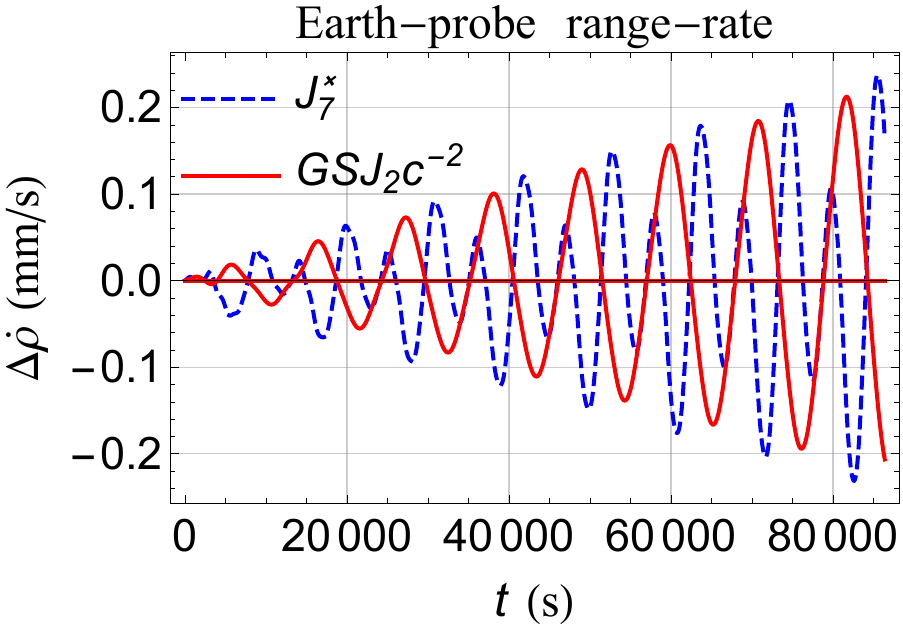}&\epsfysize= 5.4 cm\epsfbox{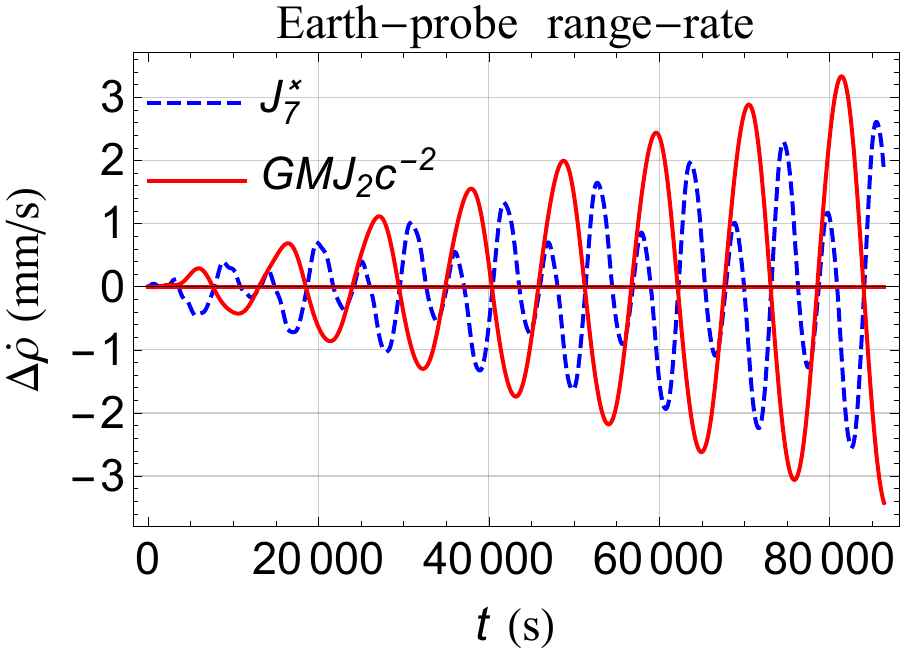}\\
\epsfysize= 5.4 cm\epsfbox{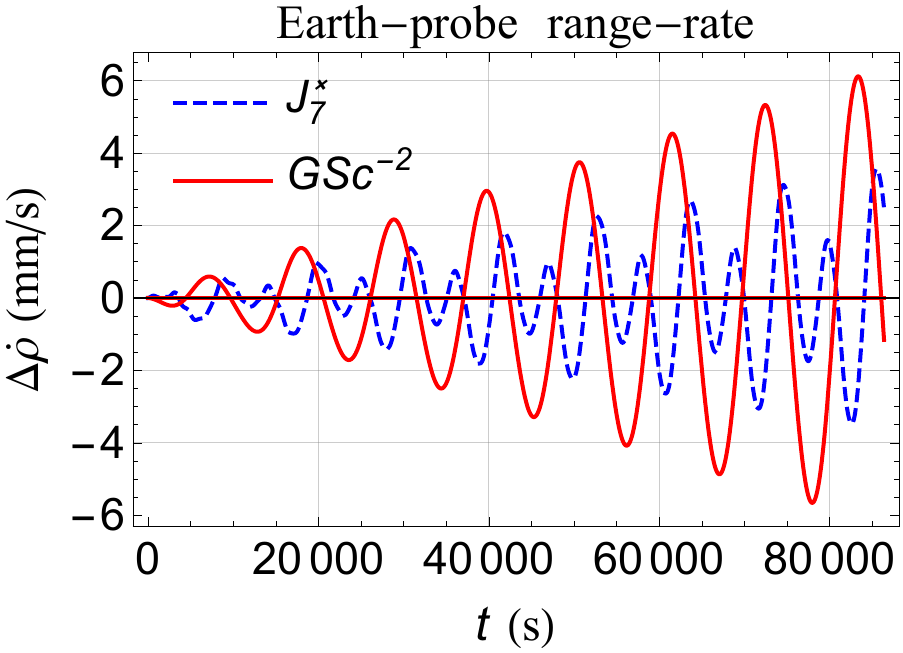}&\epsfysize= 5.4 cm\epsfbox{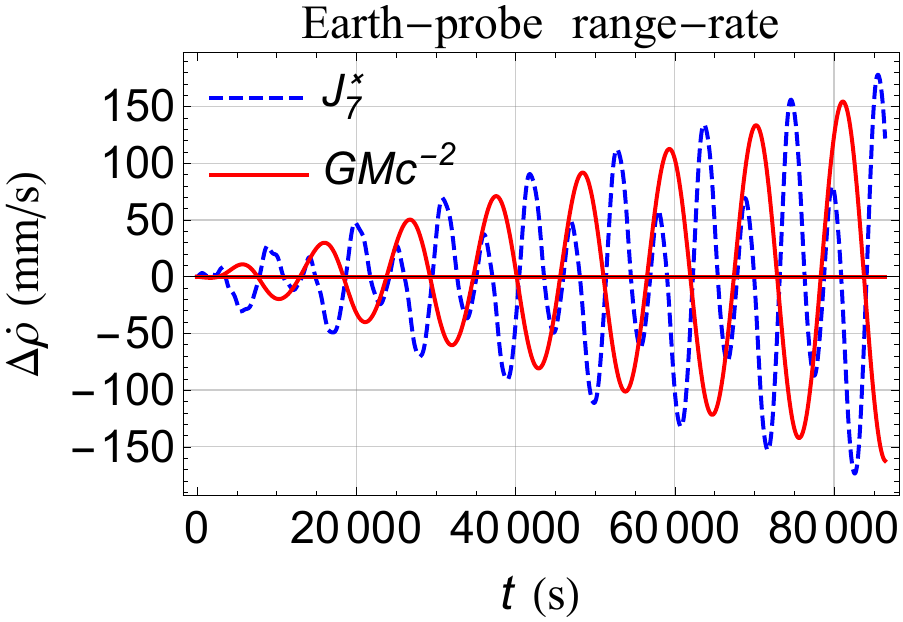}\\
\end{tabular}
}
}
\caption{Simulated range-rate signatures $\Delta\dot\rho$, in $\textrm{mm~s}^{-1}$, of a hypothetical Jovian orbiter  induced by the nominal post-Newtonian accelerations considered in the text and by the Newtonian third odd zonal harmonic $J_7$ of Jupiter after 1 d. In each panel, a fictitious value $J_7^\ast$  is used in the Newtonian signature just for illustrative and comparative purposes. Indeed, it is suitably tuned from time to time in order to bring the associated classical signature to the level of the nominal post-Newtonian effect of interest,  so to inspect the mutual (de)correlations of their temporal patterns more easily. Upper-left corner: post-Newtonian gravitomagnetic spin-octupole moment $\left(GSJ_2c^{-2};~J_7^\ast = 3.4\times 10^{-10}\right)$. Upper-right corner: post-Newtonian gravitoelectric moment $\left(GMJ_2 c^{-2};~J_7^\ast = 3.74\times 10^{-9}\right)$. Lower-left corner: Lense-Thirring effect $\left(GS c^{-2};~J_7^\ast = 5.1\times 10^{-9}\right)$. Lower-right corner: Schwarzschild $\left(GM c^{-2};~J_7^\ast = 2.55\times 10^{-7}\right)$. The present-day actual uncertainty in the Jovian third odd zonal is $\upsigma_{J_7} = 1.7\times 10^{-8}$ \citep[Tab.~1]{2018Natur.555..220I}. The adopted orbital configuration for the probe is $a_0 = 1.015~R,~e_0 = 0.0049,~I_0 = 50\deg,~\Omega_0 = 140\deg,~\omega_0 = 149.43\deg,~f_0 = 228.32\deg$ }\label{figJ7}
\end{center}
\end{figure*}
\begin{figure*}
\begin{center}
\centerline{
\vbox{
\begin{tabular}{cc}
\epsfysize= 5.4 cm\epsfbox{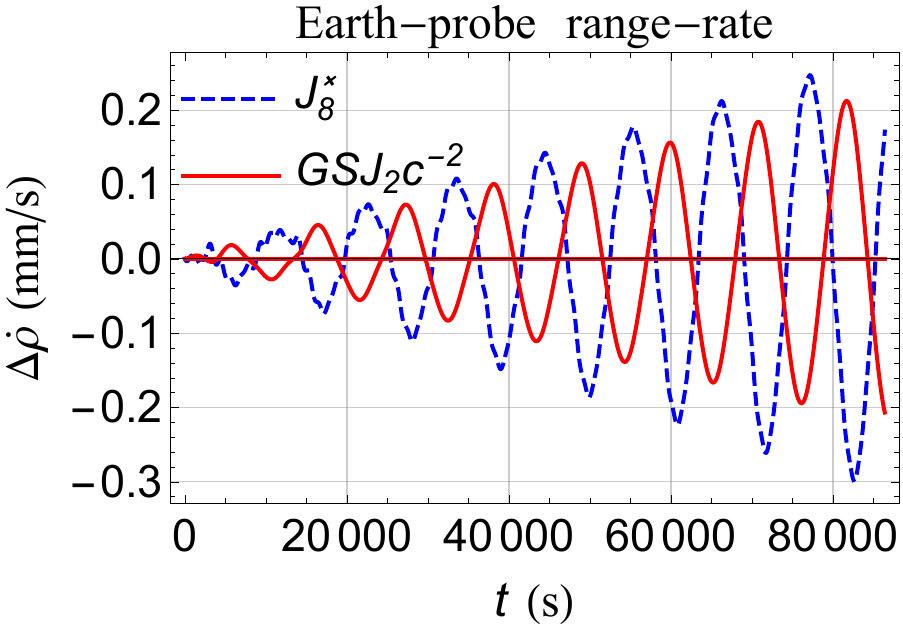}&\epsfysize= 5.4 cm\epsfbox{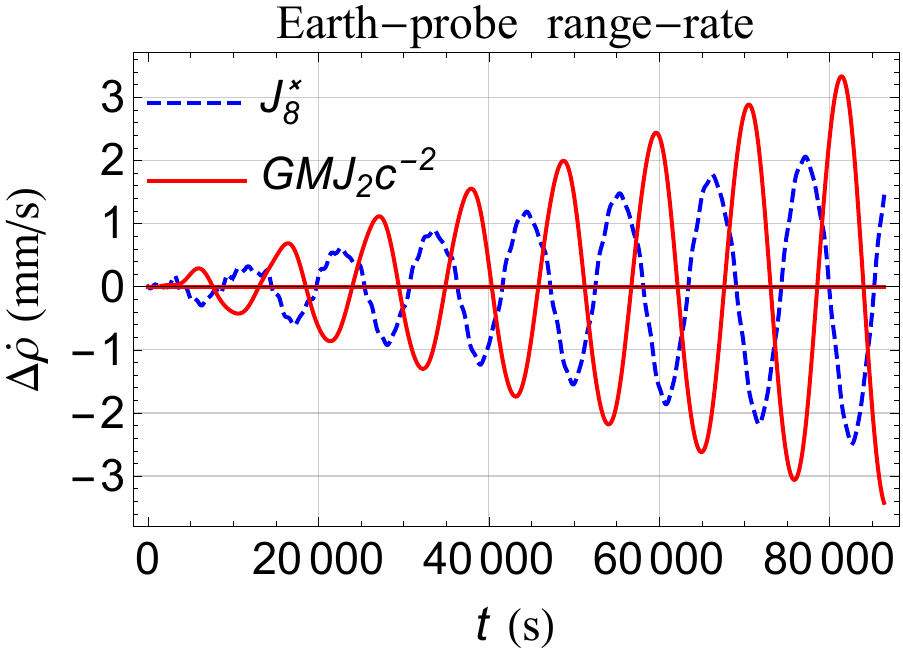}\\
\epsfysize= 5.4 cm\epsfbox{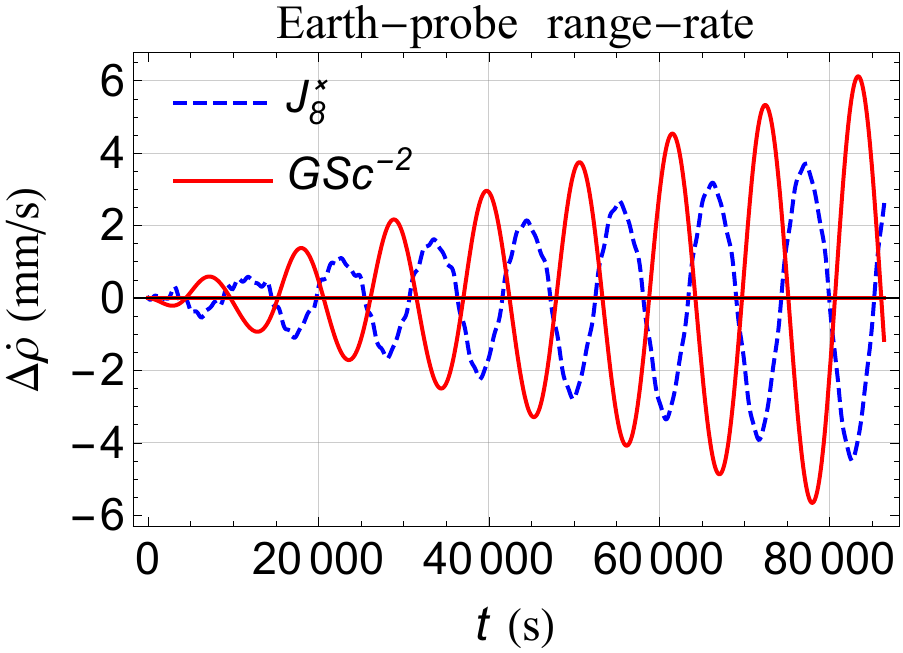}&\epsfysize= 5.4 cm\epsfbox{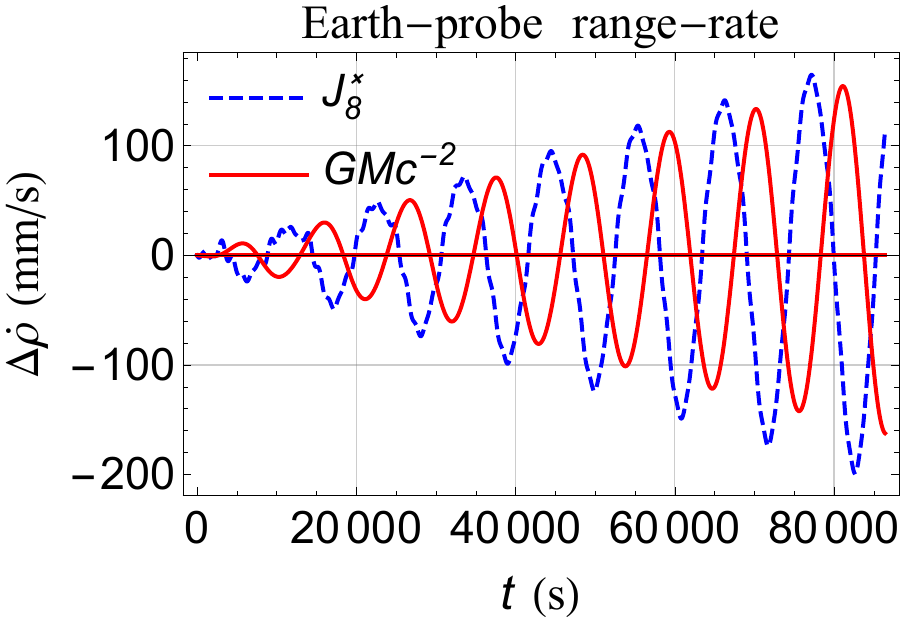}\\
\end{tabular}
}
}
\caption{Simulated range-rate signatures $\Delta\dot\rho$, in $\textrm{mm~s}^{-1}$, of a hypothetical Jovian orbiter  induced by the nominal post-Newtonian accelerations considered in the text and by the Newtonian fourth even zonal harmonic $J_8$ of Jupiter after 1 d. In each panel, a fictitious value $J_8^\ast$  is used in the Newtonian signature just for illustrative and comparative purposes. Indeed, it is suitably tuned from time to time in order to bring the associated classical signature to the level of the nominal post-Newtonian effect of interest,  so to inspect the mutual (de)correlations of their temporal patterns more easily. Upper-left corner: post-Newtonian gravitomagnetic spin-octupole moment $\left(GSJ_2c^{-2};~J_8^\ast = 7.5\times 10^{-10}\right)$. Upper-right corner: post-Newtonian gravitoelectric moment $\left(GMJ_2 c^{-2};~J_8^\ast = 6.25\times 10^{-9}\right)$. Lower-left corner: Lense-Thirring effect $\left(GS c^{-2};~J_8^\ast = 1.125\times 10^{-8}\right)$. Lower-right corner: Schwarzschild $\left(GM c^{-2};~J_8^\ast = 5.0\times 10^{-7}\right)$. The present-day actual uncertainty in the Jovian fourth even zonal is $\upsigma_{J_8} = 2.5\times 10^{-8}$ \citep[Tab.~1]{2018Natur.555..220I}. The adopted orbital configuration for the probe is $a_0 = 1.015~R,~e_0 = 0.0049,~I_0 = 50\deg,~\Omega_0 = 140\deg,~\omega_0 = 149.43\deg,~f_0 = 228.32\deg$ }\label{figJ8}
\end{center}
\end{figure*}
\begin{figure*}
\begin{center}
\centerline{
\vbox{
\begin{tabular}{cc}
\epsfysize= 5.4 cm\epsfbox{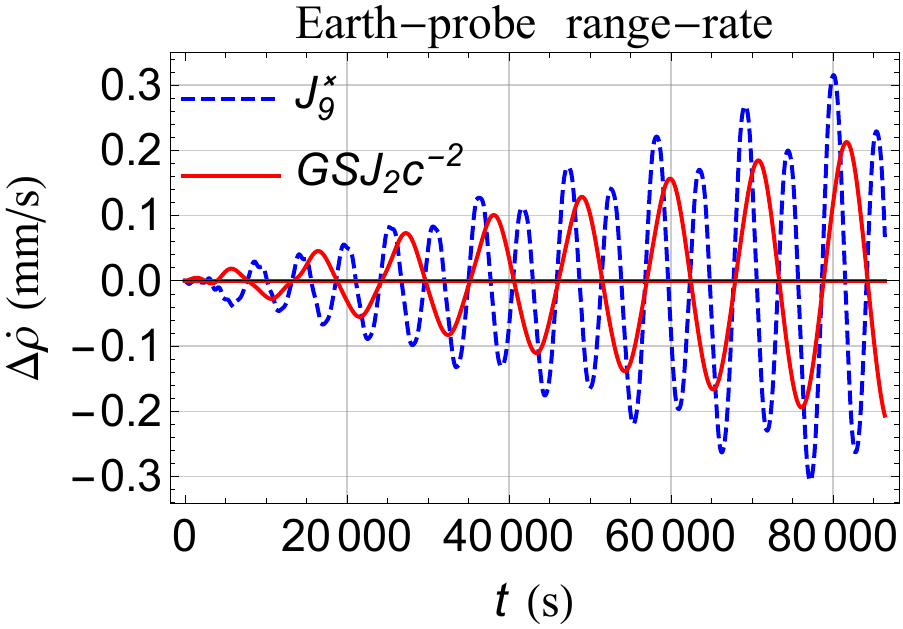}&\epsfysize= 5.4 cm\epsfbox{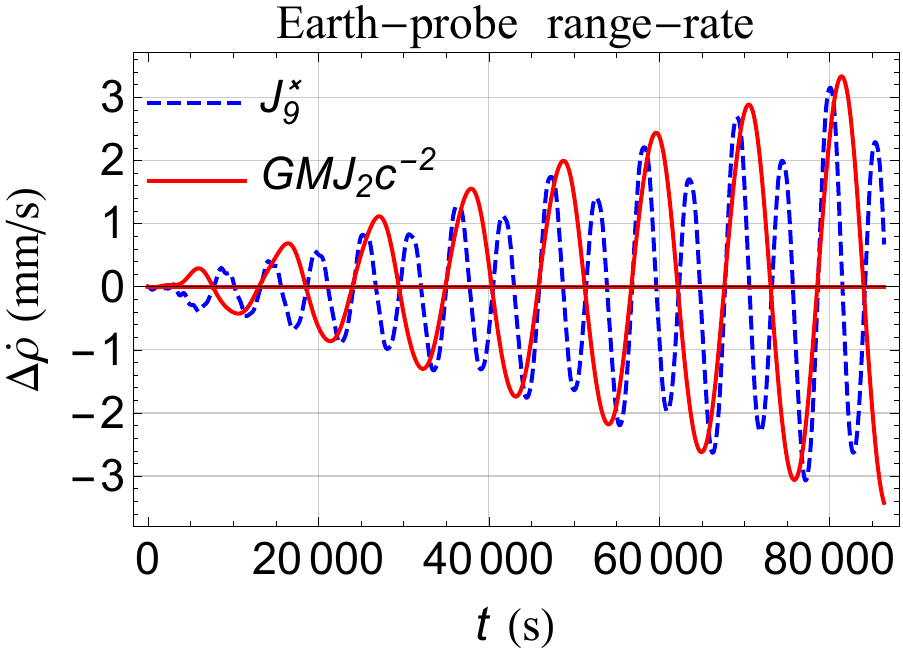}\\
\epsfysize= 5.4 cm\epsfbox{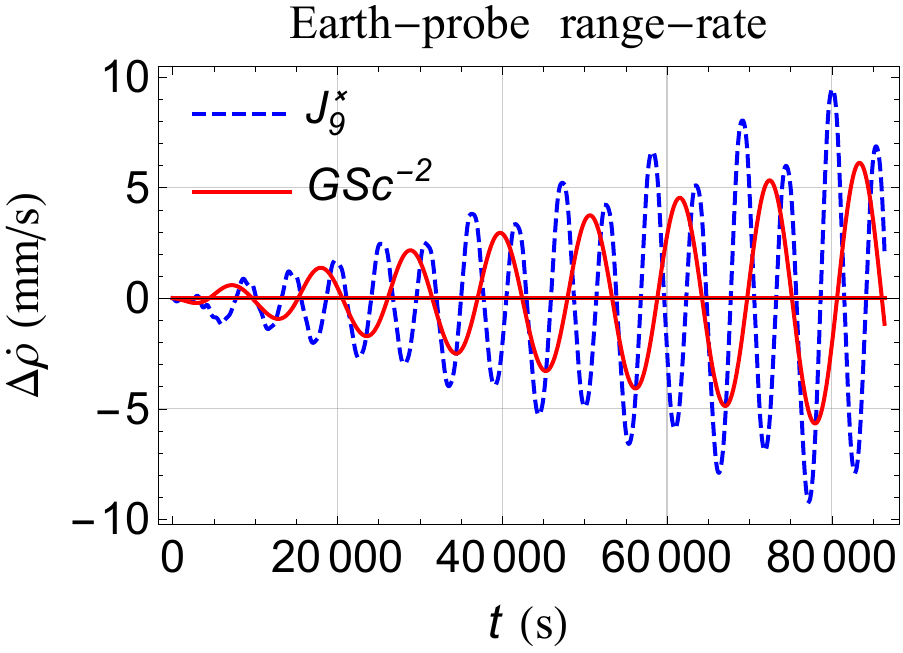}&\epsfysize= 5.4 cm\epsfbox{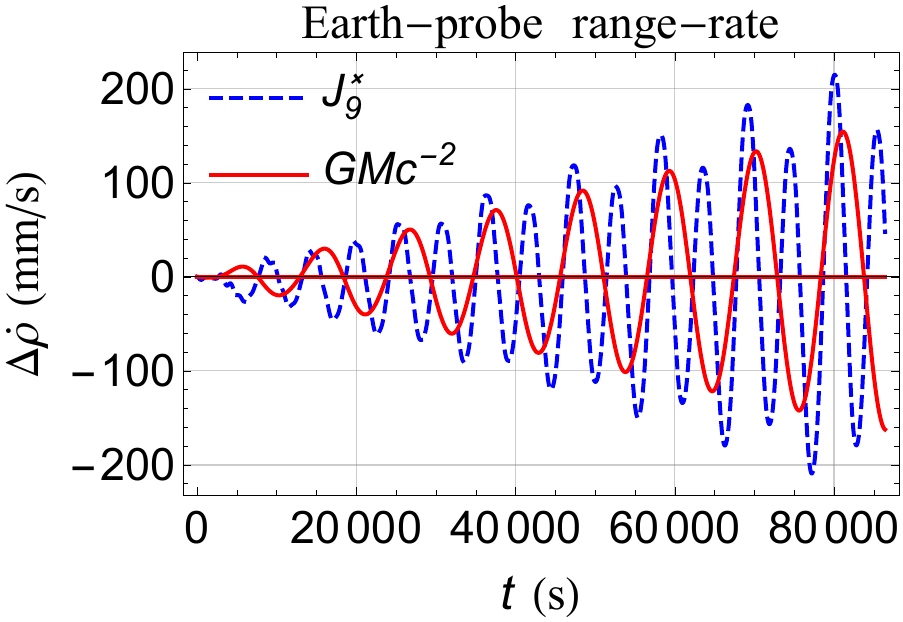}\\
\end{tabular}
}
}
\caption{Simulated range-rate signatures $\Delta\dot\rho$, in $\textrm{mm~s}^{-1}$, of a hypothetical Jovian orbiter  induced by the nominal post-Newtonian accelerations considered in the text and by the Newtonian fourth odd zonal harmonic $J_9$ of Jupiter after 1 d. In each panel, a fictitious value $J_9^\ast$  is used in the Newtonian signature just for illustrative and comparative purposes. Indeed, it is suitably tuned from time to time in order to bring the associated classical signature to the level of the nominal post-Newtonian effect of interest,  so to inspect the mutual (de)correlations of their temporal patterns more easily. Upper-left corner: post-Newtonian gravitomagnetic spin-octupole moment $\left(GSJ_2c^{-2};~J_9^\ast = 4.4\times 10^{-10}\right)$. Upper-right corner: post-Newtonian gravitoelectric moment $\left(GMJ_2 c^{-2};~J_9^\ast = 4.4\times 10^{-9}\right)$. Lower-left corner: Lense-Thirring effect $\left(GS c^{-2};~J_9^\ast = 1.32\times 10^{-8}\right)$. Lower-right corner: Schwarzschild $\left(GM c^{-2};~J_9^\ast = 3.0\times 10^{-7}\right)$. The present-day actual uncertainty in the Jovian fourth odd zonal is $\upsigma_{J_9} = 4.4\times 10^{-8}$ \citep[Tab.~1]{2018Natur.555..220I}. The adopted orbital configuration for the probe is $a_0 = 1.015~R,~e_0 = 0.0049,~I_0 = 50\deg,~\Omega_0 = 140\deg,~\omega_0 = 149.43\deg,~f_0 = 228.32\deg$ }\label{figJ9}
\end{center}
\end{figure*}
\begin{figure*}
\begin{center}
\centerline{
\vbox{
\begin{tabular}{cc}
\epsfysize= 5.4 cm\epsfbox{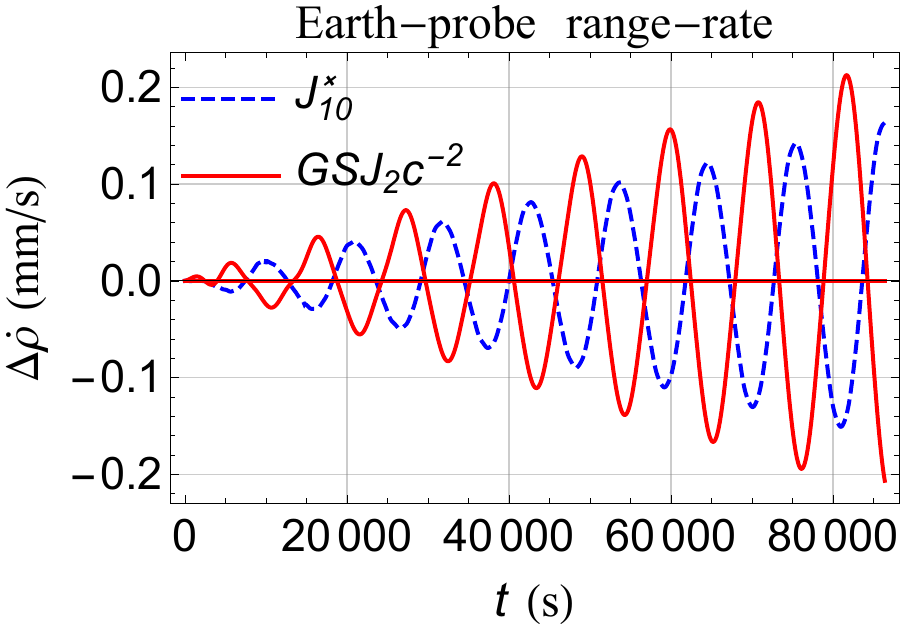}&\epsfysize= 5.4 cm\epsfbox{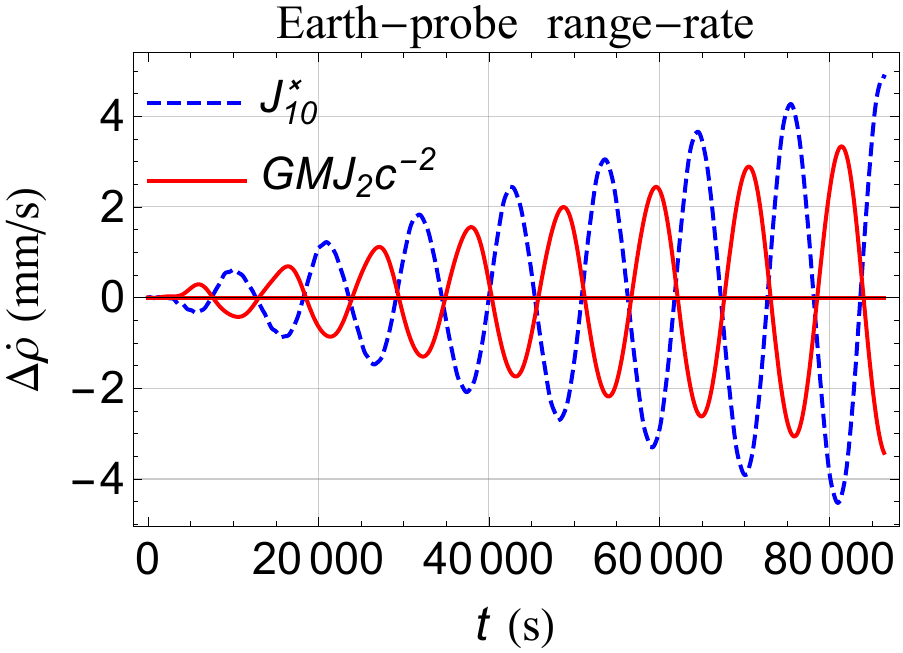}\\
\epsfysize= 5.4 cm\epsfbox{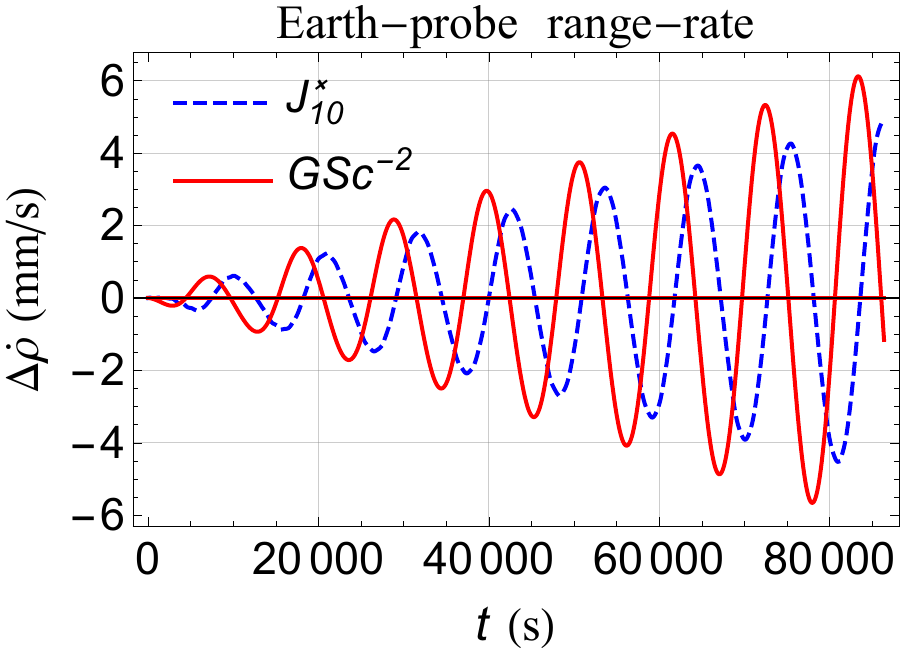}&\epsfysize= 5.4 cm\epsfbox{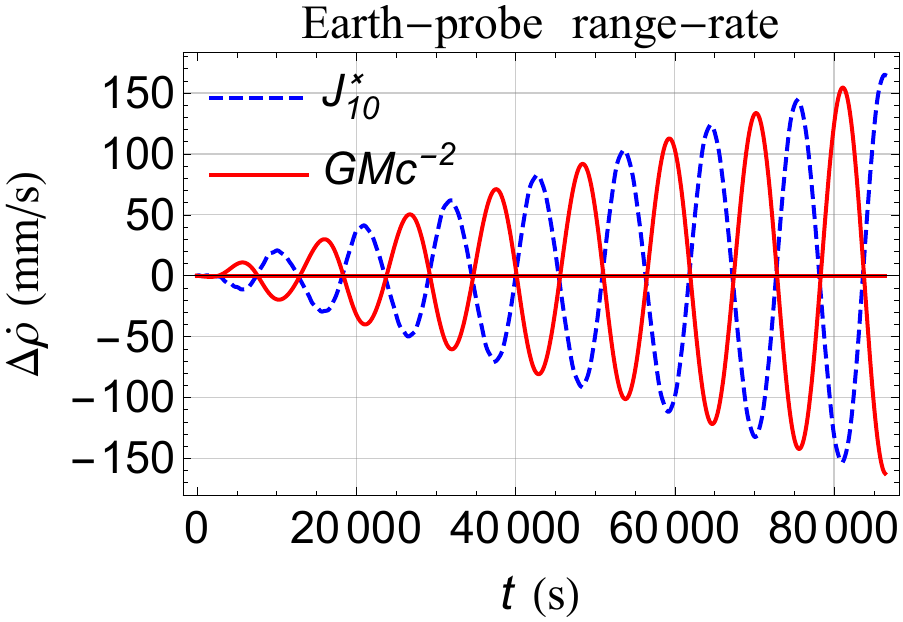}\\
\end{tabular}
}
}
\caption{Simulated range-rate signatures $\Delta\dot\rho$, in $\textrm{mm~s}^{-1}$, of a hypothetical Jovian orbiter  induced by the nominal post-Newtonian accelerations considered in the text and by the Newtonian fifth even zonal harmonic $J_{10}$ of Jupiter after 1 d. In each panel, a fictitious value $J_{10}^\ast$  is used in the Newtonian signature just for illustrative and comparative purposes. Indeed, it is suitably tuned from time to time in order to bring the associated classical signature to the level of the nominal post-Newtonian effect of interest,  so to inspect the mutual (de)correlations of their temporal patterns more easily. Upper-left corner: post-Newtonian gravitomagnetic spin-octupole moment $\left(GSJ_2c^{-2};~J_{10}^\ast = 6.9\times 10^{-11}\right)$. Upper-right corner: post-Newtonian gravitoelectric moment $\left(GMJ_2 c^{-2};~J_{10}^\ast = 2.07\times 10^{-9}\right)$. Lower-left corner: Lense-Thirring effect $\left(GSc^{-2};~J_{10}^\ast = 2.07\times 10^{-9}\right)$. Lower-right corner: Schwarzschild $\left(GM c^{-2};~J_{10}^\ast = 7\times 10^{-8}\right)$. The present-day actual uncertainty in the Jovian fifth even zonal is $\upsigma_{J_{10}} = 6.9\times 10^{-8}$ \citep[Tab.~1]{2018Natur.555..220I}. The adopted orbital configuration for the probe is $a_0 = 1.015~R,~e_0 = 0.0049,~I_0 = 50\deg,~\Omega_0 = 140\deg,~\omega_0 = 149.43\deg,~f_0 = 228.32\deg$ }\label{figJ10}
\end{center}
\end{figure*}
\begin{figure*}
\begin{center}
\centerline{
\vbox{
\begin{tabular}{cc}
\epsfysize= 5.4 cm\epsfbox{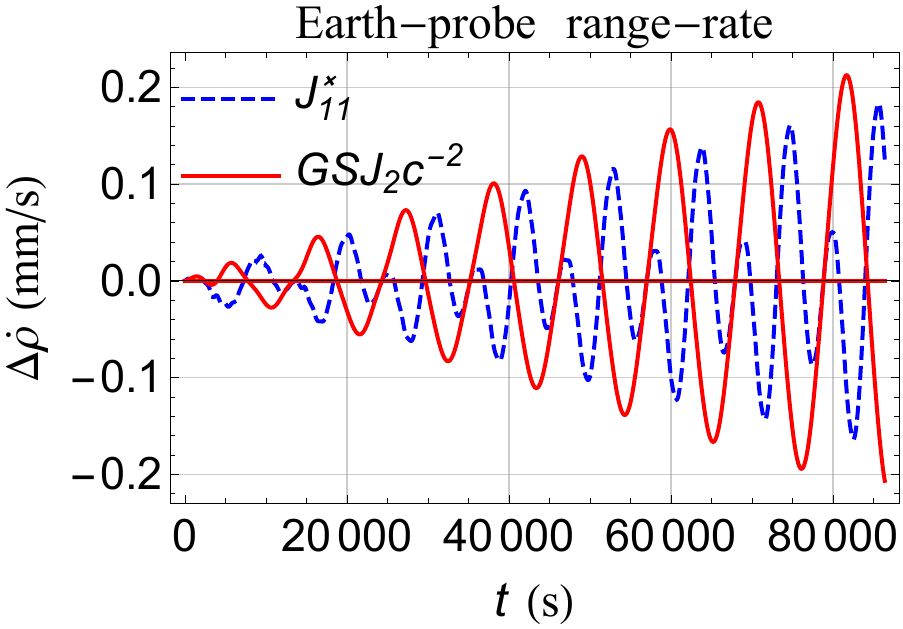}&\epsfysize= 5.4 cm\epsfbox{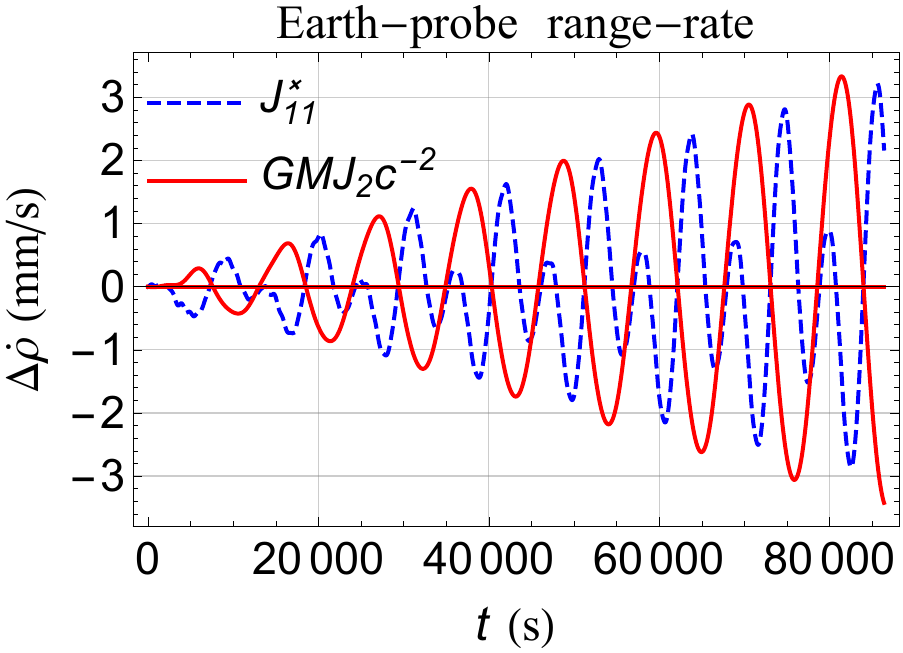}\\
\epsfysize= 5.4 cm\epsfbox{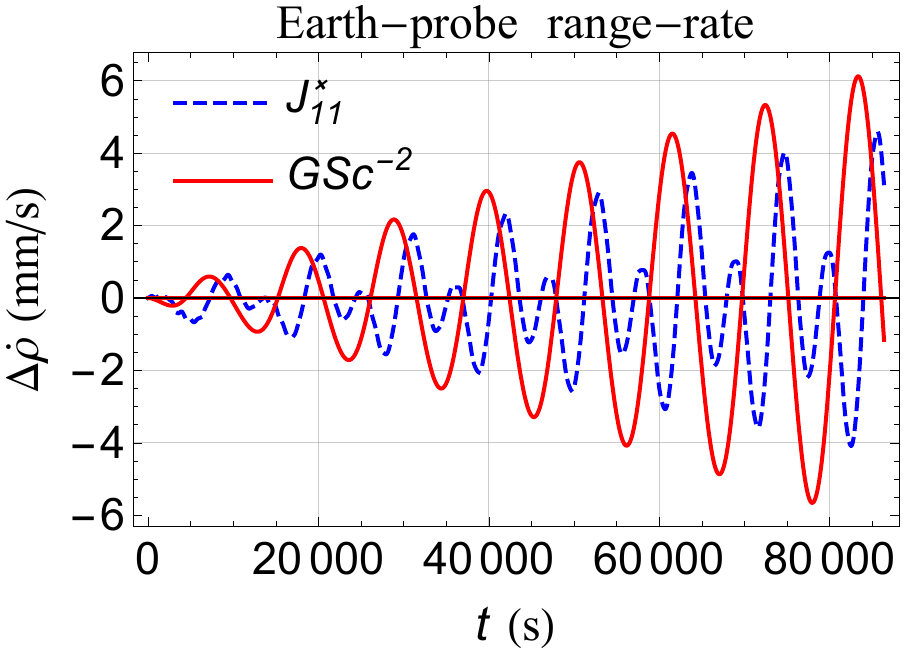}&\epsfysize= 5.4 cm\epsfbox{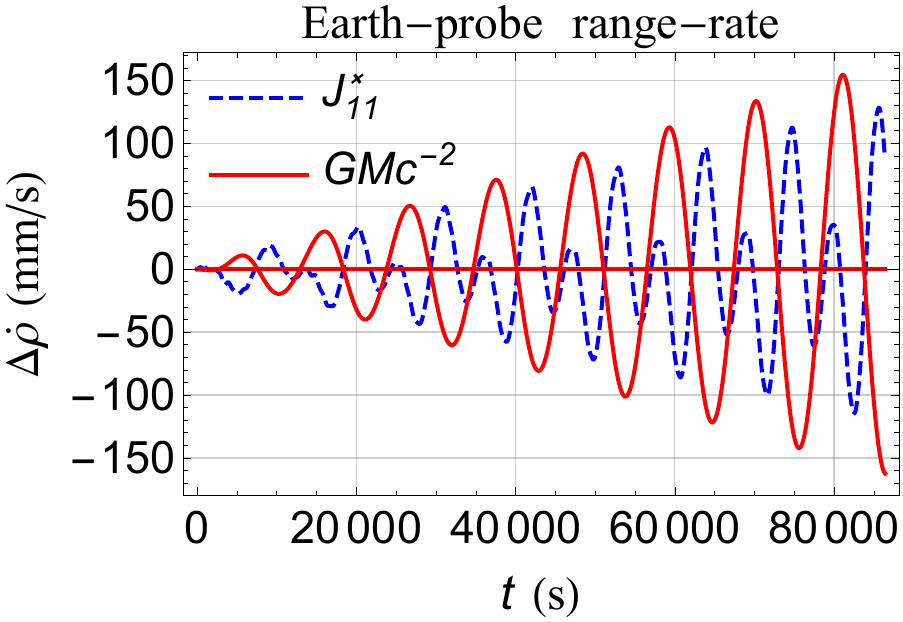}\\
\end{tabular}
}
}
\caption{Simulated range-rate signatures $\Delta\dot\rho$, in $\textrm{mm~s}^{-1}$, of a hypothetical Jovian orbiter  induced by the nominal post-Newtonian accelerations considered in the text and by the Newtonian fifth odd zonal harmonic $J_{11}$ of Jupiter after 1 d. In each panel, a fictitious value $J_{11}^\ast$  is used in the Newtonian signature just for illustrative and comparative purposes. Indeed, it is suitably tuned from time to time in order to bring the associated classical signature to the level of the nominal post-Newtonian effect of interest,  so to inspect the mutual (de)correlations of their temporal patterns more easily. Upper-left corner: post-Newtonian gravitomagnetic spin-octupole moment $\left(GSJ_2c^{-2};~J_{11}^\ast = 2.24\times 10^{-10}\right)$. Upper-right corner: post-Newtonian gravitoelectric moment $\left(GMJ_2 c^{-2};~J_{11}^\ast = 3.92\times 10^{-9}\right)$. Lower-left corner: Lense-Thirring effect $\left(GS c^{-2};~J_{11}^\ast = 5.6\times 10^{-9}\right)$. Lower-right corner: Schwarzschild $\left(GM c^{-2};~J_{11}^\ast = 1.568\times 10^{-7}\right)$. The present-day actual uncertainty in the Jovian fifth odd zonal is $\upsigma_{J_{11}} = 1.12\times 10^{-7}$ \citep[Tab.~1]{2018Natur.555..220I}. The adopted orbital configuration for the probe is $a_0 = 1.015~R,~e_0 = 0.0049,~I_0 = 50\deg,~\Omega_0 = 140\deg,~\omega_0 = 149.43\deg,~f_0 = 228.32\deg$ }\label{figJ11}
\end{center}
\end{figure*}
\begin{figure*}
\begin{center}
\centerline{
\vbox{
\begin{tabular}{cc}
\epsfysize= 5.4 cm\epsfbox{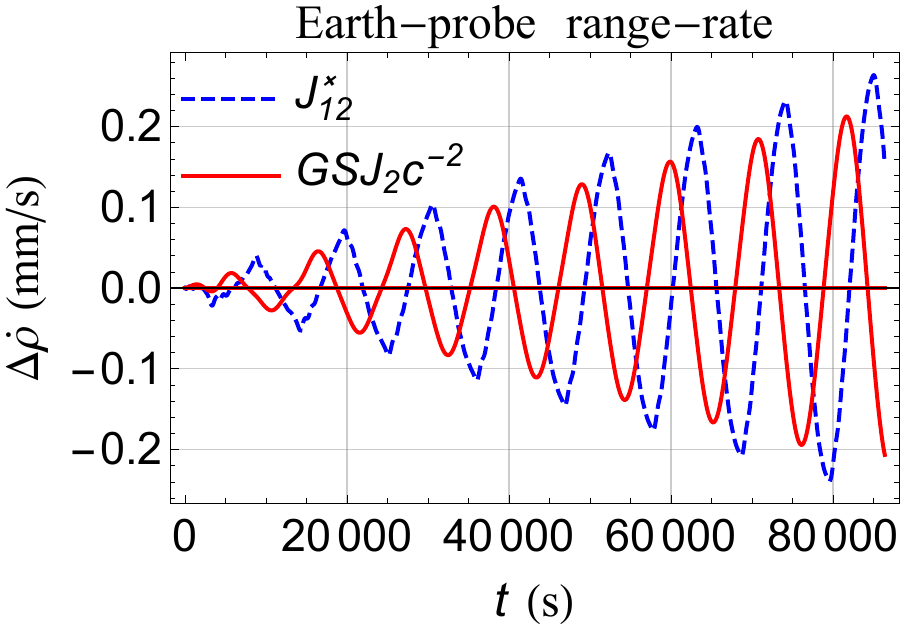}&\epsfysize= 5.4 cm\epsfbox{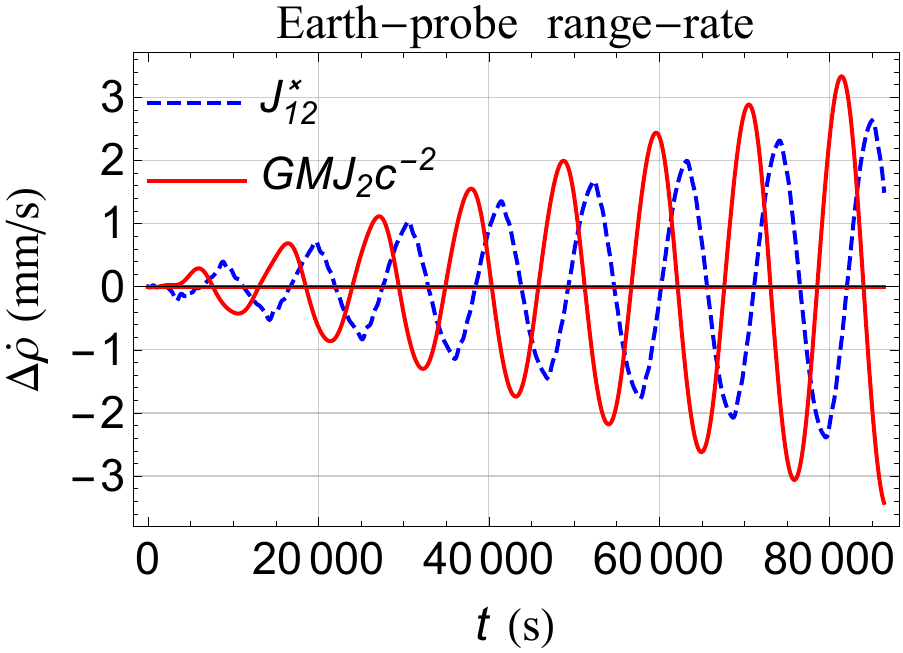}\\
\epsfysize= 5.4 cm\epsfbox{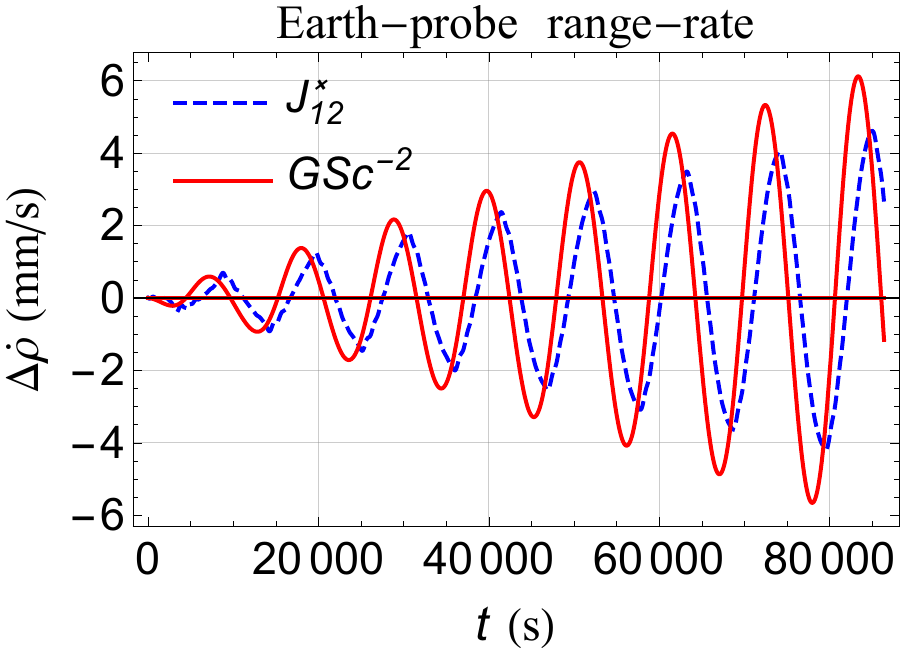}&\epsfysize= 5.4 cm\epsfbox{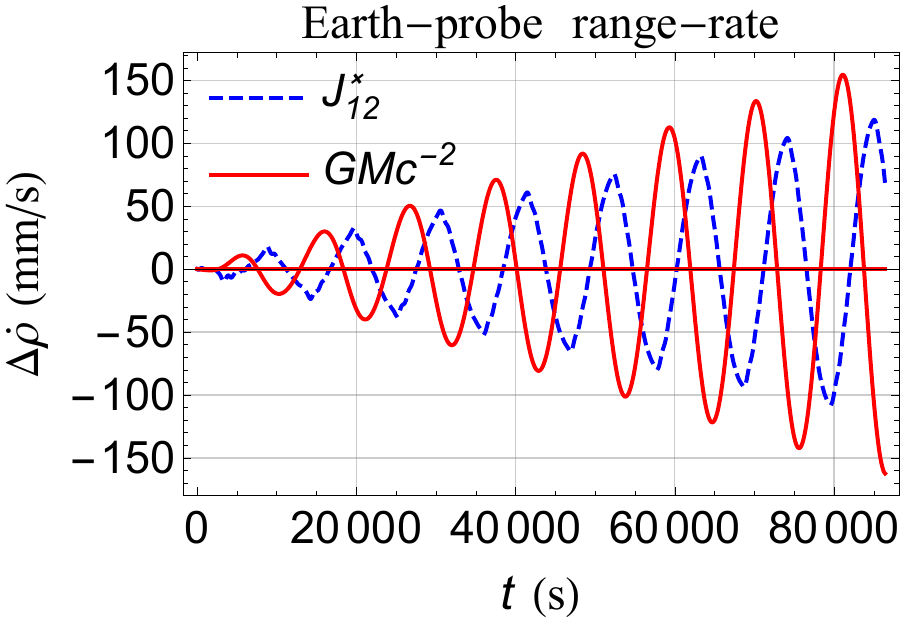}\\
\end{tabular}
}
}
\caption{Simulated range-rate signatures $\Delta\dot\rho$, in $\textrm{mm~s}^{-1}$, of a hypothetical Jovian orbiter  induced by the nominal post-Newtonian accelerations considered in the text and by the Newtonian fifth odd zonal harmonic $J_{12}$ of Jupiter after 1 d. In each panel, a fictitious value $J_{12}^\ast$  is used in the Newtonian signature just for illustrative and comparative purposes. Indeed, it is suitably tuned from time to time in order to bring the associated classical signature to the level of the nominal post-Newtonian effect of interest,  so to inspect the mutual (de)correlations of their temporal patterns more easily. Upper-left corner: post-Newtonian gravitomagnetic spin-octupole moment $\left(GSJ_2c^{-2};~J_{12}^\ast = 3.56\times 10^{-10}\right)$. Upper-right corner: post-Newtonian gravitoelectric moment $\left(GMJ_2 c^{-2};~J_{12}^\ast = 3.56\times 10^{-9}\right)$. Lower-left corner: Lense-Thirring effect $\left(GS c^{-2};~J_{12}^\ast = 6.23\times 10^{-9}\right)$. Lower-right corner: Schwarzschild $\left(GM c^{-2};~J_{12}^\ast = 1.602\times 10^{-7}\right)$. The present-day actual uncertainty in the Jovian fifth odd zonal is $\upsigma_{J_{12}} = 1.78\times 10^{-7}$ \citep[Tab.~1]{2018Natur.555..220I}. The adopted orbital configuration for the probe is $a_0 = 1.015~R,~e_0 = 0.0049,~I_0 = 50\deg,~\Omega_0 = 140\deg,~\omega_0 = 149.43\deg,~f_0 = 228.32\deg$ }\label{figJ12}
\end{center}
\end{figure*}
\begin{figure*}
\begin{center}
\centerline{
\vbox{
\begin{tabular}{cc}
\epsfysize= 5.4 cm\epsfbox{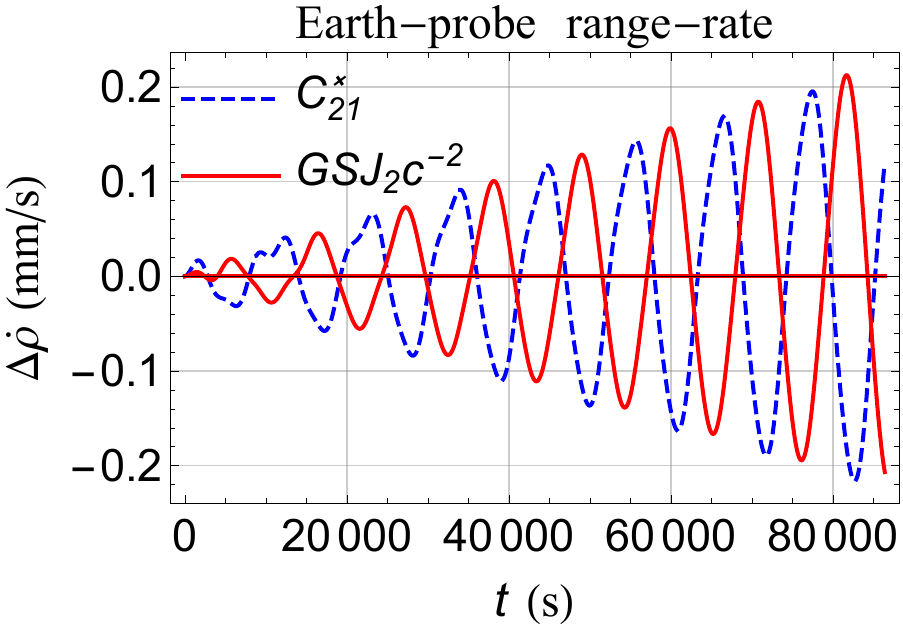}&\epsfysize= 5.4 cm\epsfbox{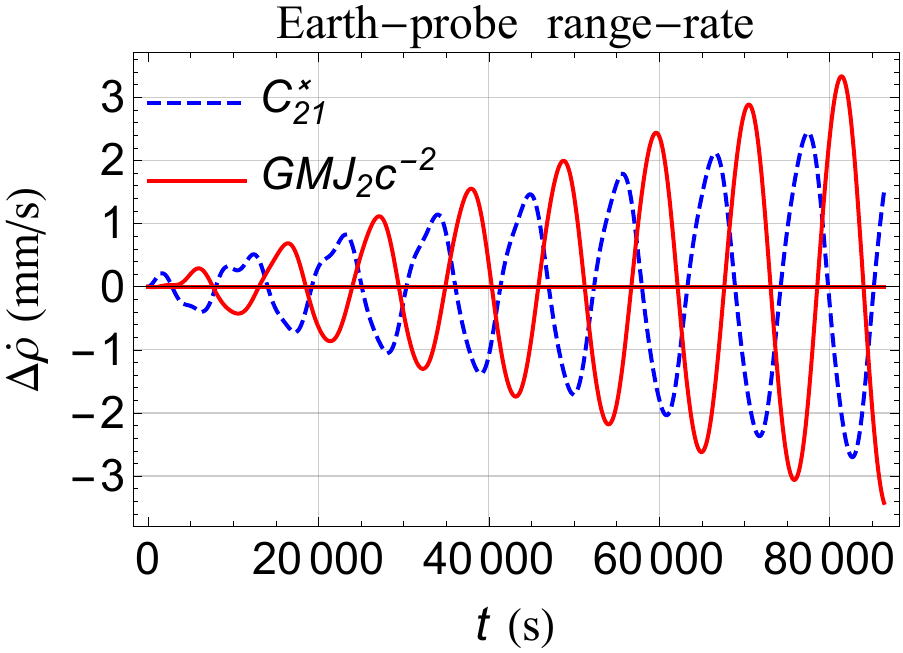}\\
\epsfysize= 5.4 cm\epsfbox{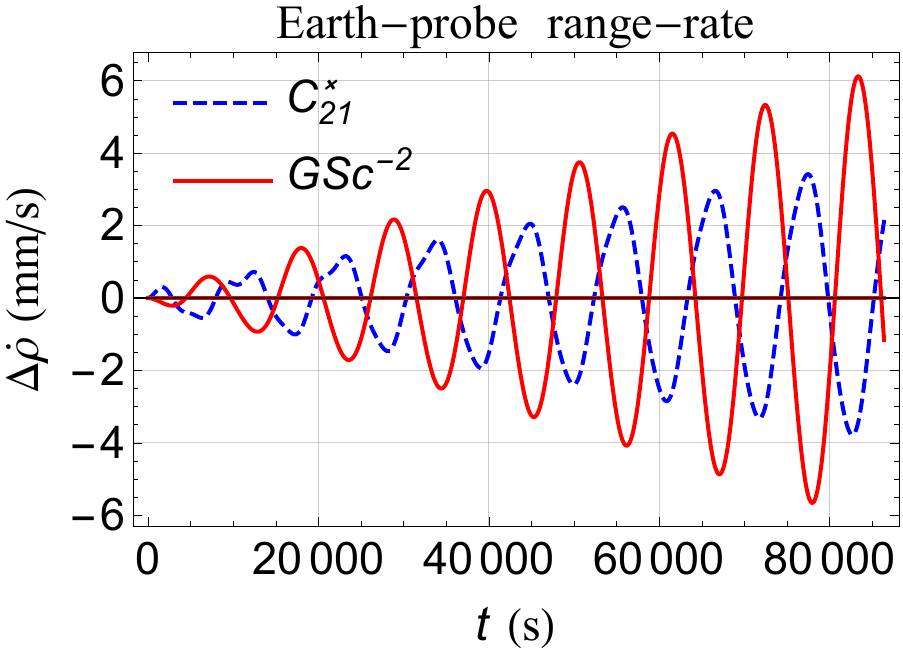}&\epsfysize= 5.4 cm\epsfbox{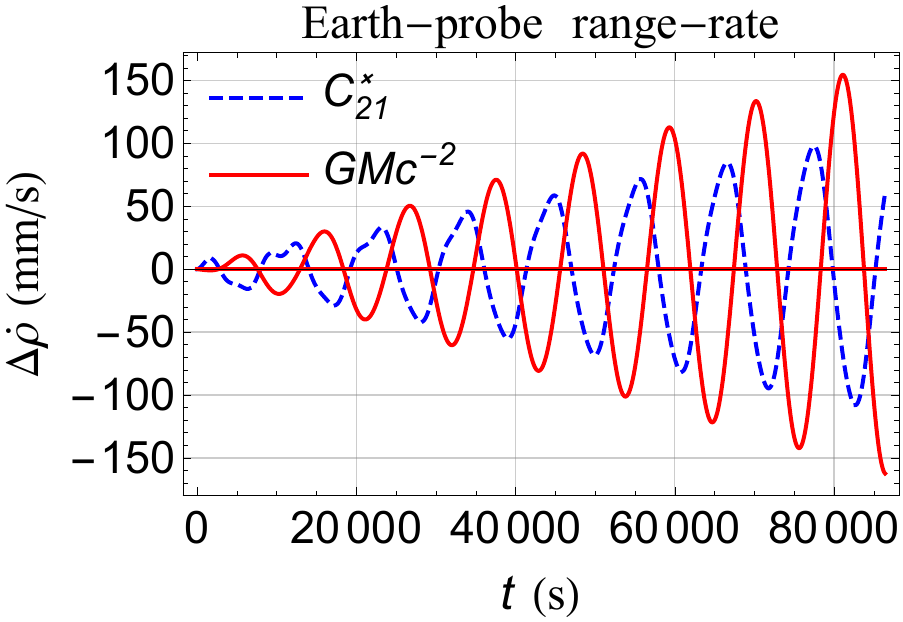}\\
\end{tabular}
}
}
\caption{Simulated range-rate signatures $\Delta\dot\rho$, in $\textrm{mm~s}^{-1}$, of a hypothetical Jovian orbiter  induced by the nominal post-Newtonian accelerations considered in the text and by the Newtonian tesseral coefficient $C_{2,1}$ of Jupiter after 1 d. In each panel, a fictitious value $C_{2,1}^\ast$  is used in the Newtonian signature just for illustrative and comparative purposes. Indeed, it is suitably tuned from time to time in order to bring the associated classical signature to the level of the nominal post-Newtonian effect of interest,  so to inspect the mutual (de)correlations of their temporal patterns more easily. Upper-left corner: post-Newtonian gravitomagnetic spin-octupole moment $\left(GSJ_2c^{-2};~C_{2,1}^\ast = 3.0\times 10^{-10}\right)$. Upper-right corner: post-Newtonian gravitoelectric moment $\left(GMJ_2 c^{-2};~C_{2,1}^\ast = 3.75\times 10^{-9}\right)$. Lower-left corner: Lense-Thirring effect $\left(GS c^{-2};~C_{2,1}^\ast = 5.25\times 10^{-9}\right)$. Lower-right corner: Schwarzschild $\left(GM c^{-2};~C_{2,1}^\ast = 1.5\times 10^{-7}\right)$. The present-day actual uncertainty in the Jovian tessreral coefficient is $\upsigma_{C_{2,1}} = 1.5\times 10^{-8}$ \citep[Tab.~1]{2018Natur.555..220I}. The adopted orbital configuration for the probe is $a_0 = 1.015~R,~e_0 = 0.0049,~I_0 = 50\deg,~\Omega_0 = 140\deg,~\omega_0 = 149.43\deg,~f_0 = 228.32\deg$ }\label{figC21}
\end{center}
\end{figure*}
\begin{figure*}
\begin{center}
\centerline{
\vbox{
\begin{tabular}{cc}
\epsfysize= 5.4 cm\epsfbox{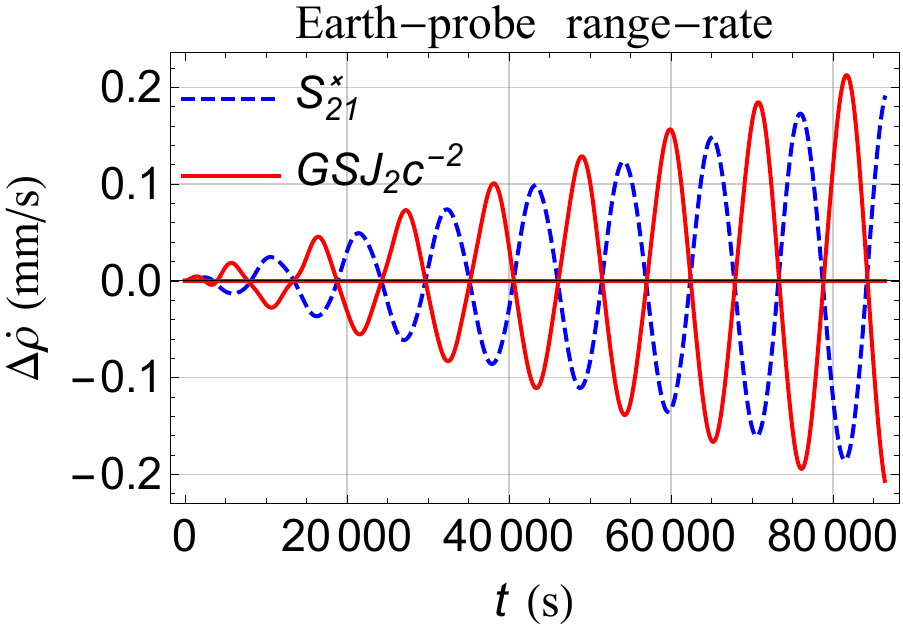}&\epsfysize= 5.4 cm\epsfbox{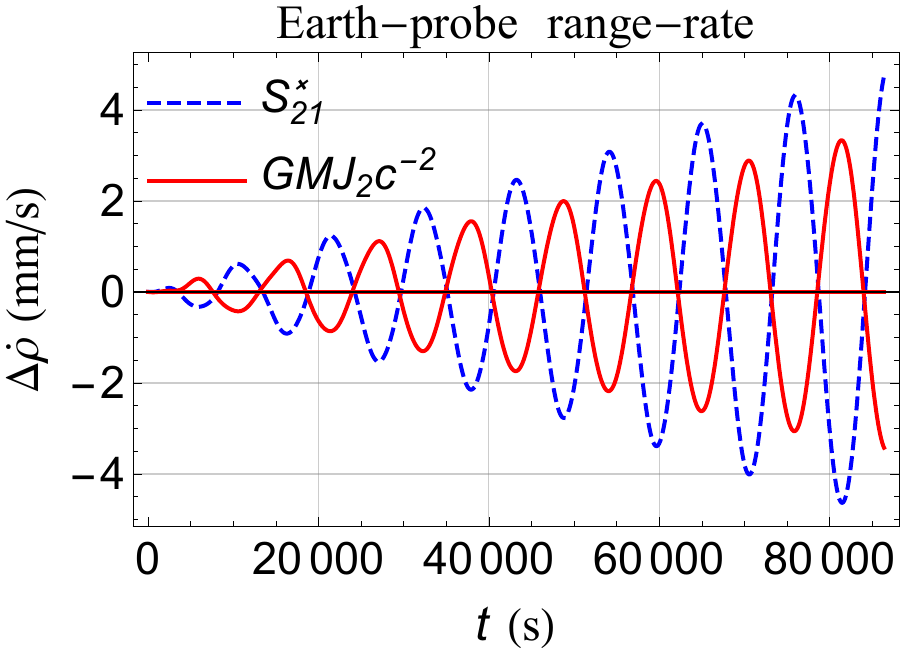}\\
\epsfysize= 5.4 cm\epsfbox{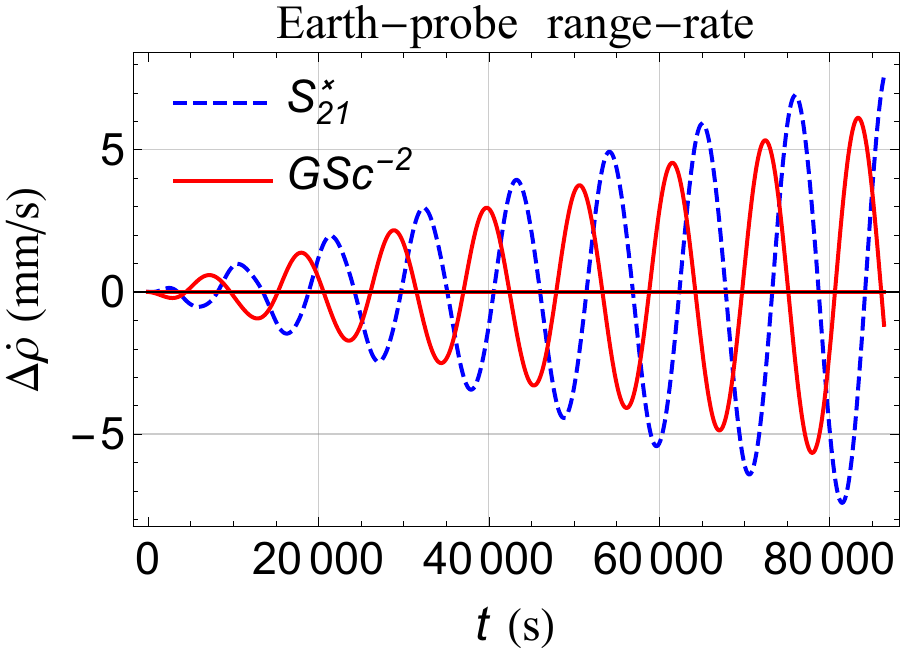}&\epsfysize= 5.4 cm\epsfbox{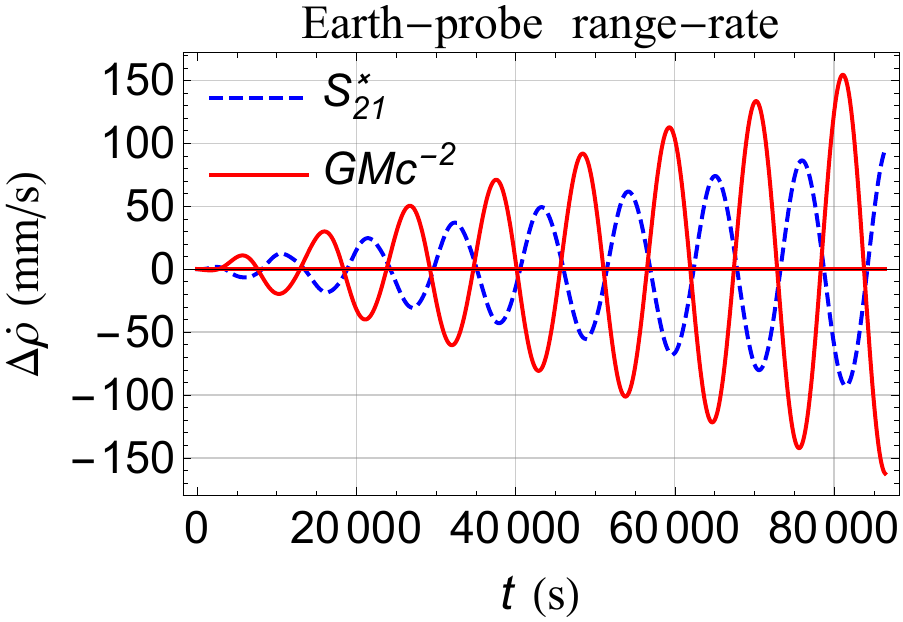}\\
\end{tabular}
}
}
\caption{Simulated range-rate signatures $\Delta\dot\rho$, in $\textrm{mm~s}^{-1}$, of a hypothetical Jovian orbiter  induced by the nominal post-Newtonian accelerations considered in the text and by the Newtonian tesseral coefficient $S_{2,1}$ of Jupiter after 1 d. In each panel, a fictitious value $S_{2,1}^\ast$  is used in the Newtonian signature just for illustrative and comparative purposes. Indeed, it is suitably tuned from time to time in order to bring the associated classical signature to the level of the nominal post-Newtonian effect of interest,  so to inspect the mutual (de)correlations of their temporal patterns more easily. Upper-left corner: post-Newtonian gravitomagnetic spin-octupole moment $\left(GSJ_2c^{-2};~S_{2,1}^\ast = 5.2\times 10^{-11}\right)$. Upper-right corner: post-Newtonian gravitoelectric moment $\left(GMJ_2 c^{-2};~S_{2,1}^\ast = 1.3\times 10^{-9}\right)$. Lower-left corner: Lense-Thirring effect $\left(GS c^{-2};~S_{2,1}^\ast = 2.08\times 10^{-9}\right)$. Lower-right corner: Schwarzschild $\left(GM c^{-2};~S_{2,1}^\ast = 2.6\times 10^{-8}\right)$. The present-day actual uncertainty in the Jovian tessreral coefficient is $\upsigma_{S_{2,1}} = 2.6\times 10^{-8}$ \citep[Tab.~1]{2018Natur.555..220I}. The adopted orbital configuration for the probe is $a_0 = 1.015~R,~e_0 = 0.0049,~I_0 = 50\deg,~\Omega_0 = 140\deg,~\omega_0 = 149.43\deg,~f_0 = 228.32\deg$ }\label{figS21}
\end{center}
\end{figure*}
\begin{figure*}
\begin{center}
\centerline{
\vbox{
\begin{tabular}{cc}
\epsfysize= 5.4 cm\epsfbox{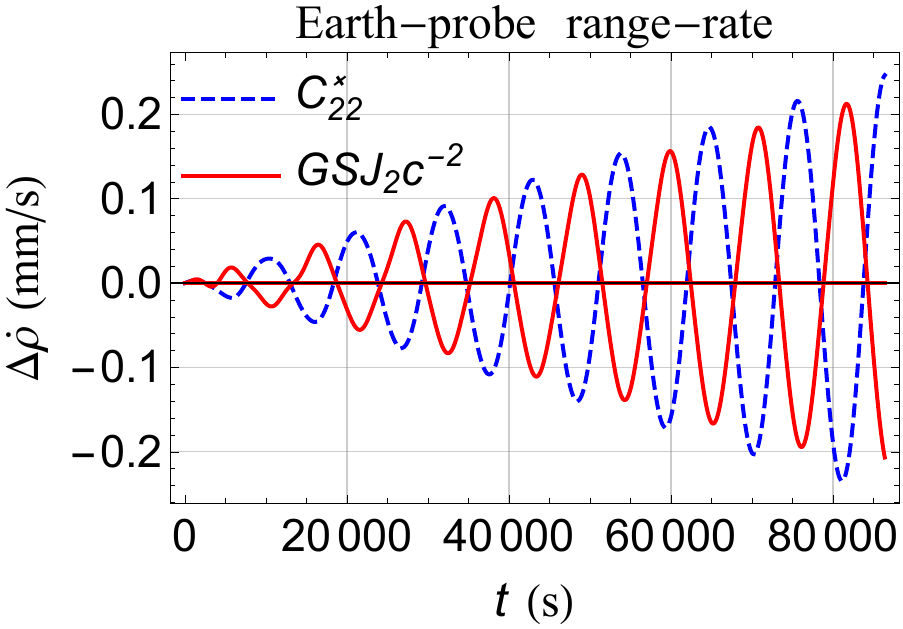}&\epsfysize= 5.4 cm\epsfbox{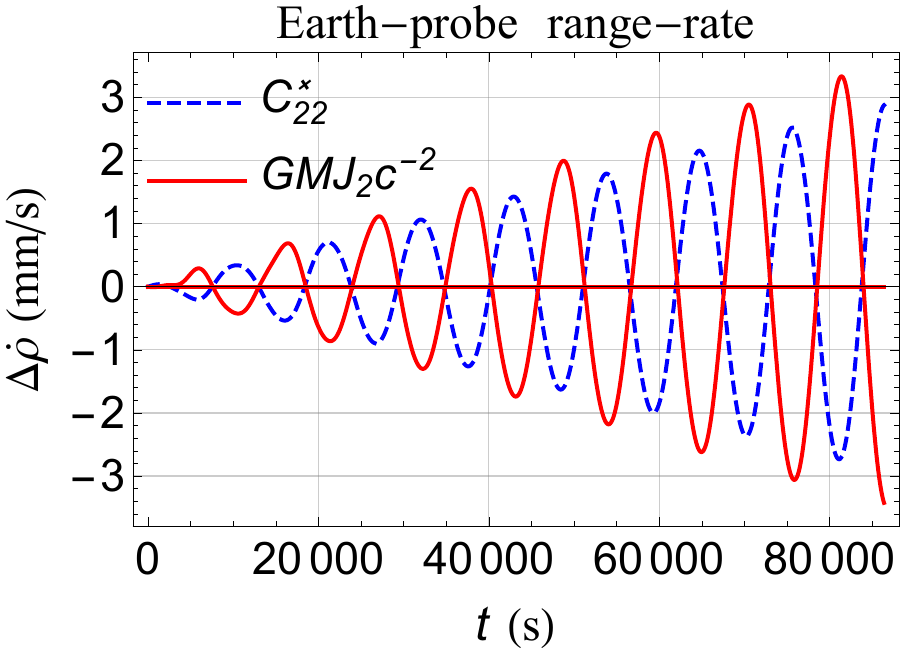}\\
\epsfysize= 5.4 cm\epsfbox{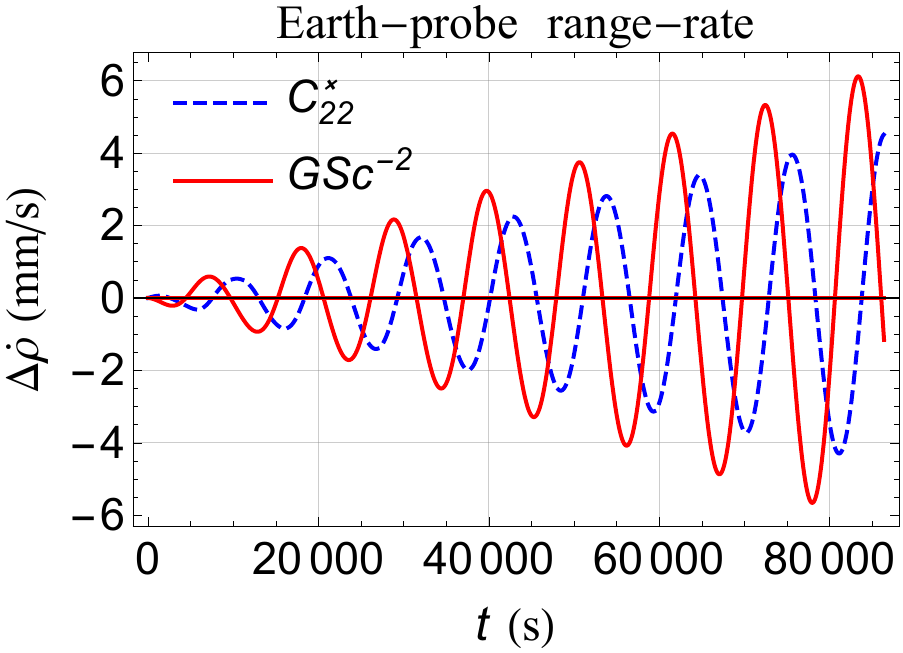}&\epsfysize= 5.4 cm\epsfbox{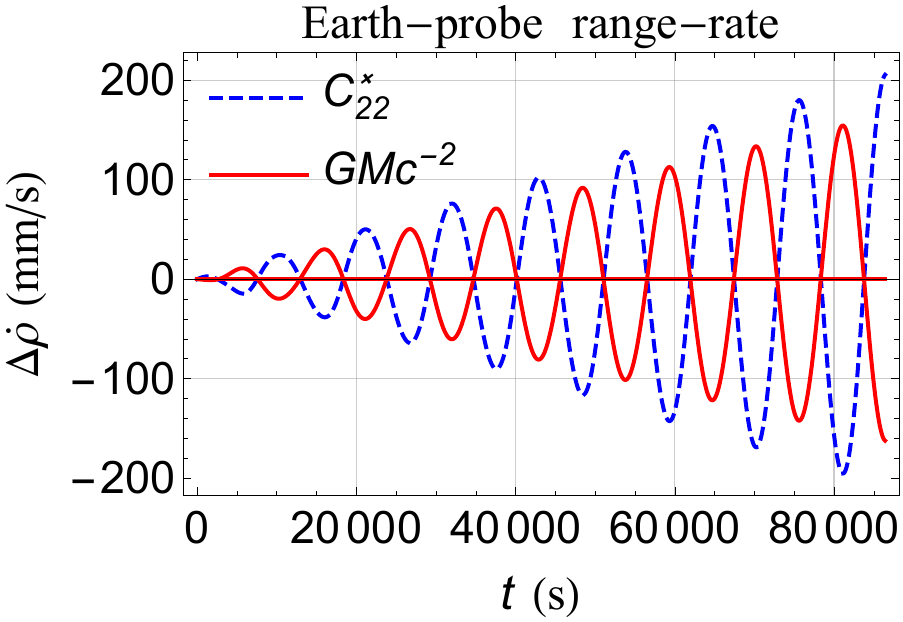}\\
\end{tabular}
}
}
\caption{Simulated range-rate signatures $\Delta\dot\rho$, in $\textrm{mm~s}^{-1}$, of a hypothetical Jovian orbiter  induced by the nominal post-Newtonian accelerations considered in the text and by the Newtonian sectorial coefficient $C_{2,2}$ of Jupiter after 1 d. In each panel, a fictitious value $C_{2,2}^\ast$  is used in the Newtonian signature just for illustrative and comparative purposes. Indeed, it is suitably tuned from time to time in order to bring the associated classical signature to the level of the nominal post-Newtonian effect of interest,  so to inspect the mutual (de)correlations of their temporal patterns more easily. Upper-left corner: post-Newtonian gravitomagnetic spin-octupole moment $\left(GSJ_2c^{-2};~C_{2,2}^\ast = 2.4\times 10^{-11}\right)$. Upper-right corner: post-Newtonian gravitoelectric moment $\left(GMJ_2 c^{-2};~C_{2,2}^\ast = 2.8\times 10^{-10}\right)$. Lower-left corner: Lense-Thirring effect $\left(GS c^{-2};~C_{2,2}^\ast = 4.4\times 10^{-10}\right)$. Lower-right corner: Schwarzschild $\left(GM c^{-2};~C_{2,2}^\ast = 2.0\times 10^{-8}\right)$. The present-day actual uncertainty in the Jovian sectorial coefficient is $\upsigma_{C_{2,2}} = 8.0\times 10^{-9}$ \citep[Tab.~1]{2018Natur.555..220I}. The adopted orbital configuration for the probe is $a_0 = 1.015~R,~e_0 = 0.0049,~I_0 = 50\deg,~\Omega_0 = 140\deg,~\omega_0 = 149.43\deg,~f_0 = 228.32\deg$ }\label{figC22}
\end{center}
\end{figure*}
\begin{figure*}
\begin{center}
\centerline{
\vbox{
\begin{tabular}{cc}
\epsfysize= 5.4 cm\epsfbox{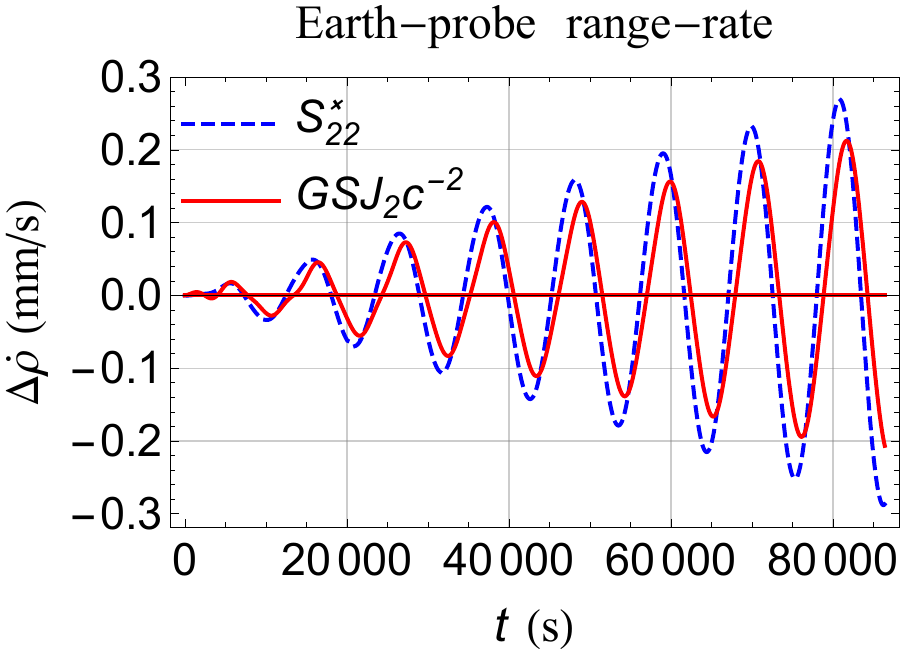}&\epsfysize= 5.4 cm\epsfbox{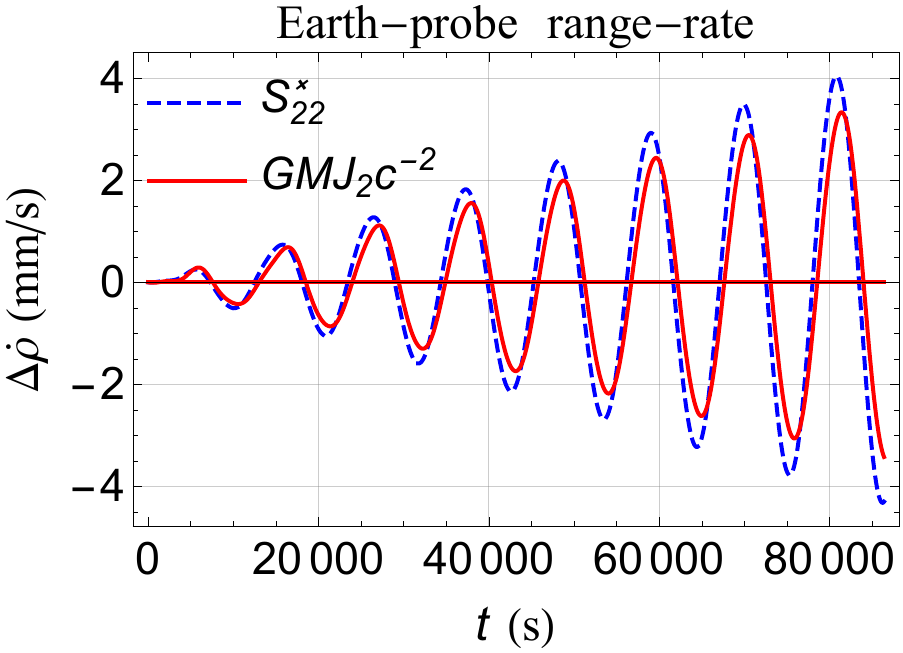}\\
\epsfysize= 5.4 cm\epsfbox{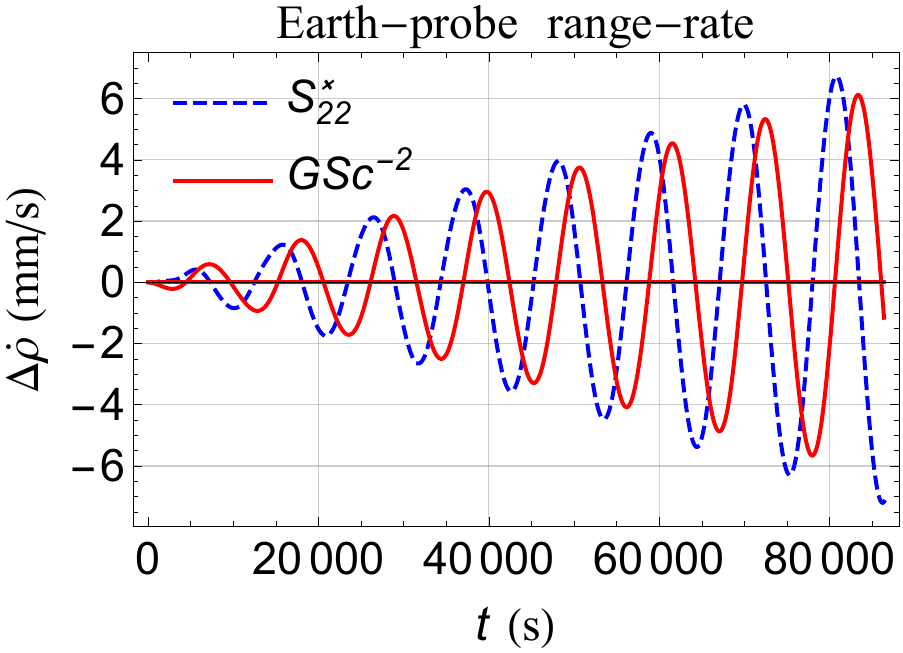}&\epsfysize= 5.4 cm\epsfbox{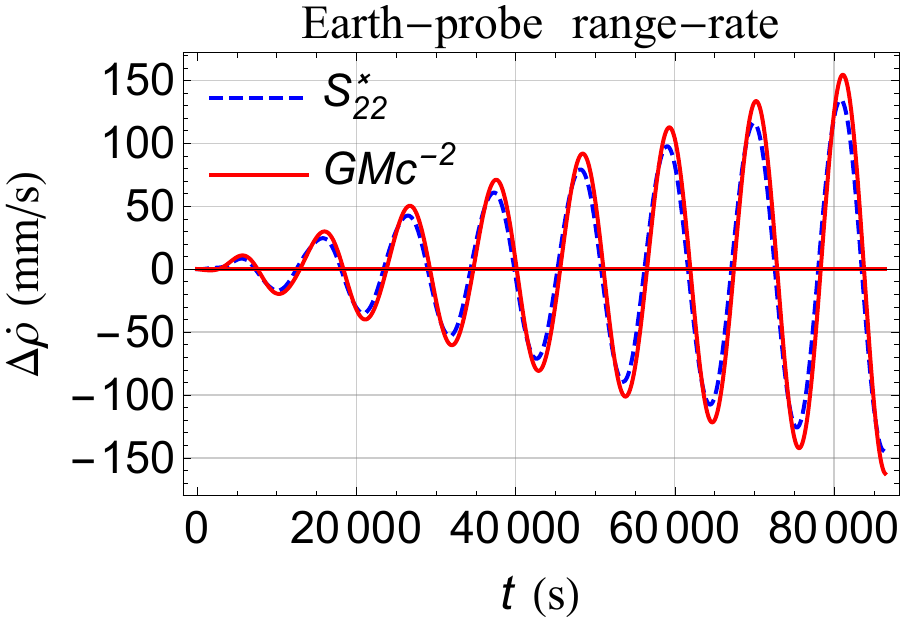}\\
\end{tabular}
}
}
\caption{Simulated range-rate signatures $\Delta\dot\rho$, in $\textrm{mm~s}^{-1}$, of a hypothetical Jovian orbiter  induced by the nominal post-Newtonian accelerations considered in the text and by the Newtonian sectorial coefficient $S_{2,2}$ of Jupiter after 1 d. In each panel, a fictitious value $S_{2,2}^\ast$  is used in the Newtonian signature just for illustrative and comparative purposes. Indeed, it is suitably tuned from time to time in order to bring the associated classical signature to the level of the nominal post-Newtonian effect of interest,  so to inspect the mutual (de)correlations of their temporal patterns more easily. Upper-left corner: post-Newtonian gravitomagnetic spin-octupole moment $\left(GSJ_2c^{-2};~S_{2,2}^\ast = 2.2\times 10^{-11}\right)$. Upper-right corner: post-Newtonian gravitoelectric moment $\left(GMJ_2 c^{-2};~S_{2,2}^\ast = 3.3\times 10^{-10}\right)$. Lower-left corner: Lense-Thirring effect $\left(GS c^{-2};~S_{2,2}^\ast = 5.5\times 10^{-10}\right)$. Lower-right corner: Schwarzschild $\left(GM c^{-2};~S_{2,2}^\ast = 1.1\times 10^{-8}\right)$. The present-day actual uncertainty in the Jovian sectorial coefficient is $\upsigma_{S_{2,2}} = 1.1\times 10^{-8}$ \citep[Tab.~1]{2018Natur.555..220I}. The adopted orbital configuration for the probe is $a_0 = 1.015~R,~e_0 = 0.0049,~I_0 = 50\deg,~\Omega_0 = 140\deg,~\omega_0 = 149.43\deg,~f_0 = 228.32\deg$ }\label{figS22}
\end{center}
\end{figure*}
\begin{figure*}
\begin{center}
\centerline{
\vbox{
\begin{tabular}{cc}
\epsfysize= 5.3 cm\epsfbox{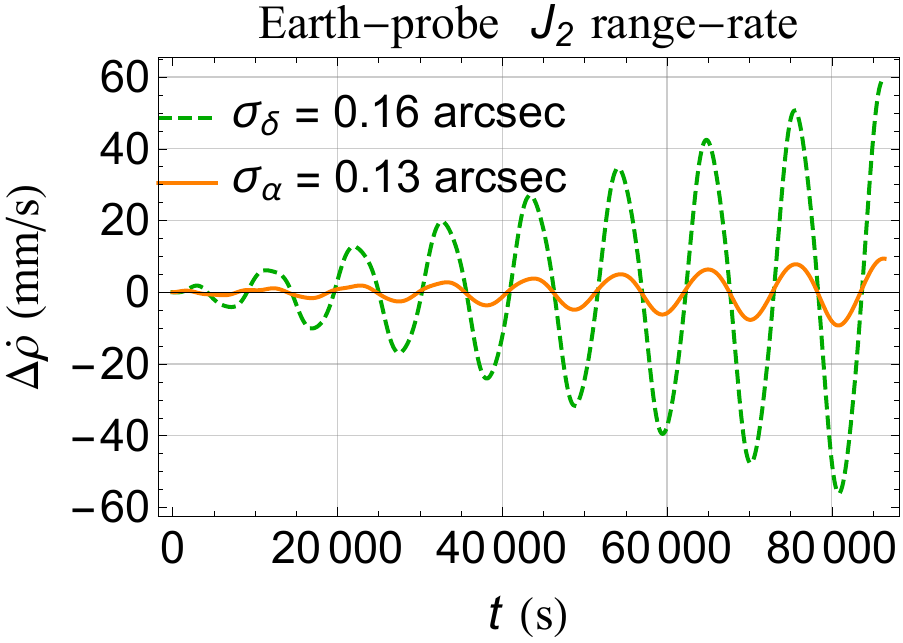}&\epsfysize= 5.3 cm\epsfbox{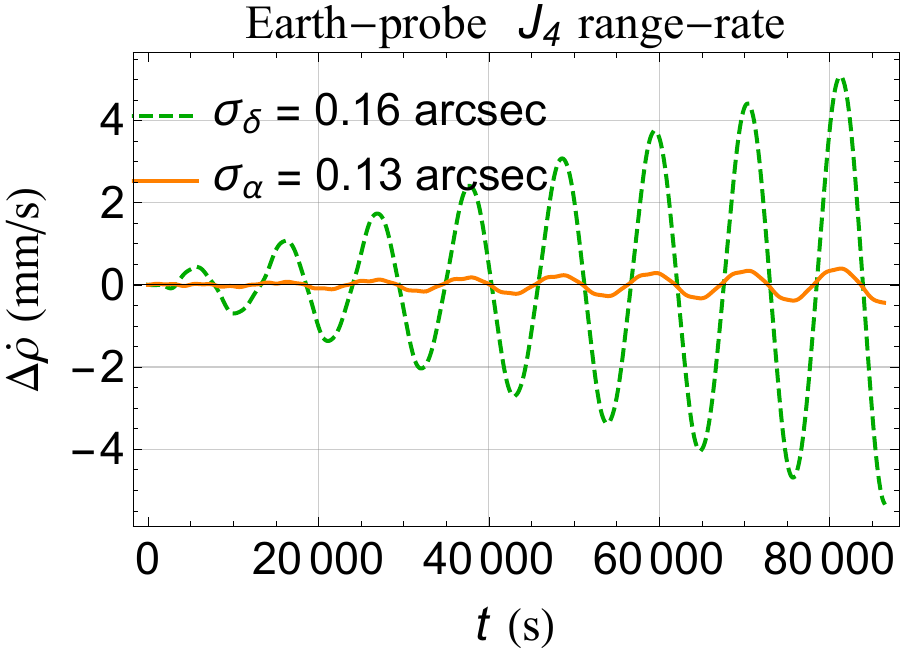}\\
\epsfysize= 5.3 cm\epsfbox{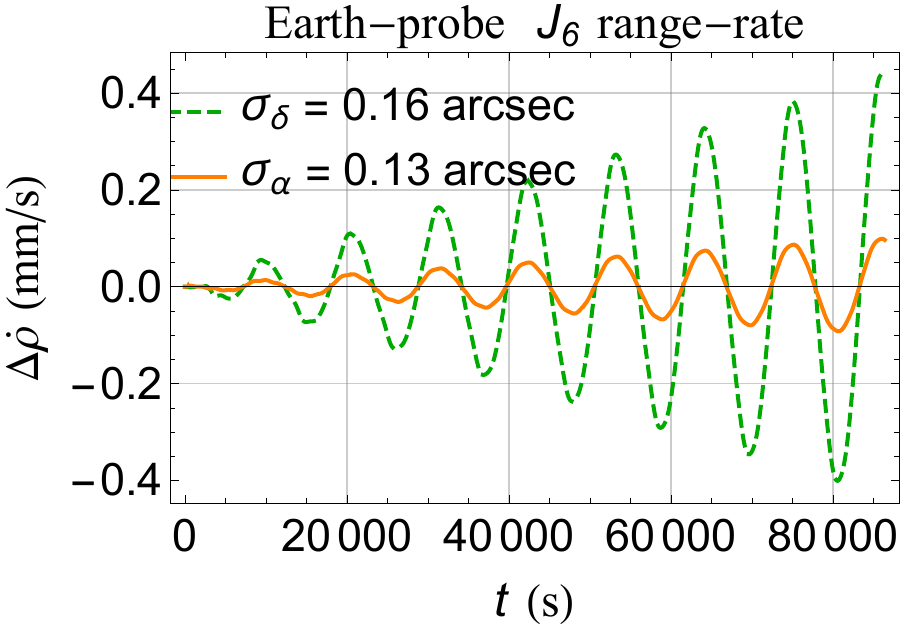}&\epsfysize= 5.3 cm\epsfbox{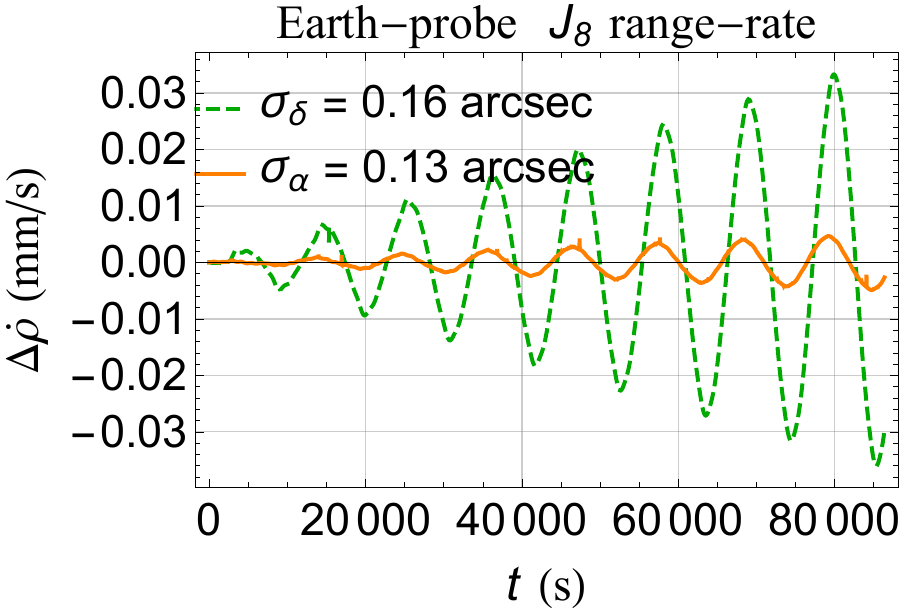}\\
\end{tabular}
}
}
\caption{Numerically simulated impact of the present-day errors $\upsigma_\alpha = 0.13~\textrm{arcsec},~\upsigma_\delta = 0.16~\textrm{arcsec}$ \citep{2018EGUGA..20.9150D} in the position of the spin axis of Jupiter on the range-rate signatures $\Delta\dot\rho$, in $\textrm{mm~s}^{-1}$, of a hypothetical Jovian orbiter  induced by the Newtonian accelerations due to the first four even zonals $J_2,~J_4,~J_6,~J_8$ after 1 d. It turns out that the uncertainties in the Jupiter's spin axis affect the odd zonals signatures in a completely negligible way. The adopted orbital configuration for the probe is $a_0 = 1.015~R,~e_0 = 0.0049,~I_0 = 50\deg,~\Omega_0 = 140\deg,~\omega_0 = 149.43\deg,~f_0 = 228.32\deg$ }\label{figRADEC}
\end{center}
\end{figure*}
\end{appendices}
\clearpage
\section*{Erratum:The post-Newtonian gravitomagnetic spin-octupole moment of an oblate rotating body and its effects on an orbiting test particle; are they measurable in the Solar system?}\lb{errazzo}
\addcontentsline{toc}{section}{\nameref{errazzo}}
In the published version \citep{2019MNRAS.484.4811I} of this paper,
%the article Lorenzo Iorio, The post-Newtonian gravitomagnetic spin-octupole moment of an oblate rotating body and its effects on an orbiting test particle; are they measurable in the Solar system?, Monthly Notices of the Royal Astronomical Society, Volume 484, Issue 4, 21 April 2019, Pages 4811-4832,%
due to the unfortunate and misleading definition\footnote{It comes from equation 27 of \citet{2014CQGra..31x5012P} for $n=1$. It reproduces incorrectly equation (56) of \citet{2003AJ....125.1580K} which, in fact, contains $(-1)^{n+1}$ instead of $(-1)^n$ entering equation 27 of \citet{2014CQGra..31x5012P}. However, it is the post-Newtonian spin-octupole acceleration relying upon $\phi_\textrm{gm}$ of equation (32) of \citet{2014CQGra..31x5012P} that matters; it is independent of all such unnecessary definitions.} $J_2 = -\varepsilon^2/5$ of \rfr{stronza}, the dimensionless quadrupole-type parameter $J_2$ entering the analytically computed post-Newtonian spin-octupole orbital precessions of \rfrs{adot}{doto} differs from the even zonal harmonic $J_2$ usually determined in standard spacecraft-based geodetic and geophysical data reductions which is, indeed, positive. Moreover, also their magnitudes are different, as can be straightforwardly noted in the case of Jupiter. Indeed,  as per the IAU 2015 Resolution B3 on Recommended Nominal Conversion Constants for Selected Solar and Planetary Properties available on the Internet at https://www.iau.org/administration/resolutions/general$\_$assemblies/, the nominal polar and equatorial radii of Jupiter amount to $66,854~\textrm{km}$ and $71,492~\textrm{km}$, respectively, yielding $-\varepsilon^2/5 = -0.0251$. Its size is larger than the Juno-based positive value $J_2=0.0147$, listed in Table~\ref{tavolaJup}, by a factor of $1.7084$. Actually, the Juno-based, positive value  $J_2=0.0147$ of Table~\ref{tavolaJup} was erroneously used in calculating the gravitomagnetic precessions $\dot e_\textrm{gm},~\dot I_\textrm{gm},~\dot\Omega_\textrm{gm},~\dot\omega_\textrm{gm}$ quoted in Tables~\ref{tavolaJuno}~to~\ref{tavola2} and mentioned throughout the paper, and in producing\footnote{In case of Figure~\ref{fig0}, it is not relevant since its purpose  was just confirming the analytical calculation of \rfrs{adot}{doto} with a numerical integration of the equations of motion in the case of a fictitious astronomical scenario: the numerical values actually adopted for the primary's physical properties are unimportant. } Figs~\ref{fig0}~to~\ref{fig3} and the post-Newtonian spin-octupole curves in the upper-left panels of
Figs~\ref{figJ2}~to~\ref{figS22} instead of $-\varepsilon^2/5$. As a consequence,  there is a minus sign mistake in $\dot e_\textrm{gm},~\dot I_\textrm{gm},~\dot\Omega_\textrm{gm},~\dot\omega_\textrm{gm}$ of Tables~\ref{tavolaJuno}~to~\ref{tavola2}, and in the post-Newtonian gravitomagnetic spin-octupole signatures displayed in  Figs~\ref{fig0}~to~\ref{fig3} and in the upper-left panels of Figs~\ref{figJ2}~to~\ref{figS22}. Moreover, the magnitudes of  $\dot e_\textrm{gm},~\dot I_\textrm{gm},~\dot\Omega_\textrm{gm},~\dot\omega_\textrm{gm}$ in Tables~\ref{tavolaJuno}~to~\ref{tavola2}, and the amplitudes in  Figs~\ref{fig0}~to~\ref{fig3} and of both the Newtonian and post-Newtonian curves in the upper-left panels of Figs~\ref{figJ2}~to~\ref{figS22} are smaller than the correct ones by a factor of $1.7084$. As such, all the curves of the post-Newtonian spin-octupole effect in Figs~\ref{fig0}~to~\ref{fig3} and in the upper-left panels of Figs~\ref{figJ2}~to~\ref{figS22} should be flipped, and  their sizes, along with those of the Newtonian signatures in the upper-left panels of
Figs~\ref{figJ2}~to~\ref{figS22}, rescaled by a factor of $1.7084$. Luckily, it strengthens our conclusions since it increases the signal-to-noise ratio of the post-Newtonian spin-octupole effect. The mutual (de)correlations of the post-Newtonian spin-octupole signatures with the classical ones in the upper-left panels of Figs~\ref{figJ2}~to~\ref{figS22} change accordingly. In the captions of Figs~\ref{figJ2}~to~\ref{figS22}, the values of $J_\ell^\ast,~\ell=2,3,\ldots 12$ and $C^\ast_{2,1},~C^\ast_{2,2},~S^\ast_{2,1},~S^\ast_{2,2}$ associated with the post-Newtonian spin-octupole effects should be rescaled by a factor of $1.7084$. All the figures in the second column from the left of Table~\ref{tabres} should be reduced by a factor of $1.7084$, which is a fortunate circumstance since it implies smaller improvements in our knowledge of the Jovian Newtonian multipoles to detect the post-Newtonian spin-octupole signatures.  Finally, it would likely be more correct to replace $GSJ_2c^{-2}$ with, say, $GS\varepsilon^2 c^{-2}$ throughout the paper to avoid further misunderstandings; in particular, it would be better to replace $J_2$ in the analytical precessions of \rfrs{adot}{doto} with $-\varepsilon^2/5$. The conclusions pertaining the other post-Newtonian effects remain unchanged.
\bibliography{Gclockbib,semimabib,PXbib}{}

\end{document}